\documentclass[preprint,prd,aps,showpacs,showkeys,nofootinbib]{revtex4}
\usepackage{amsmath}
\usepackage{graphicx}
\usepackage{caption}
\usepackage{subcaption}
\usepackage{makecell}
\usepackage{bm}
\usepackage{url}
\usepackage{amsmath}
\usepackage{xcolor}
\UseRawInputEncoding

\begin{document}

\title{125 GeV Higgs boson rare decays in a flavor-dependent $U(1)_F$ model}

\author{Zhan Cao$^{1,3,4}$\footnote{Contact author: zhancao2022@163.com}, Jin-Lei Yang$^{1,3,4}$\footnote{Contact author: jlyang@hbu.edu.cn}, Ti-Bin Hou$^{1,3,4}$, and Tai-Fu Feng$^{1,2,3,4}$\footnote{Contact author: fengtf@hbu.edu.cn}}

\affiliation{$^1$Department of Physics, Hebei University, Baoding, 071002, China\\
$^2$Department of Physics, Guangxi University, Nanning, 530004, China\\
	$^3$Key Laboratory of High-precision Computation and Application of Quantum Field Theory of Hebei Province, Baoding, 071002, China\\
	$^4$Research Center for Computational Physics of Hebei Province, Baoding, 071002, China}

\begin{abstract}
 In this work, we analyze the Higgs boson decay channels, specifically, $h{\rightarrow}\gamma\gamma$, $h{\rightarrow} VV^*$ (with $V=Z,W$), and $h{\rightarrow} f\bar{f}$ (for $f=b,c,\tau$) within the flavor-dependent $U(1)_F$ model (FDM). We also investigate processes induced by flavor-changing neutral currents, including the decays $\bar B \to X_s\gamma$ and $B_s^0 \to \mu^+\mu^-$, the top quark decays $t\to c h$ and $t\to u h$, and the lepton flavor-violating decays $\tau \to 3e$, $\tau \to 3\mu$, and $\mu \to 3e$. Furthermore, we incorporate the electroweak precision observables constraints via the S, T, and U parameters. Compared to the Standard Model, the scalar sector of the FDM is extended by two Higgs doublets and one Higgs singlet, which affects the 125 GeV Higgs properties significantly. Meanwhile, the decays $h{\rightarrow} Z\gamma$, $h\rightarrow MZ$, and $h{\rightarrow} M\gamma$ (where $M$ is a vector meson $(\rho,\omega,\phi,J/\Psi,\Upsilon)$ of the Standard-Model-like Higgs are studied, and we illustrate how changes in the scalar sector and the Yukawa coupling influence the signal strengths for the 125 GeV Higgs decay channels and the Higgs mass in the FDM.
\end{abstract}


\maketitle

\section{Introduction\label{sec1}}

 The ATLAS and CMS Collaborations have reported significant excess events for a new boson, which is interpreted as the neutral Higgs with mass around $125.20\pm 0.11$ GeV~\cite{ATLAS:2023oaq,ParticleDataGroup:2024cfk}. The $\mathit{CP}$ properties and couplings of the particle are also being established~\cite{ATLAS1,CMS1}. This implies that the Higgs mechanism breaking electroweak (EW) symmetry has a solid experimental cornerstone~\cite{Zhang:2013hga,deBlas:2016ojx}, although there is currently no solid evidence to suggest that experimental results have deviated from the predictions of the Standard Model (SM). The mass of the Higgs boson in the SM is a free parameter; its self-energy correction in the SM is not protected by any symmetry, leading to quadratic divergences, thus raising the issue of naturalness~\cite{Kim:2024yrk}. Although the $\mathit{CP}$ violation of SM explains some experimental results, it fails to account for the Universe's matter-antimatter asymmetry, necessitating new sources of $\mathit{CP}$ violation. The SM also lacks explanations for dark matter and neutrino oscillations~\cite{Batra:2022wsd,Steingasser:2024hqi}. These shortcomings strongly suggest new physics (NP) beyond the SM.

 An extension of the SM, known as the flavor-dependent model introduces an extra local gauge group $U(1)_F$, along with two scalar doublets, one scalar singlet, and two right-handed neutrinos~\cite{Yang:2024kfs,Yang:2024duo}. In this model, the additional $U(1)_F$ charges are related to the flavors of fermions. The enhanced fermion sector present in the FDM offers a simultaneous resolution to the puzzles of flavor mixings and fermions mass hierarchy. Moreover, utilizing the type-I seesaw mechanism, the model naturally accounts for the generation of nonzero Majorana neutrino masses~\cite{Yang:2024znv,Yang:2024duo,Canetti:2012kh,Abada:2007ux}. In FDM, significant changes have occurred in the extended scalar sector and the Yukawa coupling part, which will directly impact the signal strength~\cite{Belanger:2013xza,Aiko:2022gmz,Banerjee:2020tqc}. Therefore, studying the signal strength of the Higgs is of great significance for testing the coupling strength of the Higgs particle with various fundamental particles and determining the properties of the Higgs particle itself and can be used to verify whether models with different Higgs sectors satisfy experimental data~\cite{ATLAS:2015xst,ATLAS:2014aga,BhupalDev:2014bir}. 

Flavor-changing neutral currents (FCNCs) refer to processes in which particles exchange flavor (quark or lepton) during strong or weak interactions. In recent years, FCNCs have been extensively studied through neutrino scattering experiments and high-energy physics experiments such as those conducted at the Large Hadron Collider (LHC)~\cite{Park:2024apx}, as exemplified by $\bar B \to X_s\gamma$, $B_s^0 \to \mu^+\mu^-$~\cite{Allanach:2023uxz}; $t \to c h$, $t \to u h$~\cite{Thielmann:2023hor}; and $\tau \to 3e$, $\tau \to 3\mu$, $\mu \to 3e$. These results may indicate that the SM is incomplete and requires additional physics theories to explain such phenomena. New particles or interactions introduced in many extensions of the SM (e.g., supersymmetry, grand unified theories, extra dimensions, etc.) could enhance FCNC processes~\cite{Duy:2024txy}. Therefore, the study of FCNCs not only deepens our understanding of weak interactions within the SM, but also serves as a powerful tool for validating theories beyond the SM~\cite{Oliveira:2022vjo}.

In this paper, we investigate the 125 GeV Higgs decay channels $h{\rightarrow}\gamma\gamma$, $h{\rightarrow} Z\gamma$, $h{\rightarrow} VV^*$ ($V=Z,W$) and $h{\rightarrow} f\bar{f}$ ($f=b,c,\tau$), and we study the 125 GeV SM-like Higgs boson rare decays $h{\rightarrow} MZ$ and $h{\rightarrow} M\gamma$ in the framework of the FDM, where $M$ denotes vector meson $[\rho, \omega, \phi, J/\psi, \Upsilon(nS)]$. For the processes $h{\rightarrow} MZ$~\cite{HVZ-sm,HVZ-zhao,HVZ-sm1,HVZ-sm2,Konig:2015qat}, there are two types of decay topologies: One is the direct contributions induced by the Yukawa couplings of the Higgs boson to the quarks, and the other is the indirect contributions resulting from a $h{\rightarrow} \gamma^* Z$ transition followed by the conversion of the off-shell boson into a vector meson~\cite{Konig:2015qat,HVZ-sm,Konig:2016xhe}. Compared with the SM-like Higgs boson coupling to the quarks in the direct contributions, the effective $h\gamma Z$ vertex in the indirect contributions will give more NP contributions~\cite{dEnterria:2023wjq,CMS:2020ggo}. Hence, these rare decays provide complementary information to the decay $h{\rightarrow} Z\gamma$, both in and beyond the SM~\cite{Konig:2016xhe,Isidori:2013cla}. In this work, the quantum chromodynamics (QCD) factorization~\cite{QCD1,QCD2,QCD3} is used for the rare SM-like Higgs boson decays $h{\rightarrow} MZ$. $h{\rightarrow} \overline{q} q$ is a method for directly studying the coupling of first- and second-generation quarks, although these channels have relatively large branching fractions, while their experimental sensitivity is substantially obscured by large multijet backgrounds~\cite{CMS:2023gjz,ATLAS:2022qef}. Method $h{\rightarrow} M\gamma$ is considered superior, because compared to method $h{\rightarrow} \overline{c} c$ and $h{\rightarrow} \overline{b} b$, the process $h{\rightarrow}J/\psi + \gamma$ or $h{\rightarrow}\Upsilon(nS) + \gamma$ offers an advantage: Despite their relatively small branching fractions, these radiative decays exhibit a unique experimental signature, effectively suppressing the substantial multijet backgrounds that typically complicate searches for $h{\rightarrow} \overline{q} q$~\cite{Batra:2022wsd,Modak:2016cdm,Carlson:2021tes,Jia:2024ini}. Additionally, decays of the Higgs boson into charmonium states, specifically, $h{\rightarrow} J/\psi + \gamma$, provide a unique probe of the charm-quark Yukawa coupling, including both its magnitude and sign~\cite{Bodwin:2013gca,CMS:2024hhg}. Similarly, Higgs decays into bottomonium states, $h{\rightarrow} \Upsilon(nS) + \gamma$, offer insights into the real and imaginary components of the bottom quark's coupling to the Higgs~\cite{Coyle:2019hvs,ATLAS:2022rej}. These processes have also been used by CMS or ATLAS for probing the Yukawa couplings~\cite{CMS:2023gjz,ATLAS:2022qef,ATLAS:2023von}.

Our presentation is organized as follows. In Sec.~\ref{sec2},  the structure of the FDM including particle content, scalar sector, fermion masses, and gauge sector are collected. We present the decay widths and signal strengths for $h{\rightarrow}\gamma\gamma$, $h{\rightarrow} VV^*$ ($V=Z,W$), and $h{\rightarrow} f\bar{f}$ ($f=b, c, \tau$). We also discuss processes induced by FCNCs, the constraints from electroweak precision observables (the S, T, and U parameters), and the decays $h{\rightarrow} MZ$ and $h{\rightarrow} M\gamma$ in Sec.~\ref{sec3}. The numerical analysis are given in Sec.~\ref{sec4}, and Sec.~\ref{sec5} gives a summary.

\section{The flavor-dependent model}\label{sec2}

\begin{table*}
	\begin{tabular*}{\textwidth}{@{\extracolsep{\fill}}lllll@{}}
		\hline
		Multiplets & $SU(3)_C$ & $SU(2)_L$ & $U(1)_Y$ & $U(1)_F$\\
		\hline
		$l_1=(\nu_{1L},e_{1L})^T$ & 1 & 2 & $-\frac{1}{2}$ & $z$\\
		$l_2=(\nu_{2L},e_{2L})^T$ & 1 & 2 & $-\frac{1}{2}$ & $-z$\\
		$l_3=(\nu_{3L},e_{3L})^T$ & 1 & 2 & $-\frac{1}{2}$ & $0$\\
		$\nu_{1R}$                & 1 & 1 & $0$            & $-z$ \\
		$\nu_{2R}$                & 1 & 1 & $0$            & $z$ \\
		$e_{1R}$                  & 1 & 1 & $-1$           & $-z$ \\
		$e_{2R}$                  & 1 & 1 & $-1$           & $z$ \\
		$e_{3R}$                  & 1 & 1 & $-1$           & $0$ \\
		$q_1=(u_{1L},d_{1L})^T$   & 3 & 2 & $\frac{1}{6}$  & $z$\\
		$q_2=(u_{2L},d_{2L})^T$   & 3 & 2 & $\frac{1}{6}$  & $-z$\\
		$q_3=(u_{3L},d_{3L})^T$   & 3 & 2 & $\frac{1}{6}$  & $0$\\
		$d_{1R}$                  & 3 & 1 & -$\frac{1}{3}$ & $-z$ \\
		$d_{2R}$                  & 3 & 1 & -$\frac{1}{3}$ & $z$ \\
		$d_{3R}$                  & 3 & 1 & -$\frac{1}{3}$ & $0$ \\
		$u_{1R}$                  & 3 & 1 & $\frac{2}{3}$  & $-z$ \\
		$u_{2R}$                  & 3 & 1 & $\frac{2}{3}$  & $z$ \\
		$u_{3R}$                  & 3 & 1 & $\frac{2}{3}$  & $0$ \\
		$\Phi_1=(\phi_1^{+},\phi_1^{0})^T$   & 1 & 2 & $\frac{1}{2}$  & $z$\\
		$\Phi_2=(\phi_2^{+},\phi_2^{0})^T$   & 1 & 2 & $\frac{1}{2}$  & $-z$\\
		$\Phi_3=(\phi_3^{+},\phi_3^{0})^T$   & 1 & 2 & $\frac{1}{2}$  & 0\\
		$\chi$   & 1 & 1 & 0  & $2z$\\
		\hline
	\end{tabular*}
	\caption{Matter content in the FDM, where the nonzero constant $z$ denotes the extra $U(1)_F$ charge.}
	\label{tab2}
\end{table*}

The gauge group of the FDM is $SU(3)_C\otimes SU(2)_L\otimes U(1)_Y\otimes U(1)_F$, where the extra $U(1)_F$ local gauge group is related to the particles' flavor~\cite{Yang:2024znv}. In the FDM, the third generation of fermions obtain masses through the tree-level Yukawa couplings, and the first two generations of fermions achieve masses through the tree-level mixings with the third generation~\cite{Yang:2024kfs}. Hence, two additional scalar doublets are introduced in the FDM to realize the tree-level mixings of the first two generations and the third generation. In addition, to coincide with the observed neutrino oscillations, two right-handed neutrinos and one scalar singlet are introduced. Then the right-handed neutrinos obtain large Majorana masses after the scalar singlet achieving large vacuum expectation value, and the tiny neutrino masses can be obtained by the type-I seesaw mechanism~\cite{Yang:2024znv}.

All fields in the FDM and the corresponding gauge symmetry charges are presented in Tab.~\ref{tab2}, the nonzero constant $z$ denotes the extra $U(1)_F$ charge. It can be noted in Tab.~\ref{tab2} that there are only two generations of right-handed neutrinos in the FDM, because both $U(1)_F$ and $U(1)_Y$ charges of the third generation of right-handed neutrinos $\nu_{R_3}$ are zero, which is trivial. In addition, it is obvious that the chiral anomaly cancellation can be guaranteed for the fermionic charges presented in Tab.~\ref{tab2}.

\subsection{The scalar sector of the FDM}\label{sec2-1}

The scalar potential in the FDM can be written as~\cite{Yang:2024znv,Yang:2024kfs}
\begin{eqnarray}
	&&V=-M_{\Phi_1}^2 \Phi_1^\dagger\Phi_1-M_{\Phi_2}^2 \Phi_2^\dagger\Phi_2-M_{\Phi_3}^2 \Phi_3^\dagger\Phi_3-M_{\chi}^2\chi^*\chi+\lambda_{\chi} (\chi^*\chi)^2+\lambda_1 (\Phi_1^\dagger\Phi_1)^2\nonumber\\
	&&\qquad+\lambda_2 (\Phi_2^\dagger\Phi_2)^2+\lambda_3 (\Phi_3^\dagger\Phi_3)^2+\lambda'_4 (\Phi_1^\dagger\Phi_1)(\Phi_2^\dagger\Phi_2)+\lambda_4'' (\Phi_1^\dagger\Phi_2)(\Phi_2^\dagger\Phi_1)\nonumber\\
	&&\qquad+\lambda_5' (\Phi_1^\dagger\Phi_1)(\Phi_3^\dagger\Phi_3)+\lambda_5'' (\Phi_1^\dagger\Phi_3)(\Phi_3^\dagger\Phi_1)+\lambda_6' (\Phi_2^\dagger\Phi_2)(\Phi_3^\dagger\Phi_3)+\lambda_6'' (\Phi_2^\dagger\Phi_3)(\Phi_3^\dagger\Phi_2)\nonumber\\
	&&\qquad+\lambda_7 (\Phi_1^\dagger\Phi_1)(\chi^*\chi)+\lambda_{8} (\Phi_2^\dagger\Phi_2)(\chi^*\chi)+\lambda_{9} (\Phi_3^\dagger\Phi_3)(\chi^*\chi)+[\lambda_{10} (\Phi_3^\dagger\Phi_1)(\Phi_3^\dagger\Phi_2)\nonumber\\
	&&\qquad+\kappa(\Phi_1^\dagger\Phi_2)\chi+H.c.],\label{eqsca}
\end{eqnarray}
where
\begin{align}
	\Phi_1&=\left(\begin{array}{c}\phi_1^+\\ \frac{1}{\sqrt2}(i A_1+S_1+v_1)\end{array}\right), &
	\Phi_2&=\left(\begin{array}{c}\phi_2^+\\ \frac{1}{\sqrt2}(i A_2+S_2+v_2)\end{array}\right), \nonumber\\
	\Phi_3&=\left(\begin{array}{c}\phi_3^+\\ \frac{1}{\sqrt2}(i A_3+S_3+v_3)\end{array}\right), &
	\chi&=\frac{1}{\sqrt2}(i A_{\chi}+S_{\chi}+v_\chi),
\end{align}
and $v_i\;(i=1,\;2,\;3)$ and $\;v_\chi$ are the VEVs of $\Phi_i,\;$ and $\chi$ respectively.

Based on the scalar potential in Eq.~(\ref{eqsca}), the tadpole equations in the FDM can be expressed as
\begin{eqnarray}
	&&M_{\Phi_1}^2=\lambda_1v_1^2+\frac{1}{2}\Big[(\lambda_4'+\lambda_4'') v_2^2+(\lambda_5'+\lambda_5'') v_3^2+\frac{v_2}{v_1}v_3^2 {\rm Re}(\lambda_{10})+\sqrt2\frac{v_2}{v_1} v_\chi {\rm Re}(\kappa)+\lambda_7v_\chi^2\Big],\nonumber\\
	&&M_{\Phi_2}^2=\lambda_2v_2^2+\frac{1}{2}\Big[(\lambda_4'+\lambda_4'') v_1^2+(\lambda_6'+\lambda_6'') v_3^2+\frac{v_1}{v_2}v_3^2 {\rm Re}(\lambda_{10})+\sqrt2\frac{v_1}{v_2} v_\chi {\rm Re}(\kappa)+\lambda_8v_\chi^2\Big],\nonumber\\
	&&M_{\Phi_3}^2=\lambda_3v_3^2+{\rm Re}(\lambda_{10})v_1v_2+\frac{1}{2}[(\lambda_5'+\lambda_5'') v_1^2+(\lambda_6'+\lambda_6'') v_2^2+\lambda_9 v_c^2],\nonumber\\
	&&M_{\chi}^2=\lambda_\chi v_\chi^2+\frac{1}{2}\Big[\lambda_7 v_1^2+\lambda_8 v_2^2+\lambda_9 v_3^2+\sqrt2\frac{v_1v_2}{v_\chi}  {\rm Re}(\kappa)\Big].\label{eqtad}
\end{eqnarray}

On the basis $(S_1,\;S_2,\;S_3,\;S_\chi)$, the $\mathit{CP}$-even Higgs squared mass matrix in the FDM is
\begin{eqnarray}
	&&M_{h}^2=\left(\begin{array}{*{20}{cccc}}
		M_{h,11}^2 & M_{h,12}^2 & M_{h,13}^2 & M_{h,14}^2 \\ [6pt]
		M_{h,12}^2 & M_{h,22}^2 & M_{h,23}^2 & M_{h,24}^2 \\ [6pt]
		M_{h,13}^2 & M_{h,23}^2 & M_{h,33}^2 & M_{h,34}^2 \\ [6pt]
		M_{h,14}^2 & M_{h,24}^2 & M_{h,34}^2 & M_{h,44}^2 \\ [6pt]
	\end{array}\right),\label{eq4h}
\end{eqnarray}
where
\begin{eqnarray}
	&&M_{h,11}^2=2\lambda_1v_1^2-\frac{v_2}{2v_1}\Big[v_3^2 {\rm Re}(\lambda_{10})+\sqrt2 v_\chi {\rm Re}(\kappa)\Big],\nonumber\\
	&&M_{h,22}^2=2\lambda_2v_2^2-\frac{v_1}{2v_2}\Big[v_3^2 {\rm Re}(\lambda_{10})+\sqrt2 v_\chi {\rm Re}(\kappa)\Big],\nonumber\\
	&&M_{h,33}^2=2\lambda_3v_3^2,\;\;M_{h,44}^2=2\lambda_\chi v_\chi^2-\frac{\sqrt2v_1v_2}{2v_\chi}  {\rm Re}(\kappa),\nonumber\\
	&&M_{h,12}^2=(\lambda_4'+\lambda_4'') v_1v_2+\frac{1}{2}{\rm Re}(\lambda_{10})v_3^2+\frac{\sqrt2}{2}{\rm Re}(\kappa)v_\chi,\nonumber\\
	&&M_{h,13}^2=(\lambda_5'+\lambda_5'')v_1v_3+{\rm Re}(\lambda_{10})v_2v_3,\;\;M_{h,14}^2=\lambda_7v_1v_\chi+\frac{\sqrt2}{2}{\rm Re}(\kappa)v_2,\nonumber\\
	&&M_{h,23}^2=(\lambda_6'+\lambda_6'')v_2v_3+{\rm Re}(\lambda_{10})v_1v_3,\;\;M_{h,24}^2=\lambda_8v_2v_\chi+\frac{\sqrt2}{2}{\rm Re}(\kappa)v_1,\nonumber\\
	&&M_{h,34}^2=\lambda_9v_3v_\chi.\label{eqmh}
\end{eqnarray}

The tadpole equations in Eq.~(\ref{eqtad}) are used to obtain the matrix elements above.

On the basis $(\phi_1^+,\;\phi_2^+,\;\phi_3^+)$ and $(\phi_1^-,\;\phi_2^-,\;\phi_3^-)^T$, the squared mass matrix of singly charged Higgs in the FDM can be written as
\begin{eqnarray}
	&&M_{H^\pm}^2=\left(\begin{array}{*{20}{ccc}}
		M_{H^\pm,11}^2 & M_{H^\pm,12}^2 & M_{H^\pm,13}^2 \\ [6pt]
		(M_{H^\pm,12}^2)^* & M_{H^\pm,22}^2 & M_{H^\pm,23}^2 \\ [6pt]
		(M_{H^\pm,13}^2)^* & (M_{H^\pm,23}^2)^* & M_{H^\pm,33}^2 \\ [6pt]
	\end{array}\right),
\end{eqnarray}
where
\begin{eqnarray}
	&&M_{H^\pm,11}^2=-\frac{v_2}{2v_1}[{\rm Re}(\lambda_{10})v_3^2+\sqrt2 v_\chi {\rm Re} (\kappa)]-\frac{1}{2}(\lambda_4'' v_2^2+\lambda_5'' v_3^2),\nonumber\\
	&&M_{H^\pm,22}^2=-\frac{v_1}{2v_2}[{\rm Re}(\lambda_{10})v_3^2+\sqrt2 v_\chi {\rm Re} (\kappa)]-\frac{1}{2}(\lambda_4'' v_1^2+\lambda_6'' v_3^2),\nonumber\\
	&&M_{H^\pm,33}^2=-{\rm Re}(\lambda_{10})v_1v_2-\frac{1}{2}(\lambda_5'' v_1^2+\lambda_6'' v_2^2),\nonumber\\
	&&M_{H^\pm,12}^2=\frac{\sqrt 2}{2}v_\chi\kappa+\frac{1}{2}\lambda_4'' v_1 v_2,\;\;M_{H^\pm,13}^2=\frac{1}{2}v_3(\lambda_5'' v_1+\lambda_{10}^*v_2),\nonumber\\
	&&M_{H^\pm,23}^2=\frac{1}{2}v_3(\lambda_6'' v_2+\lambda_{10}^*v_1).\label{eqmCH}
\end{eqnarray}
It is easy to verify that there are two neutral Goldstones and one singly charged Goldstone in the FDM.

\subsection{The fermion masses in the FDM}\label{sec2-2}
Based on the matter content listed in Tab.~\ref{tab2}, the Yukawa couplings in the FDM can be written as
\begin{eqnarray}
	&&\mathcal{L}_Y=Y_u^{33}\bar q_3 \tilde \Phi_3 u_{R_3}+Y_d^{33}\bar q_3 \Phi_3 d_{R_3}+Y_u^{32}\bar q_3 \tilde{\Phi}_1 u_{R_2}+Y_u^{23}\bar q_2 \tilde \Phi_1 u_{R_3}+Y_d^{32}\bar q_3 \Phi_2 d_{R_2}\nonumber\\
	&&\qquad\; +Y_d^{23}\bar q_2 \Phi_2 d_{R_3}+Y_u^{21}\bar q_2 \tilde{\Phi}_3 u_{R_1}+Y_u^{12}\bar q_1 \tilde \Phi_3 u_{R_2}+Y_d^{21}\bar q_2 \Phi_3 d_{R_1}+ Y_d^{12}\bar q_1 \Phi_3 d_{R_2}\nonumber\\
	&&\qquad\; +Y_u^{31}\bar q_3 \tilde{\Phi}_2 u_{R_1}+Y_u^{13}\bar q_1 \tilde \Phi_2 u_{R_3}+Y_d^{31}\bar q_3 \Phi_1 d_{R_1}+Y_d^{13}\bar q_1 \Phi_1 d_{R_3}\nonumber\\
	&&\qquad\; +Y_e^{33}\bar l_3 \Phi_3 e_{R_3}+Y_e^{32}\bar l_3 \Phi_2 e_{R_2}+Y_e^{23}\bar l_2 \Phi_2 e_{R_3}+Y_e^{21}\bar l_2 \Phi_3 e_{R_1}+ Y_e^{12}\bar l_1 \Phi_3 e_{R_2}\nonumber\\
	&&\qquad\; +Y_e^{31}\bar l_3 \Phi_1 e_{R_1}+Y_e^{13}\bar l_1 \Phi_1 e_{R_3}+Y_R^{11}\bar\nu^c_{R_1}\nu_{R_1}\chi+Y_R^{22}\bar\nu^c_{R_2}\nu_{R_2} \chi^*+Y_D^{21}\bar l_2 \tilde \Phi_3 \nu_{R_1}\nonumber\\
	&&\qquad\; +Y_D^{12}\bar l_1 \tilde \Phi_3 \nu_{R_2}+Y_D^{31}\bar l_3 \tilde \Phi_2 \nu_{R_1}+Y_D^{32}\bar l_3 \tilde \Phi_1 \nu_{R_2}+H.c..\label{eq9}
\end{eqnarray}
Then, the mass matrices of quarks and leptons can be written as~\cite{Yang:2024znv,Yang:2024kfs}
\begin{eqnarray}
	&&m_q=\left(\begin{array}{ccc} 0 & m_{q,12} & m_{q,13}\\
		m_{q,12}^* & 0 & m_{q,23}\\
		m_{q,13}^* & m_{q,23}^* & m_{q,33}\end{array}\right),m_e=\left(\begin{array}{ccc} 0 & m_{e,12} & m_{e,13}\\
		m_{e,12}^* & 0 & m_{e,23}\\
		m_{e,13}^* & m_{e,23}^* & m_{e,33}\end{array}\right),m_\nu=\left(\begin{array}{cc} 0 & M_D^T\\
		M_D & M_R\end{array}\right),\label{eq2}
\end{eqnarray}
where $q=u,d$, the parameters $m_{q,33}$ and $m_{e,33}$ are real, $M_D$ is $2\times3$ Dirac mass matrix, and $M_R$ is $2\times2$ Majorana mass matrix (the nonzero neutrino masses are obtained by the type-I seesaw mechanism). The elements of the matrices in Eq.~(\ref{eq2}) are
\begin{eqnarray}
	&&m_{u,11}=m_{u,22}=0,\;m_{u,33}=\frac{1}{\sqrt2}Y_u^{33}v_3,\;m_{u,12}=\frac{1}{\sqrt2}Y_u^{12}v_3,\;m_{u,13}=\frac{1}{\sqrt2}Y_u^{13}v_1,\nonumber\\
	&&m_{u,23}=\frac{1}{\sqrt2}Y_u^{23}v_2,\label{eqmu}\\
	&&m_{d,11}=m_{d,22}=0,\;m_{d,33}=\frac{1}{\sqrt2}Y_d^{33}v_3,\;m_{d,12}=\frac{1}{\sqrt2}Y_d^{12}v_3,\;m_{d,13}=\frac{1}{\sqrt2}Y_d^{13}v_1,\nonumber\\
	&&m_{d,23}=\frac{1}{\sqrt2}Y_d^{23}v_2,\label{eqmd}\\
	&&m_{e,11}=m_{e,22}=0,\;m_{e,33}=\frac{1}{\sqrt2}Y_e^{33}v_3,\;m_{e,12}=\frac{1}{\sqrt2}Y_e^{12}v_3,\;m_{e,13}=\frac{1}{\sqrt2}Y_e^{13}v_1,\nonumber\\
	&&m_{e,23}=\frac{1}{\sqrt2}Y_e^{23}v_2,\label{eqme}\\
	&&M_{D,11}=M_{D,22}=0,\;\;M_{D,12}=\frac{1}{\sqrt2}Y_D^{12}v_3,\;M_{D,31}=\frac{1}{\sqrt2}Y_D^{31}v_1,\nonumber\\
	&&M_{D,32}=\frac{1}{\sqrt2}Y_D^{32}v_2,\;M_{R,12}=M_{R,21}=0,\;M_{R,11}=\frac{1}{\sqrt2}Y_R^{11}v_\chi,\;M_{R,22}=\frac{1}{\sqrt2}Y_R^{22}v_\chi.
\end{eqnarray}

\subsection{The gauge sector of the FDM}\label{sec2-3}

Because of the introduction of an extra $U(1)_F$ local gauge group in the FDM, the covariant derivative corresponding to $SU(2)_L\otimes U(1)_Y\otimes U(1)_F$ is defined as~\cite{Yang:2024znv,Yang:2024kfs}
\begin{eqnarray}
	&&D_\mu=\partial_\mu+i g_2 T_j A_{j\mu}+i g_1 Y B_\mu+i g_{_F} F B'_\mu+i g_{_{YF}} Y B'_\mu,\;(j=1,\;2,\;3),\label{eqCD}
\end{eqnarray}
where $(g_2,\;g_1,\; g_{_F})$, $(T_j,\;Y,\;F)$, $(A_{j\mu},\;B_\mu,\; B'_\mu)$ denote the gauge coupling constants, generators and gauge bosons of groups $(SU(2)_L,\;U(1)_Y,\;U(1)_F)$, respectively, and $g_{_{YF}}$ is the gauge coupling constant arises from the gauge kinetic mixing effect which presents in the models with two Abelian groups. Then, the $W$ boson mass can be written as
\begin{eqnarray}
	&&M_W=\frac{1}{2} g_2 (v_1^2+v_2^2+v_3^2)^{1/2},
\end{eqnarray}
where $(v_1^2+v_2^2+v_3^2)^{1/2}=v\approx246.22\;{\rm GeV}$ and we have $v_1=\;v_2 < v_3$ in the FDM. The $\gamma$, $Z$ and $Z'$ boson masses in the FDM can be written as
\begin{eqnarray}
	&&M_\gamma=0,\;M_Z\approx\frac{1}{2}(g_1^2+g_2^2)^{1/2} v,\;M_{Z'}\approx 2|zg_{_F}| v_\chi,\label{eq19}
\end{eqnarray}
and
\begin{eqnarray}
	&&\gamma=c_W B+s_W A_3,\;Z=-s_W B+c_W A_3+s'_W B',\;Z'=s_W'(s_WB-c_W A_3)+c'_W B',
\end{eqnarray}
respectively, where $\gamma,\;Z,\;$and $Z'$ are the mass eigenstates, $c_W\equiv \cos \theta_W,\;s_W\equiv \sin \theta_W$ with $\theta_W$ denoting the Weinberg angle, and $s_W'\equiv \sin \theta'_W$, $c_W'\equiv \cos \theta'_W$ with $\theta_W'$ representing the $Z-Z'$ mixing effect.

\section{Experimental constraints and rare decay of the 125 GeV Higgs boson\label{sec3}}
At the LHC, the gluon fusion process is the main neutral Higgs particle production channel, and the Higgs boson is an unstable particle with a very short lifespan, which quickly decays after being produced. In the FDM, the changes in the scalar sector and the Yukawa coupling have a significant impact on the decays.

\subsection{The decay processes $h{\rightarrow} gg,\gamma\gamma$, $h{\rightarrow} VV^*$ ($V=Z,W$) and $h{\rightarrow} f\bar{f}$ ($f=c,b,\tau$)}

The decay process $h {\rightarrow} gg$ amplitude can be expressed by incorporating two form factors
\begin{eqnarray} \label{higg}
	{\cal M}_{gg h}^{ab}=-\frac{\alpha_s(m_h)\,m_{h}^2\,\delta^{ab}}{4\pi\,v}
	\bigg\{S^g(m_{h})
	\left(\epsilon^*_{1\perp}\cdot\epsilon^*_{2\perp}\right)
	-P^g(m_{h})\frac{2}{m_{h}^2}
	\langle\epsilon^*_1\epsilon^*_2 k_1k_2\rangle
	\bigg\}\,,
\end{eqnarray}
where $a$ and $b$ (with $a, b = 1$ to 8) are indices corresponding to the eight generators in
the $\mathit{SU(3)}$ adjoint representation, $k_{1,2}$ represent the four-momenta of the
two gluons, and $\epsilon_{1,2}$ denote the wave vectors of the respective gluons. $m_{h}$ is the mass of the lightest $\mathit{CP}$-even Higgs boson.
$\epsilon^\mu_{1\perp} = \epsilon^\mu_1 - 2k^\mu_1 (k_2 \cdot
\epsilon_1) / m^2_{h}$, $\epsilon^\mu_{2\perp} = \epsilon^\mu_2 -
2k^\mu_2 (k_1 \cdot \epsilon_2) / m^2_{h}$, and $\langle \epsilon_1
\epsilon_2 k_1 k_2 \rangle \equiv \epsilon_{\mu\nu\rho\sigma}\,
\epsilon_1^\mu \epsilon_2^\nu k_1^\rho k_2^\sigma$,
considering solely the leading contributions originating from third-generation quarks. The scalar and pseudoscalar form factors are then determined based on the methodology outlined in Ref.~\cite{Choi:2021nql}:
\begin{eqnarray}
	S^g(m_{h})
	= \sum_{f=b,t}
	\kappa_{f}\,F_{sf}(\tau_{f})
	+ \Delta S^g\,; \ \ \
	P^g(m_{h})
	= \sum_{f=b,t}
	\tilde{\kappa_{f}}\,F_{pf}(\tau_{f})
	+ \Delta P^g\,,
\end{eqnarray}
with $\tau_{f}=m_{{h}}^2/(4m_q^2)$ ($q=t,\:b$) are defined by using the pole masses of the bottom and top quarks. $\kappa_{f}$ represent $CP$-even effective Higgs couplings to the quarks and the leptons, respectively, and $\tilde{\kappa_{f}}$ represent $CP$-odd. The loop integral functions $F_{sf}(\tau_{f})$ [and $ F_{pf}(\tau_{f})\:,A_1\:,A_0$ below] are defined in Appendix~\ref{app-form}.

In the framework of NP, integration of QCD and EW corrections is performed. These corrections are factorized based on the universal infrared and collinear properties of QCD corrections and the universality of the predominant component of EW corrections. The QCD correction $\delta^{g:S}_{\rm QCD}$ is computed up to next-to-leading order (NLO), considering the complete quark mass dependence \cite{Spira:1995rr} and up to next-to-next-to-next-to-leading order (N$^3$LO) in the heavy top quark limit~\cite{Djouadi:1991tka,Choi:2021nql,Baikov:2006ch,Chetyrkin:1998mw}. The normalized decay widths of a neutral Higgs boson with mass $m_h$ into $gg$ are calculated by assuming $\Delta S^g=0$ and $\tilde{\kappa_{f}}=\Delta P^g=0$:
\begin{eqnarray}
	\label{eq:ghgg}
	\Gamma(h{\rightarrow} gg)\ =\ \frac{m_{h}^3\alpha^2_S}{32\pi^3\,v^2}
	\left[
	\left|S^g(m_{h})\right|^2\,\left(1+\delta^{g:S}_{\rm QCD}\right)
	\left(1+\delta^{g:S}_{\rm ew}\right)\right]\,.
\end{eqnarray}
The concrete expressions of $\kappa^S_{t} $and$ \:\kappa^S_{b}$ are formulated, respectively, as:
\begin{eqnarray}
		\kappa_{t}&& = \frac{v_{\rm{EW}}}{m_t}C_{hu_3\bar u_3},\nonumber\\
		\kappa_{b}&& = \frac{v_{\rm{EW}}}{m_b}C_{hd_3\bar d_3},\label{eq-tb}
\end{eqnarray}
$\delta^{g:S}_{\rm QCD}$ and $\delta^{g:S}_{\rm ew}$  are formulated, respectively, as~\cite{Spira:1995rr,Chetyrkin:1997iv,Baikov:2006ch}
\begin{eqnarray}
	\label{eq:ghgg_qcd_s}
	\delta^{g:S}_{\rm QCD} & = &
	\left(\frac{95}{4}-\frac{7}{6} N_F^L + \Delta^{g:S}_m\right)
	\frac{\alpha_s^{(N_F^L)}(m_h)}{\pi}  \nonumber \\
	& + & \left[370.20- 47.19 N_F^L + 0.902 (N_F^L)^2
	+ (2.375 + 0.667 N_F^L)
	\log\frac{m_h^2}{M_t^2}\right] \left( \frac{\alpha_s^{(N_F^L)}(m_h)}{\pi}
	\right)^2 \nonumber \\
	& + & \left[ 4533.46 - 1062.82 N_F^L + 52.62 (N_F^L)^2 - 0.5378 (N_F^L)^2
	\phantom{\log\frac{m_h^2}{M_t^2}} \right.
	\nonumber \\
	&&\, + (66.66 + 14.60 N_F^L - 0.6887 (N_F^L)^2) \log\frac{m_h^2}{M_t^2}
	\nonumber  \\
	&&\left. + (6.53 + 1.44 N_F^L - 0.111 (N_F^L)^2) \log^2\frac{m_h^2}{M_t^2} \right]
	\left( \frac{\alpha^{(N_F^L)}_{s}(m_h)}{\pi} \right)^3\,,\\
	\label{eq:gSew}
	\delta^{g:S}_{\rm ew} & = &\frac{G_F M_t^2}{8\sqrt{2}\pi^2}\,.
\end{eqnarray}
Considering $N_F^L=5$ to account for the number of light quark flavors, and we use $\Delta^{g:S}_m\approx 0.7$ \footnote{This value is obtained from Fig.~7(b) in Ref.~\cite{Spira:1995rr}.} for the NLO quark-mass effects contributed by the top, bottom, and charm quarks \cite{Spira:1995rr}. At $m_h=125.2$ GeV, the QCD corrections are computed as $\delta^{g:S}_{\rm QCD} = 0.668+0.197+0.021$ for the NLO,~\footnote{In this chapter, we denote the pole mass of the fermion $f$ by $M_f$ and its running mass by $m_f(\mu)$.} next-to-next-to-leading order(NNLO), and N$^3$LO levels. The EW correction is approximately 0.3\%, making it negligibly small.

 Under the framework of NP, supplementary enhancements to this leading-order (LO) decay width arise from loops involving third-generation fermions ($f = t, b, \tau$) and $W$ bosons. This phenomenon can be mathematically represented as~\cite{Zhang:2013hga}
\begin{eqnarray}
&&\Gamma_{{\rm{NP}}}(h{\rightarrow}\gamma\gamma)=\frac{G_{F}\alpha^2m_{{h}}^3}{128\sqrt{2}\pi^3}
\Big|\sum\limits_f N_c Q_{f}^2 \kappa_{f} A_{1/2}(x_f)+\kappa_{W}A_1(x_{W})
\nonumber\\
&&\hspace{3.0cm}
+\sum\limits_{\alpha=1}^2 \kappa _{H_\alpha^\pm }\frac{m_{Z}^2}{m_{H_\alpha^\pm}^2}A_0(x_{H_\alpha^\pm})\Big|^2,
\end{eqnarray}
 Because of charge conjugation invariance and color conservation, gluon radiation from the colored quark loop involved in the $h\gamma\gamma$ vertex is prohibited. Therefore, the complete two-loop QCD corrections can be effectively considered by incorporating scaling factors into the form factors corresponding to the $b$- and $t$-quark contributions, as\cite{Spira:1995rr,Muhlleitner:2006wx,Bonciani:2007ex,Djouadi:1990aj,Choi:2021nql,Djouadi:1996pb}
\footnote{For a detailed description of the scaling factors of
	$C_{sf}(x)$ see Appendix \ref{app-form}.}
\begin{eqnarray}
	\label{eq:ghaa_qcd}
	A_{1/2}(x_f) &\longrightarrow A_{1/2}(x_f)\,
	\left[1+C_{sf}(x_f)\frac{\alpha_s(M_h)}{\pi}\right] \,;
\end{eqnarray}
in the limit $x_f{\rightarrow} 0$, the scaling factors $C_{sf}$ approach $-1$. And the expressions of $\kappa_{\tau}$, $\kappa_{W}$, $\kappa_{Z}$ and $\kappa _{H_\alpha^\pm }$ are, respectively, 
\begin{eqnarray}
&&\kappa_{\tau}=\frac{\upsilon_{_{\rm{EW}}}}{m_{\tau}}C_{e_3\bar{e}_3h},   \nonumber\\
&&\kappa_{W}=-\frac{\upsilon_{_{\rm{EW}}}}{2 m_{W}^2} C_{hW_\sigma ^+W_\mu ^-} ,   \nonumber\\
&&\kappa_{Z}=-\frac{\upsilon_{_{\rm{EW}}}}{2 m_{Z}^2} C_{hZ_\sigma Z_\mu },   \nonumber\\
&&\kappa _{H_\alpha^\pm }=-\frac{\upsilon_{_{\rm{EW}}}}{2 m_{Z}^2} C_{hH_\alpha ^+H_\alpha ^-} \quad (\alpha=1,2),\label{eq-wz}
\end{eqnarray}
 where the couplings $C_{e_3\bar{e}_3h}$, $C_{hW_\sigma ^+W_\mu ^-}$, $C_{hZ_\sigma Z_\mu }$, and $C_{hH_\alpha ^+H_\alpha ^-}$ are defined in Appendix \ref{app-coupling}.

The light doubletlike Higgs with $125\:{\rm GeV}$ mass can decay through the channels  $h{\rightarrow} WW^*$ and $h{\rightarrow} ZZ^*$, where $W^*$ and $Z^*$ denote the off-shell EW gauge bosons. Summing over all modes available to the $W^*$ or $Z^*$, the decay widths are given by~\cite{Zhang:2013hga}
\begin{eqnarray}
&&\Gamma_{{\rm{NP}}}(h{\rightarrow} ZZ^*)=\frac{e^4m_h}{2048\pi^3s_W^4c_W^4}|\kappa_{Z}|^2
\Big(7-\frac{40}{3}s_W^2+\frac{160}{9}s_W^4\Big)F\left(\frac{m_Z}{m_h}\right), \nonumber\\
&&\Gamma_{{\rm{NP}}}(h{\rightarrow} WW^*)=\frac{3e^4m_{{h}}}{512\pi^3s_{W}^4}|\kappa_{W}|^2
F\left(\frac{m_{W}}{m_h}\right),
\end{eqnarray}
and the form factor $F(x)$ is formulated in Appendix~\ref{app-form}. $s_{W}$ is the Weinberg angle.

Including the radiative corrections known up to now, the decay width of Higgs to
fermions can be organized as~\cite{Choi:2021nql,Spira:2016ztx}
\begin{eqnarray}
	\label{eq:ghff}
	\Gamma(h{\rightarrow} f \bar{f}) &= &
	N_C^f\frac{m_f^2}{v^2}\frac{\beta_f m_h}{8\pi}
	\left[
	\beta_f^2|\kappa_{f}|^2
	\left(1+\delta_{\rm QCD}+\delta^{f:S}_t
	+\delta^f_{\rm mixed}\right)
	\left(1+\delta^f_{\rm ew}\right)
	\right.
	\nonumber \\[2mm]
	&&\hspace{2.40cm} \ + \
	\left.
	|\tilde{\kappa_{f}}|^2
	\left(1+\delta_{\rm QCD}+\delta^{f:P}_t\right) \right]\,,
\end{eqnarray}
where $\beta_f \equiv \sqrt{1-4\tau_f}$ with
$\tau_f=M_f^2/m_h^2$ and the color factor $N_C^f=3$ for quarks and 1 for
leptons.
The lepton mass $m_f$ use their pole mass, while for quarks, the $\overline{\rm MS}$ quark mass $\overline{m}_q(m_h)$ is used~\cite{Spira:2016ztx}.

The pure QCD corrections to the Higgs decays into a quark pair $q\bar{q}$ consist of a universal part $\delta_{\rm QCD}$, as well as two types of flavor- and parity-dependent contributions $\delta_t^{q:S}$ and $\delta_t^{q:P}$, which are expressed as~\cite{Braaten:1980yq,Kataev:1993be,Melnikov:1995yp,Choi:2021nql,Larin:1995sq}
\begin{eqnarray}
	\label{eq:ghff_qcd}
	\delta_{\rm QCD}&=&
	5.67 \frac{\alpha_s (m_h)}{\pi} + (35.94 - 1.36
	N_F) \left( \frac{\alpha_s (m_h)}{\pi} \right)^2
	\nonumber \\ &&
	+ (164.14 - 25.77 N_F + 0.259 N_F^2) \left( \frac{\alpha_s(m_h)}{\pi} \right)^3
	\nonumber \\ &&
	+(39.34-220.9 N_F+9.685 N_F^2-0.0205 N_F^3) \left(
	\frac{\alpha_s(m_h)}{\pi} \right)^4\,, \nonumber \\[2mm]
	\delta^{q:S}_t &=&\frac{\kappa_t}{\kappa_q}\,
	\left(\frac{\alpha_s (m_h)}{\pi}\right)^2 \left[ 1.57 -
	\frac{2}{3} \log \frac{m_h^2}{M_t^2} + \frac{1}{9} \log^2
	\frac{\overline{m}_q^2 (m_h)}{m_h^2} \right]\,, \nonumber \\[2mm]
	\delta^{q:P}_t &=&\frac{\tilde{\kappa_t}}{\tilde{\kappa_q}}\,
	\left(\frac{\alpha_s (m_h)}{\pi}\right)^2 \left[ 3.83 -
	\log \frac{m_h^2}{M_t^2} + \frac{1}{6} \log^2
	\frac{\overline{m}_q^2 (m_h)}{m_h^2} \right]\,,
\end{eqnarray}
where $N_F$ counts the flavor number of quarks
lighter than $m_h$. The QCD coupling constant $\alpha_s$ and the running $\overline{\rm MS}$ quark mass $\overline{m}_q(m_h)$ are defined at the Higgs mass scale to account for the absorption of significant mass logarithms, and we use program package CRunDec3.1 to compute these parameters~\cite{Herren:2017osy,Chetyrkin:2015mxa}.%

For the EW corrections~\cite{Fleischer:1980ub,Bardin:1990zj,Kniehl:1991ze},
we adopt the approximation~\cite{Djouadi:1991uf,Spira:2016ztx}
\begin{equation}
	\label{eq:ghff_ew}
	\begin{split}
		\delta^f_{\rm ew} = &\frac{3}{2} \frac{\alpha}{\pi}Q_f^2 \left(\frac{3}{2} -
		\log \frac{m_h^2}{M_f^2} \right) \\
		+& \frac{G_F}{8 \sqrt{2} \pi^2}
		\left\{ k_f M_t^2 + m_W^2 \left[ - 5 + \frac{3}{s_W^2} \log c_W^2 \right]
		- 8\,m_Z^2 (6 v_{Z\bar f f}^2 - a_{Z\bar f f}^2) \right\}\,.
	\end{split}
\end{equation}
The vector and axial-vector couplings of the $Z$ boson to fermions are given by $v_{Z\bar{f}f} = \frac{1}{2}I_3^f - Q_f s_W^2$ and $a_{Z\bar{f}f} = \frac{1}{2}I_3^f$, respectively, where $I_3^f$ denotes the third component of the EW isospin and $Q_f$ represents the electric charge of the fermion $f$. The large logarithm $\log(m_h^2/M_f^2)$ can be absorbed into the running fermion mass, analogous to the treatment in QCD corrections. For decays involving leptons and light quarks, the coefficient $k_f$ is 7, whereas for $b$ and $t$ quarks, $k_f$ is 1. The EW corrections are below the 1\% level for $f = b, c$, while they are the order of ${\cal O}(1\%-5\%)$ for $f = \tau, \mu$~\cite{Choi:2021nql}.

The mixed corrections, determined using low-energy theorems, can be formulated into the expressions~\cite{Kwiatkowski:1994cu,Chetyrkin:1996wr}
\begin{eqnarray}
	\delta^q_{\rm mixed}&=&-\frac{G_FM_t^2}{8\sqrt{2}\pi^2}
	\left(\frac{3}{2}+\zeta_2\right)
	\frac{\alpha_s(M_t)}{\pi}
	\hspace{1cm}\mbox{for light quarks}\,,\nonumber \\[2mm]
	\delta^{b,t}_{\rm mixed}&=&-\frac{G_FM_t^2}{8\sqrt{2}\pi^2}\
	4\left(1+\zeta_2\right)\frac{\alpha_s(M_t)}{\pi}
	\hspace{1cm}\mbox{for $b$ and $t$}\,,
\end{eqnarray}
at NNLO with $\zeta_2 = \pi^2/6$.

Normalized to the SM expectation, the signal strengths for the Higgs decay channels are quantified by the ratios~\cite{Zhang:2013hga,Ge:2024rdr}
\begin{eqnarray}
\mu_{h\gamma\gamma,hVV^*}&&= \frac{\sigma_{{\rm{NP}}}({\rm{ggF}})}
{\sigma_{{\rm{SM}}}({\rm{ggF}})} \, \frac{{\rm{BR}}_{{\rm{NP}}}(h{\rightarrow}\gamma\gamma,VV^*)}
{{\rm{BR}}_{{\rm{SM}}}(h{\rightarrow}\gamma\gamma,VV^*)}   \qquad (V=Z,W), \nonumber\\
\mu_{hf\bar{f}}&&= \frac{\sigma_{{\rm{NP}}}({\rm{VBF}})}
{\sigma_{{\rm{SM}}}({\rm{VBF}})} \, \frac{{\rm{BR}}_{{\rm{NP}}}(h{\rightarrow}{f\bar{f}})}
{{\rm{BR}}_{{\rm{SM}}}(h{\rightarrow}{f\bar{f}})}  \qquad (f=b,c,\tau).
\label{eq-ratios}
\end{eqnarray}

One can evaluate the Higgs decay branch ratios by
\begin{eqnarray}
	\mathrm{BR}\left(h_{125}^{\mathrm{NP}} {\rightarrow} \gamma \gamma\right) &=& \frac{\Gamma_{{\rm{NP}}}(h{\rightarrow}\gamma\gamma)}{\Gamma_{{\rm{SM}}}(h{\rightarrow}\gamma\gamma)}  \mathrm{BR}\left(h_{125}^{\mathrm{SM}} {\rightarrow} \gamma \gamma\right) \frac{\Gamma_{\mathrm{tot}, 125}^{\mathrm{SM}}}{\Gamma_{\mathrm{tot}, 125}^{\mathrm{NP}}}, \nonumber\\
	\mathrm{BR}\left(h_{125}^{\mathrm{NP}} {\rightarrow} VV^*\right) &=& \frac{\Gamma_{{\rm{NP}}}(h{\rightarrow}VV^*)}{\Gamma_{{\rm{SM}}}(h{\rightarrow}VV^*)}  \mathrm{BR}\left(h_{125}^{\mathrm{SM}} {\rightarrow} VV^*\right) \frac{\Gamma_{\mathrm{tot}, 125}^{\mathrm{SM}}}{\Gamma_{\mathrm{tot}, 125}^{\mathrm{NP}}}, \nonumber\\
	\mathrm{BR}\left(h_{125}^{\mathrm{NP}} {\rightarrow} f \bar{f}\right) &=& \frac{\Gamma_{{\rm{NP}}}(h{\rightarrow}f \bar{f})}{\Gamma_{{\rm{SM}}}(h{\rightarrow}f \bar{f})}  \mathrm{BR}\left(h_{125}^{\mathrm{SM}} {\rightarrow} f \bar{f}\right) \frac{\Gamma_{\mathrm{tot}, 125}^{\mathrm{SM}}}{\Gamma_{\mathrm{tot}, 125}^{\mathrm{NP}}},
\label{eq-Branch}
\end{eqnarray}
and, moreover,
\begin{eqnarray}
	\frac{\sigma_{{\rm NP}}({\rm ggF})}{\sigma_{{\rm SM}}({\rm ggF})} 
	&\approx& \frac{\Gamma_{{\rm NP}}(h\rightarrow gg)}{\Gamma_{{\rm SM}}(h\rightarrow gg)}
	= \frac{\Gamma_{{\rm NP}}^h}{\Gamma_{{\rm SM}}^h}\:
	\frac{\Gamma_{{\rm NP}}(h\rightarrow gg)/\Gamma_{{\rm NP}}^h}{\Gamma_{{\rm SM}}(h\rightarrow gg)/\Gamma_{{\rm SM}}^h}\nonumber\\
	&=&\frac{\Gamma_{{\rm NP}}^h}{\Gamma_{{\rm SM}}^h}\:
	\frac{{\rm BR}_{{\rm NP}}(h\rightarrow gg)}{{\rm BR}_{{\rm SM}}(h\rightarrow gg)},\nonumber\\
	\frac{\sigma_{{\rm NP}}({\rm VBF})}{\sigma_{{\rm SM}}({\rm VBF})} &\approx& \frac{\Gamma_{{\rm NP}}(h\rightarrow{VV^*})}{\Gamma_{{\rm SM}}(h\rightarrow{VV^*})}
	=\frac{\Gamma_{{\rm NP}}^h}{\Gamma_{{\rm SM}}^h}\:
	\frac{\Gamma_{{\rm NP}}(h\rightarrow{VV^*})/\Gamma_{{\rm NP}}^h}{\Gamma_{{\rm SM}}(h\rightarrow{VV^*})/\Gamma_{{\rm SM}}^h}\nonumber\\
	&=&\frac{\Gamma_{{\rm NP}}^h}{\Gamma_{{\rm SM}}^h}\:
	\frac{{\rm BR}_{{\rm NP}}(h\rightarrow{VV^*})}{{\rm BR}_{{\rm SM}}(h\rightarrow{VV^*})},
	\label{eq-cross}
\end{eqnarray}
with the 125 GeV Higgs total decay width for the NP
\begin{eqnarray}
&&\Gamma_{{\rm{NP}}}^h\approx \sum\limits_{f=b,\tau,c} \Gamma_{{\rm{NP}}}(h{\rightarrow} f\bar{f})+ \sum\limits_{V=Z,W} \Gamma_{{\rm{NP}}}(h{\rightarrow} VV^*) \nonumber\\
&&\qquad\quad +\: \Gamma_{{\rm{NP}}}(h{\rightarrow} gg) +\Gamma_{{\rm{NP}}}(h{\rightarrow} \gamma\gamma)+\Gamma_{{\rm{NP}}}(h{\rightarrow} Z\gamma).
\end{eqnarray}

Here, we disregard the influences from rare or invisible decays, with $\Gamma_{{\rm{SM}}}^h$ representing the total decay width of the SM Higgs. By using Eqs.~(\ref{eq-ratios})-(\ref{eq-cross}), we are able to assess the signal strengths for the Higgs decay modes within the FDM~\cite{Ge:2024rdr} as
\begin{eqnarray}
	\mu_{h\gamma\gamma} &\approx& \frac{\Gamma_{{\rm{NP}}}(h{\rightarrow} gg)}{\Gamma_{{\rm{SM}}}(h{\rightarrow} gg)} \cdot \frac{\Gamma_{{\rm{NP}}}(h{\rightarrow}\gamma\gamma)/\Gamma_{{\rm{NP}}}^h}{\Gamma_{{\rm{SM}}}(h{\rightarrow}\gamma\gamma)/\Gamma_{{\rm{SM}}}^h} \nonumber\\
	\qquad &=& \frac{\Gamma_{{\rm{SM}}}^h}{\Gamma_{{\rm{NP}}}^h} \cdot \frac{\Gamma_{{\rm{NP}}}(h{\rightarrow} gg)}{\Gamma_{{\rm{SM}}}(h{\rightarrow} gg)} \cdot \frac{\Gamma_{{\rm{NP}}}(h{\rightarrow}\gamma\gamma)}{\Gamma_{{\rm{SM}}}(h{\rightarrow}\gamma\gamma)}, \nonumber\\
	\mu_{hVV^*} &\approx& \frac{\Gamma_{{\rm{NP}}}(h{\rightarrow} gg)}{\Gamma_{{\rm{SM}}}(h{\rightarrow} gg)} \cdot \frac{\Gamma_{{\rm{NP}}}(h{\rightarrow} VV^*)/\Gamma_{{\rm{NP}}}^h}{\Gamma_{{\rm{SM}}}(h{\rightarrow} VV^*)/\Gamma_{{\rm{SM}}}^h} \nonumber\\
	\qquad &=& \frac{\Gamma_{{\rm{SM}}}^h}{\Gamma_{{\rm{NP}}}^h} \cdot \frac{\Gamma_{{\rm{NP}}}(h{\rightarrow} gg)}{\Gamma_{{\rm{SM}}}(h{\rightarrow} gg)} \cdot |\kappa_{V}|^2, \nonumber\\
	\mu_{hf\bar{f}} &\approx& \frac{\Gamma_{{\rm{NP}}}(h{\rightarrow} VV^*)}{\Gamma_{{\rm{SM}}}(h{\rightarrow} VV^*)} \cdot \frac{\Gamma_{{\rm{NP}}}(h{\rightarrow} f\bar{f})/\Gamma_{{\rm{NP}}}^h}{\Gamma_{{\rm{SM}}}(h{\rightarrow} f\bar{f})/\Gamma_{{\rm{SM}}}^h} \nonumber\\
	\qquad &=& \frac{\Gamma_{{\rm{SM}}}^h}{\Gamma_{{\rm{NP}}}^h} \cdot |\kappa_{V}|^2 \cdot |\kappa_{f}|^2 \qquad (V=Z,W; f=b,c,\tau),
	\label{signals}
\end{eqnarray}
where $\kappa_{V}$ ($\mathit{V=Z,W}$) and $\kappa_{f}$ ($\mathit{f=b,c},\tau$) are defined in Eqs.~(\ref{eq-tb}) and (\ref{eq-wz}). The corresponding experimental values are listed in Table~\ref{tab1}.
\begin{table*}
	\begin{tabular*}{\textwidth}{@{\extracolsep{\fill}}lllll@{}}
		\hline
		Signal & Value from PDG ~\cite{ParticleDataGroup:2024cfk}\\
		\hline
		$\mu_{\gamma\gamma}$ & $1.10\pm0.06$ \\
		$\mu_{ZZ^*}$ & $1.02\pm0.08$ \\
		$\mu_{WW^*}$ & $1.00\pm0.08$  \\
		$\mu_{b\bar{b}}$ & $0.99\pm0.12$ \\
		$\mu_{c\bar{c}}$ & $<14$ \\
		$\mu_{\tau\bar{\tau}}$ & $0.91\pm0.09$  \\
		$\mu_{Z\gamma}$ & $2.2\pm0.7$  \\
		\hline
	\end{tabular*}
	\caption{Experimental values for the Higgs decay rates.}
	\label{tab1}
\end{table*}

\subsection{The flavor-changed neutral currents in the FDM}\label{subFCNC}

This work examines effects induced by FCNCs within the FDM, where such currents are mediated by a newly defined $\mathit{Z}$ boson, notably the $Z^\prime$. The rare decay processes considered include $\bar B \to X_s\gamma$ and $B_s^0 \to \mu^+\mu^-$ in the $\mathit{B}$ meson sector, $t \to c h$ and $t \to u h$ in the top quark sector, as well as the charged lepton flavor violation decays $\tau \to 3e$, $\tau \to 3\mu$, and $\mu \to 3e$. For simplicity, the non-zero $U_F(1)$ charge is fixed at $z = 1$ throughout the analysis.

The $B$ meson rare decay processes $\bar B \to X_s\gamma$ and $B_s^0 \to \mu^+\mu^-$ are related closely to the NP contributions, and the average experimental data on the branching ratios of $\bar B \to X_s\gamma$ and $B_s^0 \to \mu^+\mu^-$ are~\cite{ParticleDataGroup:2024cfk}
\begin{eqnarray}
&&{\rm Br}(\bar B \to X_s\gamma)=(3.49\pm0.19)\times 10^{-4},\nonumber\\
&&{\rm Br}(B_s^0 \to \mu^+\mu^-)=(3.01\pm0.35)\times 10^{-9}.\label{eqBDd}
\end{eqnarray}

The newly introduced scalars in the FDM including $CP$-even Higgs, $CP$-odd Higgs and
charged Higgs can make contributions to these two processes, the analytical calculations of the contributions are collected in the Appendix in Ref.~\cite{Yang:2018fvw}.

The branching ratios of the top quark rare decay processes $t\to ch$ and $t\to uh$ can be written as~\cite{Yang:2018utw}
\begin{eqnarray}
&&{\rm Br}(t\rightarrow q_u h)=\frac{|\mathcal{M}_{t q_u h}|^2\sqrt{((m_t+m_h)^2-m_{q_u}^2)((m_t-m_h)^2-m_{q_u}^2)}}{32\pi m_t^3\Gamma^t_{{\rm total}}},
\end{eqnarray}
where $q_u=u,\;c$, the amplitude $\mathcal{M}_{tq_uh}$ can be read directly from the Yukawa couplings in Eq.~(\ref{eq9}), and $\Gamma^t_{{\rm total}}=1.42\;$GeV~\cite{ParticleDataGroup:2024cfk} is the total decay width of top quark. %The measured quark masses, CKM matrix and $B$ meson rare decay processes $\bar B \to X_s\gamma$, $B_s^0 \to \mu^+\mu^-$ should be considered in the calculations of top quark rare decay processes $t\to ch$ and $t\to uh$, hence we take the points obtained in Fig.~\ref{Bdecay} as inputs.
The experimental upper bounds on the branching ratios of $t\to ch$ and $t\to uh$ are, respectively, 
\begin{eqnarray}
&&{\rm Br}(t\to ch)<3.4\times 10^{-4},\nonumber\\
&&{\rm Br}(t\to uh)<1.9\times 10^{-9}.\label{eqBD1}
\end{eqnarray}

Finally, we focus on the lepton flavor violation processes $\tau\to 3e$ and $\tau\to 3\mu$, $\mu\to 3e$ predicted in the FDM. The corresponding amplitude can be written as~\cite{Hisano:1995cp}
\begin{eqnarray}
&&\mathcal{M}(e_j\rightarrow e_i e_i\bar e_i)=C_1^L\bar u_{e_i}(p_2)\gamma_\mu P_L u_{e_j}(p_1) u_{e_i}(p_3)\gamma^\mu P_L \nu_{e_i}(p_4)\nonumber\\
&&\qquad\quad+C_1^R\bar u_{e_i}(p_2)\gamma_\mu P_R u_{e_j}(p_1) u_{e_i}(p_3)\gamma^\mu P_R \nu_{e_i}(p_4)\nonumber\\
&&\qquad\quad+[C_2^L\bar u_{e_i}(p_2)\gamma_\mu P_L u_{e_j}(p_1) u_{e_i}(p_3)\gamma^\mu P_R \nu_{e_i}(p_4)\nonumber\\
&&\qquad\quad+C_2^R\bar u_{e_i}(p_2)\gamma_\mu P_R u_{e_j}(p_1) u_{e_i}(p_3)\gamma^\mu P_L \nu_{e_i}(p_4)-(p_2\leftrightarrow p_3)]\nonumber\\
&&\qquad\quad+[C_3^L\bar u_{e_i}(p_2) P_L u_{e_j}(p_1) u_{e_i}(p_3) P_L \nu_{e_i}(p_4)\nonumber\\
&&\qquad\quad+C_3^R\bar u_{e_i}(p_2) P_R u_{e_j}(p_1) u_{e_i}(p_3) P_R \nu_{e_i}(p_4)-(p_2\leftrightarrow p_3)],
\end{eqnarray}
where $i=1,\;2$ for $j=3$, $i=1$ for $j=2$, $u_{e_i}$ denotes the spinor of lepton, $\nu_{e_i}$ denotes the spinor of antilepton, $P_L=(1-\gamma_5)/2$, $P_R=(1+\gamma_5)/2$, and $p_k$ denotes the momentum of charged lepton with $k=1,2,3,4$. The coefficients $C_{1,2,3}^{L,R}$ from the contributions of Higgs bosons and $Z,\;Z'$ bosons, can be obtained through the Yukawa couplings in Eq.~(\ref{eq9}) and the definition of covariant derivative in Eq.~(\ref{eqCD}). Then we can calculate the decay rate~\cite{Hisano:1995cp}
\begin{eqnarray}
&&\Gamma(e_j\rightarrow e_i e_i\bar e_i)=\frac{m_{e_j}^5}{1536\pi^3}\Big[\frac{1}{2}(|C_1^L|^2+|C_1^R|^2)+|C_2^L|^2+|C_2^R|^2+\frac{1}{8}(|C_3^L|^2+|C_3^R|^2)\Big].
\end{eqnarray}
The total decay widthes of $\mu $and$\;\tau$ are taken as $\Gamma^\mu_{{\rm total}}=2.996\times 10^{-19}\;$GeV and $\Gamma^\tau_{{\rm total}}=2.265\times 10^{-12}\;$GeV~\cite{ParticleDataGroup:2024cfk}.

The experimental upper bounds on decay widths are~\cite{ParticleDataGroup:2024cfk}
\begin{eqnarray}
&&{\rm Br}(\mu^-\rightarrow e^- e^+e^-)< 1.0\times 10^{-12},\nonumber\\
&&{\rm Br}(\tau^-\rightarrow e^- e^+e^-)< 2.7\times 10^{-8},\nonumber\\
&&{\rm Br}(\tau^-\rightarrow \mu ^- \mu^+\mu^-)< 1.9\times 10^{-8}.
\end{eqnarray}

\subsection{Constraints from electroweak precision observables}\label{subEWPM}
The masses and decay properties of the electroweak bosons, combined with low-energy data, can be used to probe and constrain possible deviations from the SM. At the electroweak one-loop level, such effects on precision observables are predominantly captured by the three oblique parameters S, T, and U, which characterize corrections to the gauge boson self-energies. The parameter T is proportional to the difference between the $W$ and $Z$ self-energies at $Q^2 = 0$, while S (S +U) is associated with the difference between the $Z (W)$ self-energy at  $Q^2 = M^2_{Z,W}$ and $Q^2 = 0$. The current values of these parameters, taken from Ref.~\cite{ParticleDataGroup:2024cfk}, are
\begin{equation}
\mathrm{S}=-0.04\pm0.10,\mathrm{T}=0.01\pm0.12,\mathrm{U}=-0.01\pm0.11.\label{eqstu}
\end{equation}

We use SPheno~\cite{Porod:2003um} to compute the mass spectrum and the three oblique parameters.

\subsection{The decay processes $h{\rightarrow} Z\gamma$ and $h{\rightarrow} MZ$}

\begin{figure}
	\setlength{\unitlength}{1mm}
	\begin{minipage}[c]{0.25\textwidth}
		\centering
		\includegraphics[width=1.5in]{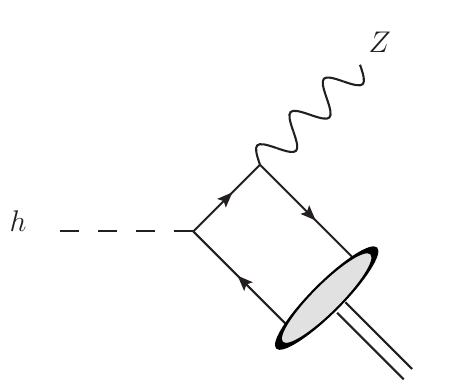}
	\end{minipage}%
	\begin{minipage}[c]{0.25\textwidth}
		\centering
		\includegraphics[width=1.5in]{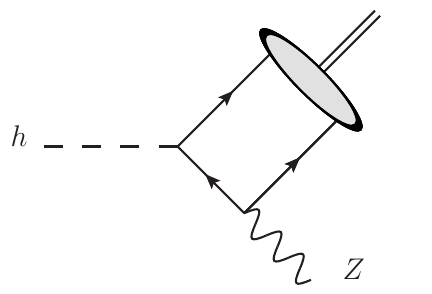}
	\end{minipage}%
	\begin{minipage}[c]{0.25\textwidth}
		\centering
		\includegraphics[width=1.5in]{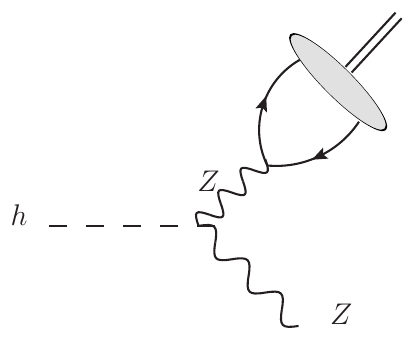}
	\end{minipage}
	\begin{minipage}[c]{0.24\textwidth}
		\centering
		\includegraphics[width=1.5in]{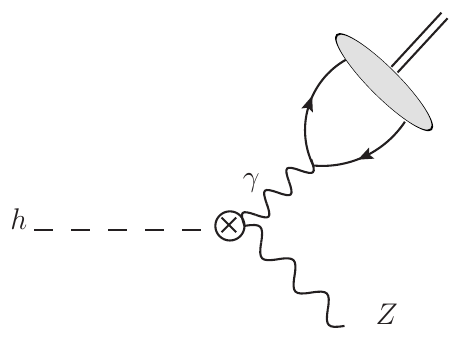}
	\end{minipage}
	\caption[]{The dominating Feynman diagrams for $h{\rightarrow} MZ$, where $M$ is a vector meson ($\rho, \omega, \phi, J/\psi, \Upsilon$)~\cite{HVZ-sm}.}
	\label{HVZ-Feynman}
\end{figure}

\begin{figure}
	\setlength{\unitlength}{1mm}
	\begin{minipage}[c]{0.8\textwidth}
		\centering
  		\includegraphics[width=5.2in]{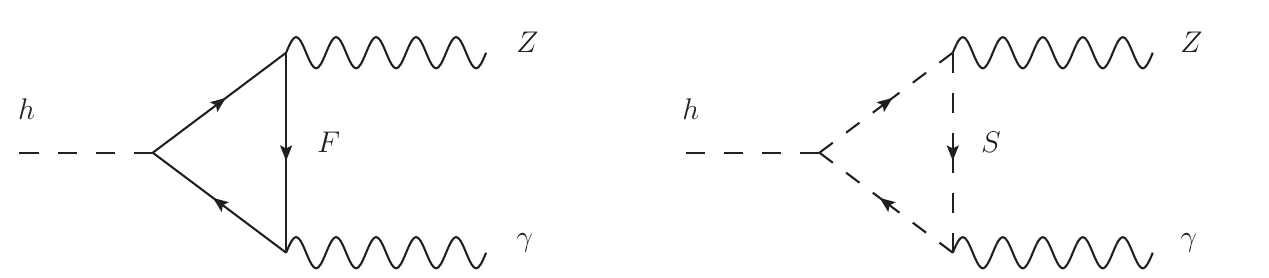}
		\vspace{0.5cm}
		\includegraphics[width=2.4in]{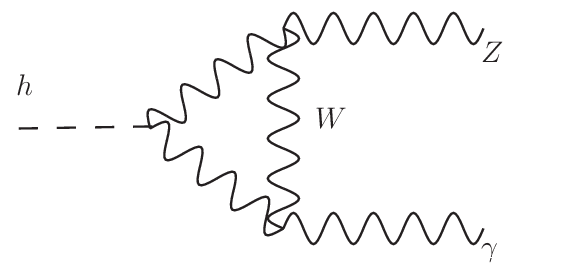}
	\end{minipage}%
	\caption[]{The one-loop diagrams for $h{\rightarrow} \gamma Z$ in the FDM, with $F$ denoting charged fermions and $S=H^{\pm}$ denoting charged Higgs, $W$ is $W^\pm$  boson.}
	\label{oneloop}
\end{figure}

The weak hadronic decays of the lightest Higgs boson $h{\rightarrow} MZ$ hold significant interest due to the potential involvement of a longitudinally polarized massive final-state gauge boson~\cite{HVZ-sm}. In Fig.~\ref{HVZ-Feynman}, we illustrate the predominant Feynman diagrams for $h{\rightarrow} MZ$. The first two diagrams in Fig.~\ref{HVZ-Feynman} correspond to the direct contributions denoted as $F_{\text{direct}}$, while the last two diagrams depict the indirect contributions labeled as $F_{\text{ind}}$. In the final diagram, the crossed circle symbolizes the effective vertex $h{\rightarrow} Z\gamma^*$ originating from the one-loop diagrams. Notably, the $hZZ^*$ vertex is present at the tree level, whereas the $hZ\gamma$ vertex arises through one-loop processes~\cite{HVZ-sm} (see Fig.~\ref{oneloop}). The process $h{\rightarrow} Z \gamma^{\ast}$ serves as a potential probe for NP; thus, we will primarily delve into the discussion of $h{\rightarrow} Z \gamma^{\ast}$.

The experimental investigation conducted in \cite{CMS:2022fsq} focused on the decay process $h {\rightarrow} J/\psi Z$, while the exploration of the decays $h {\rightarrow} \rho Z$ and $h {\rightarrow} \phi Z$ was elaborated in Ref.~\cite{experiment2}. The upper limits on the branch ratio ${\rm{Br}}(h {\rightarrow} \rho Z)$ is below 1.21\%, corresponding to less than 868 times the SM prediction. Correspondingly, the upper limit for ${\rm{Br}}(h {\rightarrow} \phi Z)$ is below 0.36\%, equivalent to 855 times the SM anticipation. These upper bounds are intricately connected to the polarization scenarios as discussed in Ref.~\cite{experiment2}.

The decay width of $h{\rightarrow} Z\gamma$ and $h{\rightarrow} MZ$ can be given as~\cite{HVZ-sm}
\begin{eqnarray}
\Gamma(h {\rightarrow} Z \gamma)= \frac{{\alpha}^2 m^3_h}{32 {\pi}^3 v^2 \sin^2\theta_{W} \cos^2 \theta_{W}}\Big(1-\frac{m^2_Z}{m^2_h}\Big)^3 (|C_{\gamma Z}|^2+|{\tilde{C}}_{\gamma Z}|^2).
\end{eqnarray}
\begin{eqnarray}
	&&\Gamma(h{\rightarrow} MZ) = \frac{m_h^3}{4\pi \upsilon^4} \lambda^{1/2}(1,r_{Z},r_{M})(1-r_{Z}-r_{M})^2  \nonumber\\
	&&\qquad\qquad\qquad\,\:\times \left[|F_{\parallel}^{MZ}|^2 + \frac{8r_{M}r_{Z}}{(1-r_{Z}-r_{M})^2}(|F_{\perp}^{MZ}|^2 + |\tilde{F}_{\perp}^{MZ}|^2)\right].
	\label{decay-width}
\end{eqnarray}
Here $\lambda(1,r_{Z},r_{M})=(1-r_{Z}-r_{M})^2-4r_{Z}r_{M}$, $r_{Z}=\frac{m_{Z}^2}{m_{h}^2}$, $r_{M}=\frac{m_{M}^2}{m_{h}^2}$, and $m_{M}$ is the mass of vector meson. $F_{\parallel}^{MZ}$ represents the $CP$-even longitudinal form factors, and $F_{\perp}^{MZ}$ and $\tilde{F}_{\perp}^{MZ}$ represent the $CP$-even and $CP$-odd transverse form factors respectively. The mass ratio $r_{M}$ is very small for all mesons, but it can make the contributions to the transverse polarization states to the $h{\rightarrow} MZ$ rates significantly, so we still keep the mass ratio $r_{M}$ in our analysis~\cite{HVZ-sm,HVZ-zhao,Liu:2020mev}.

Equation.~(\ref{decay-width}) comprises two components of form factors: the direct and the indirect contributions. We initiate the analysis with the indirect contributions, which play a significant role in probing NP. These contributions entail hadronic matrix elements of local currents, enabling calculations to be extended to all orders in QCD~\cite{HVZ-sm}. Consequently, the indirect contributions are
\begin{eqnarray}
	&&F_{||\,ind}^{MZ}=\frac{g_{hZZ}}{1-{r_{M}/r_{Z}}}\sum_{q}f_{M}^q\upsilon_{q}+C_{\gamma Z}\frac{\alpha(m_{M})}{4\pi}\frac{4r_{Z}}{1-r_{Z}-r_{M}}\sum_{q}f_{M}^q Q_{q},
	\label{form-factors1}\\
	&&F_{\perp\,ind}^{MZ}=\frac{g_{hZZ}}{1-{r_{M}/r_{Z}}}\sum_{q}f_{M}^q\upsilon_{q}+C_{\gamma Z}\frac{\alpha(m_{M})}{4\pi}\frac{1-r_{Z}-r_{M}}{r_{M}}\sum_{q}f_{M}^q Q_{q},
	\label{form-factors2}\\
	&&\widetilde{F}_{\perp\,ind}^{MZ}=\widetilde{C}_{\gamma Z}\frac{\alpha(m_{M})}{4\pi}\frac{\lambda^{1/2}(1,r_{Z},r_{M})}{r_{M}}\sum_{q}f_{M}^q Q_{q},
	\label{form-factors}
\end{eqnarray}
where $\upsilon_q=T_{3}^q/2-Q_q s_{W}^2 $ are the vector couplings of the $Z$ boson to the quark $q$, $T_{3}^{q}$ and $Q_{q}$ represent the weak isospin and charge of quark $q$ respectively, and $s_W=\sin\theta_W$ with $\theta_W$ denoting the Weinberg angle. $\alpha$ is the electromagnetic coupling constant can be calculated by AlphaQED~\cite{Jegerlehner:2011mw}. The flavor-specific decay constants $f_M^q$ are defined in terms of the local matrix elements~\cite{HVZ-sm,HVZ-zhao,Konig:2015qat}
\begin{eqnarray}
	\langle M(k,\varepsilon)|\bar{q}\,\gamma^\mu q|0\rangle=-i f_{M}^q m_{V} \varepsilon^{*\mu}.
\end{eqnarray}
We use the following relations to simplify our calculation
\begin{eqnarray}
	\sum_{q}f_{M}^{q}Q_{q}=f_{M}Q_{M},\qquad\qquad\sum_{q}f_{M}^{q}\upsilon_{q}=f_{M}\upsilon_{M}.
	\label{qf}
\end{eqnarray}
The mesons decay constants $f_{M}, Q_{M} $ and $ \upsilon_{M}$ for the vector meson $M=[\rho,\omega,\phi,J/\Psi,\Upsilon(1S)]$ can be seen in Table~\ref{tab3}.

\begin{table}
	\begin{tabular}{|cccccc|}
		\hline
		Mesons $M$ & $m_{M}$/GeV & $f_{M}$/GeV & $Q_{M}$ & $\upsilon_{M}$ & $f_{M}^{\bot}/f_{M}=f_{M}^{q\bot}/f_{M}^{q}$ \\
		\hline
		$\rho$ & 0.77 & 0.216 & $\frac{1}{\sqrt{2}}$ & $\frac{1}{\sqrt{2}} \left( \frac{1}{2} - s_W^2 \right)$ & 0.72 \\
		$\omega$ & 0.782 & 0.194 & $\frac{1}{3\sqrt{2}}$ & $-\frac{s_W^2}{3\sqrt{2}}$ & 0.71 \\
		$\phi$ & 1.02 & 0.223 & $-\frac{1}{3}$ & $-\frac{1}{4} + \frac{s_W^2}{3}$ & 0.76 \\
		$J/\psi$ & 3.097 & 0.403 & $\frac{2}{3}$ & $\frac{1}{4} - \frac{2s_W^2}{3}$ & 0.91 \\
		$\Upsilon(1S)$ & 9.46 & 0.648 & $-\frac{1}{3}$ & $-\frac{1}{4} + \frac{s_W^2}{3}$ & 1.09 \\
		\hline
	\end{tabular}
	\caption{The mesons decay constants $f_{M}, Q_{M} $ and $ \upsilon_{M}$ will be used in the numerical analysis, $f_{M}^{\perp}$ and $f_{M}^{q\perp}$ represent the transverse decay constants and the flavor-specific transverse decay constants, respectively~\cite{Konig:2015qat}.}
	\label{tab3}
\end{table}

The concrete forms of $C_{\gamma Z}$ and $\widetilde{C}_{\gamma Z}$ in Eqs.~(\ref{form-factors1})-(\ref{form-factors}) are given by~\cite{HVZ-sm,HVZ-zhao}
\begin{eqnarray}
	C_{\gamma Z}=C_{\gamma Z}^{SM}+C_{\gamma Z}^{NP},\qquad\qquad \widetilde{C}_{\gamma Z}=\widetilde{C}_{\gamma Z}^{SM}+\widetilde{C}_{\gamma Z}^{NP},
\end{eqnarray}
\begin{eqnarray}
	C_{\gamma Z}^{SM}&=&\sum_{q}\kappa_{q}\frac{2N_cQ_q\upsilon_q}{3}A_f(\tau_q,r_Z)+\sum_{l}\kappa_{l}\frac{2Q_l\upsilon_l}{3}A_f(\tau_l,r_Z)-\frac{1}{2}\kappa_{W}A_{W}^{\gamma Z}(\tau_{W},r_Z), \\
	\widetilde{C}_{\gamma Z}^{SM}&=&\sum_{q}\tilde{\kappa}_qN_cQ_q\upsilon_q B_f(\tau_q,r_Z)+\sum_l\tilde{\kappa}_lQ_l\upsilon_lB_f(\tau_l,r_Z),
\end{eqnarray}
where $\tau _f=4m_f^2/m_h^2$ $( f=q,l)$. We use the running quark masses  $m_q$, evaluated at the hadronic scale $\mu_{hZ}$, which is given by $\mu_{hZ} = (m_h^2 - m_Z^2) / m_h \approx 58.78$ GeV~\cite{Konig:2015qat}. $\upsilon_l$ are the vector couplings of the $Z$ boson to the leptons and $Q_l$ represent the charge of leptons. $\tilde{\kappa}_q$ and $\tilde{\kappa}_l$ are the $CP$-odd effective Higgs couplings to the quarks and the leptons respectively. $C_{\gamma Z}^{SM}$ and $\widetilde{C}_{\gamma Z}^{SM}$ are the corresponding SM $CP$-even and $CP$-odd contributions to $h {\rightarrow} \gamma Z$ respectively. $C_{\gamma Z}^{NP}$ and $\widetilde{C}_{\gamma Z}^{NP}$ are the NP $CP$-even and $CP$-odd contributions to $h {\rightarrow} \gamma Z$ respectively. The loop functions $A_f$, $A_W^{\gamma Z}$ and $B_f$ can be found in Refs.~\cite{HVZ-sm1,HVZ-sm2,Konig:2015qat}.

In the SM, $\tilde{\kappa}_q = \tilde{\kappa}_l = 0$, so the $CP$-odd coupling $\widetilde{C}_{\gamma Z}^{SM}$ is 0~\cite{HVZ-sm}. In the NP, $h\gamma Z$ interaction with $CP$-even and $CP$-odd parts can be written as $\bar{F}_2 i (A + B \gamma_{5}) F_1 h$, where $A$ is the $CP$-even part and $B$ is the $CP$-odd part~\cite{HVZ-sm,HVZ-zhao}. Considering the interaction $\bar{F}_2 i (C^{L} P_{L} + C^{R} P_{R}) F_1 h$ with $P_{L} = \frac{1 - \gamma_{5}}{2}$ and $P_{R} = \frac{1 + \gamma_{5}}{2}$, the $CP$-even part can be $A = \frac{1}{2} (C^{L} + C^{R})$ and the $CP$-odd part is $B = \frac{1}{2} (C^{L} - C^{R})$. In the FDM, $C_{hH_{\alpha}^{+}H_{\alpha}^{-}}^{L} = C_{hH_{\alpha}^{+}H_{\alpha}^{-}}^{R}$. So the $CP$-odd coupling $\widetilde{C}_{\gamma Z}^{NP}$ in the FDM can be neglected approximately. The expression of $CP$-even coupling $C_{\gamma Z}^{NP}$ in the FDM is
\begin{eqnarray}
	&&C_{\gamma Z}^{NP}=\frac{c_{W}}{2}[(2c_{W}^{2}-1)\sum_{\alpha=1}^{2} \kappa_{H_\alpha^\pm } \frac{m_{Z}^{2}}{m_{H_{\alpha}^{\pm}}^{2}}
	A_{0}(x_{H_{\alpha}^{\pm}},\lambda_{H_{\alpha}^{\pm}})],
\end{eqnarray}
where $x_{i}={4m_{i}^2/ m_{h}^2}$, $\lambda_{i}={4m_{i}^2/ m_{Z}^2}$ and 	$A_{0}(x_{H_{\alpha}^{\pm}},\lambda_{H_{\alpha}^{\pm}})$ can be found in Appendix~\ref{app-form}.

Compared to the indirect contributions, the direct contributions to the decay amplitudes can be calculated in a power series in $(\Lambda_{QCD}/m_{h})^2$ or $(m_{q}/m_{h})^2$ \cite{HVZ-sm,HVZ-zhao}. Here, the $\Lambda_{QCD}$ is a hadronic scale where the authors in Ref.~\cite{QCD5} provide $\Lambda_{QCD}/m_Z\sim0.01$, and $m_{q}$ are the effective masses of the constituent quarks of a given meson, where the asymptotic function $\phi_{M}^{\perp}(x)=6x(1-x)$ \cite{HVZ-zhao,phi-function3} is needed; then the direct contributions are as follows:
\begin{eqnarray}
	F_{\bot \text{direct}}^{MZ} &=& \sum\limits_{q} f_{M}^{q\bot} \upsilon_{q} \kappa_{q} \frac{3m_q}{2m_M} \frac{1 - r_{Z}^2 + 2r_{Z} \ln{r_{Z}}}{(1 - r_{Z})^2}, \\
	\tilde{F}_{\bot \text{direct}}^{MZ} &=& \sum\limits_{q} f_{M}^{q\bot} \upsilon_{q} \tilde{\kappa}_{q} \frac{3m_q}{2m_M} \frac{1 - r_{Z}^2 + 2r_{Z} \ln{r_{Z}}}{(1 - r_{Z})^2}.
\end{eqnarray}
$f_{M}^{q\bot}$ represent the flavor-specific transverse decay constants of the meson \cite{Konig:2015qat}. Other studies~\cite{Liu:2020nsm} indicate that the direct contribution is significantly smaller than the indirect contribution. Our calculations also show that the direct contribution is relatively small compared to the indirect contributions. Because of the difficulty of observing NP in the direct contribution and the prominence of NP effects in the indirect contribution, we neglect the former in our analysis.

Normalized to the SM expectation, the signal strengths for the Higgs decay channels are quantified by the ratios \cite{signal,Liu:2020mev}
\begin{eqnarray}
	\mu_{hMZ} &=& \frac{\sigma_{\text{NP}}(\text{ggF}) \text{Br}_{\rm{NP}}(h{\rightarrow} MZ)}{\sigma_{\rm{SM}}(\text{ggF}) \text{Br}_{\rm{SM}}(h{\rightarrow} MZ)},
	\label{eq}
	\\
	\mu_{hZ\gamma} &=& \frac{\sigma_{\rm{NP}}(\text{ggF}) \text{Br}_{\rm{NP}}(h{\rightarrow} Z\gamma)}{\sigma_{\rm{SM}}(\text{ggF}) \text{Br}_{\rm{SM}}(h{\rightarrow} Z\gamma)}.
	\label{eq3}
\end{eqnarray}

Through Eqs.~(\ref{eq}), (\ref{eq3}) and Eq.~(\ref{eq-cross}), we quantify the signal strengths for $h{\rightarrow} {MZ}$ and $h{\rightarrow} Z\gamma$:
\begin{eqnarray}
	\mu_{hMZ} &\approx& \frac{\Gamma_{\rm{NP}}(h{\rightarrow} gg)}{\Gamma_{\rm{SM}}(h{\rightarrow} gg)} \frac{\Gamma_{\rm{NP}}(h{\rightarrow} MZ)/\Gamma_{\rm{NP}}^{h}}{\Gamma_{\rm{SM}}(h{\rightarrow} MZ)/\Gamma_{\rm{SM}}^{h}}
	\nonumber\\
	&=& \frac{\Gamma_{\rm{SM}}^{h}}{\Gamma_{\rm{NP}}^{h}} \frac{\Gamma_{\rm{NP}}(h{\rightarrow} gg)}{\Gamma_{\rm{SM}}(h{\rightarrow} gg)} \frac{\Gamma_{\rm{NP}}(h{\rightarrow} MZ)}{\Gamma_{\rm{SM}}(h{\rightarrow} MZ)},
	\label{muss}\\
	\mu_{hZ\gamma} &\approx& \frac{\Gamma_{\rm{NP}}(h{\rightarrow} gg)}{\Gamma_{\rm{SM}}(h{\rightarrow} gg)} \frac{\Gamma_{\rm{NP}}(h{\rightarrow} Z\gamma)/\Gamma_{\rm{NP}}^{h}}{\Gamma_{\rm{SM}}(h{\rightarrow} Z\gamma)/\Gamma_{\rm{SM}}^{h}}
	\nonumber\\
	&=& \frac{\Gamma_{\rm{SM}}^{h}}{\Gamma_{\rm{NP}}^{h}} \frac{\Gamma_{\rm{NP}}(h{\rightarrow} gg)}{\Gamma_{\rm{SM}}(h{\rightarrow} gg)} \frac{\Gamma_{\rm{NP}}(h{\rightarrow} Z\gamma)}{\Gamma_{\rm{SM}}(h{\rightarrow} Z\gamma)}.
\end{eqnarray}

\subsection{Higgs radiative decay to vector quarkonium}

\begin{figure}
	\begin{center}
		\includegraphics[width=0.28\textwidth]{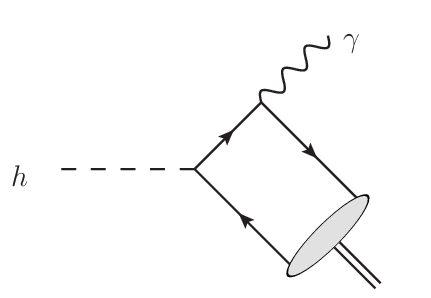}
		\includegraphics[width=0.28\textwidth]{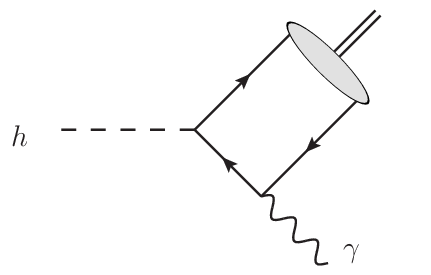}
		\includegraphics[width=0.25\textwidth]{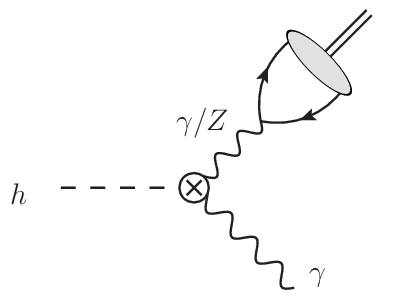}
		\parbox{15.5cm}
		{\caption{\label{fig:diags}
				Direct (left and center) and indirect (right) contributions to the $h\to M\gamma$ decay amplitude. The crossed circle in the third diagram denotes the off-shell $h\to\gamma\gamma^*$ and $h\to\gamma Z^*$ amplitudes, which in the SM arise first at one-loop order.}}
	\end{center}
\end{figure}

Analogous to the $h{\rightarrow} {MZ}$ process, the $h{\rightarrow} M\gamma$ decay rate is governed by the destructive interference of two amplitudes, one of which involves the Higgs coupling to the quark-antiquark pair inside the vector meson, and the other amplitude arises from the loop-induced effective $h\gamma \gamma^* $ and $h\gamma Z^*$  couplings, where the off-shell gauge boson converts into the vector meson~\cite{Konig:2015qat,Bodwin:2016edd,Dong:2022bkd}. The LO Feynman diagrams are shown in Fig.~\ref{fig:diags}, the first two graphs are the so-called direct contributions, and the third one is the indirect contributions. In this paper, we will use the results in Ref.~\cite{Brambilla:2019fmu} to calculate the indirect and direct amplitudes by computing the order-$v^4$ correction to the decay rate in the NRQCD factorization formalism.

For the process $h{\rightarrow} M\gamma$ $\left ( M=\rho, \omega, \phi \right ) $, the most general parametrization of the decay amplitude is~\cite{Konig:2015qat,Bodwin:2016edd}
\begin{eqnarray}
	i A(h {\rightarrow} M \gamma)=-\frac{e f_{V}}{2}\left[\left(\varepsilon_{V}^{*} \cdot \varepsilon_{\gamma}^{*}-\frac{q \cdot \varepsilon_{V}^{*} k \cdot \varepsilon_{\gamma}^{*}}{k \cdot q}\right) F_{1}^{V}-i \epsilon_{\mu \nu \alpha \beta} \frac{k^{\mu} q^{\nu} \varepsilon_{V}^{* \alpha} \varepsilon_{\gamma}^{* \beta}}{k \cdot q} F_{2}^{V}\right],
\end{eqnarray}
with $k$ representing the momentum of $M$ boson and $q$ representing the momentum of $\gamma$, so that $\varepsilon_{V}^{*} \cdot k = \varepsilon_{\gamma}^{*} \cdot q=0$. $\varepsilon_{V}^{*}$ and $\varepsilon_{\gamma}^{*}$ are, respectively, the polarization vectors of vector mesons. $F_1^V$ and $F_2^V$ will be defined below. $f_V$ is a decay constant and can be seen in Table~\ref{tab3}. All of them are influenced by the direct and indirect contributions. The decay width is
\begin{eqnarray}
	\Gamma(h {\rightarrow} M\gamma) = \frac{\alpha f_V^2}{8 m_h} \left( |F_1^V|^2 + |F_2^V|^2 \right).
\end{eqnarray}

Here, $\alpha $ = 1/137.036 is the fine-structure constant evaluated at $q^2 = 0$. 

For direct contributions, we have 
\begin{eqnarray}
	F_{1,\text{direct}}^V &&= \bar{\kappa}_V \, Q_V F_V \,, \qquad F_{2,\text{direct}}^V = i \bar{ \widetilde{\kappa}}_V \, Q_V F_V \,,\\
	F_V &&= \frac{6 m_b(\mu)}{v} \frac{f_V^\perp(\mu)}{f_V} \left[ 1 - \frac{C_F \alpha_s(\mu)}{\pi} \ln \frac{m_h^2}{\mu^2} \right] I_V(m_h) \,,\\
	I_V(m_h) &&= \sum_{n=0}^{\infty} C_{2n}(m_h, \mu) \, a_{2n}^{V\perp}(\mu) \,,
\end{eqnarray}
where
\begin{eqnarray}
	C_n(m_h, \mu) = 1 + \frac{C_F \alpha_s(\mu)}{4\pi} \left[ -4 \left( H_{n+1} - 1 \right) \left( \ln \frac{m_h^2}{\mu^2} - i\pi \right) + 4 H_{n+1}^2 - 3 + 4i\pi \right] + \mathcal{O}(\alpha_s^2) \,,
\end{eqnarray}
and $H_{n+1}=\sum\nolimits_{k=1}^{n+1} \frac{1}{k}$ are the harmonic numbers, $a_{2n}^{V\perp}(\mu)$ are the Gegenbauer moments, $\bar{\kappa}_V \,$ characterizes the $CP$-even gauge boson couplings in scenarios where the Higgs field interactions deviate from the SM expectations, $\bar{ \widetilde{\kappa}}_V \,$ is $CP$-odd.
For indirect contributions, we follow the treatment in Ref.~\cite{Konig:2015qat}:
\begin{eqnarray}
	F_{1,\text{indirect}}^V & = & \frac{\alpha(m_M)}{\pi} \frac{m_h^2 - m_M^2}{m_M v} \left[ Q_V C_{\gamma\gamma}(r_M) - \frac{v_M}{(s_W c_W)^2} \frac{m_M^2}{m_Z^2 - m_M^2} C_{\gamma Z}(r_M) \right], \nonumber \\\cr
	F_{2,\text{indirect}}^V & = & i \frac{\alpha(m_M)}{\pi} \frac{m_h^2 - m_M^2}{m_M v} \left[ Q_V \tilde{C}_{\gamma\gamma}(r_M) - \frac{v_M}{(s_W c_W)^2} \frac{m_M^2}{m_Z^2 - m_M^2} \tilde{C}_{\gamma Z}(r_M) \right],
\end{eqnarray}
\begin{eqnarray}
	C_{\gamma\gamma}(r_M) = \sum_q \kappa_{q} \frac{2 N_c Q_q^2}{3} A_f(\tau_q, r_M) + \sum_l \kappa_{l} \frac{2 Q_l^2}{3} A_f(\tau_l, r_M) - \frac{\kappa_{W}}{2} A_W^{\gamma\gamma}(\tau_W, r_M),\nonumber\\	
	C_{\gamma Z}(r_M) = \sum_q \kappa_{q} \frac{2 N_c Q_q v_q}{3} A_f(\tau_q, r_M) + \sum_l \kappa_{l} \frac{2 Q_l v_l}{3} A_f(\tau_l, r_M) - \frac{\kappa_{W}}{2} A_W^{\gamma Z}(\tau_W, r_M),
\end{eqnarray}
the first two terms in each coefficient are the contributions from the quarks and leptons, and the third term in $C_{\gamma\gamma}$ and $C_{\gamma Z}$ arises from gauge-boson loops,
\begin{eqnarray}
	\tilde{C}_{\gamma\gamma}(r_M) &=& \sum_q \tilde{\kappa}_q N_c Q_q^2 B_f(\tau_q, r_M) + \sum_l \tilde{\kappa}_l Q_l^2 B_f(\tau_l, r_M),\nonumber\\
	\tilde{C}_{\gamma Z}(r_M) &=& \sum_q \tilde{\kappa}_q N_c Q_q v_q B_f(\tau_q, r_M) + \sum_l \tilde{\kappa}_l Q_l v_l B_f(\tau_l, r_M),\label{kkk}
\end{eqnarray}
where $\tau _W=4m_W^2/m_h^2$; in the $\tau _q$ or $\tau _l$ we use the running masses $m_q\left ( m_h \right ) $ or $m_l\left ( m_h \right ) $. All loop functions $A_f, B_f, A_W^{\gamma\gamma}, A_W^{\gamma Z}$ can be found in Appendix~\ref{app-form}.

As discussed in $h{\rightarrow} MZ$, the $CP$-odd coupling $\widetilde{C}_{\gamma Z}^{SM}$ and $\widetilde{C}_{\gamma\gamma}^{SM}$ are 0 in the SM. Through the same discussion, the $CP$-odd coupling $\widetilde{C}_{\gamma Z}^{NP}$ and $\widetilde{C}_{\gamma\gamma}^{NP}$ in the FDM can be neglected approximatively. In the SM we have $\kappa_q = \kappa_l = \kappa_W = 1$. The effective Higgs couplings $\tilde{\kappa}_i$ entering in Eq.~(\ref{kkk}) all vanish in the SM. The expression of $CP$-even coupling $C_{\gamma Z}^{NP}$ and $C_{\gamma\gamma}^{NP}$ in the FDM are
\begin{eqnarray}
	C_{\gamma Z}^{NP} &=& \frac{c_{W}}{2} \left[ (2c_{W}^{2} - 1) \sum_{\alpha=1}^{2} \kappa_{H_\alpha^\pm} \frac{m_{Z}^{2}}{m_{H_{\alpha}^{\pm}}^{2}} A_{0}(x_{H_{\alpha}^{\pm}}, \lambda_{H_{\alpha}^{\pm}}) \right], \\
	C_{\gamma\gamma}^{NP} &=& \sum_{\alpha=1}^{2} \kappa_{H_\alpha^\pm} \frac{m_{Z}^2}{m_{H_\alpha^\pm}^2} A_0(x_{H_\alpha^\pm}),
\end{eqnarray}
where $x_{H_{\alpha}^{\pm}}={4m_{H_{\alpha}^{\pm}}^2/ m_{h}^2}$, $\lambda_{H_{\alpha}^{\pm}}={4m_{H_{\alpha}^{\pm}}^2/ m_{Z}^2}$ and $A_{0}(x_{H_{\alpha}^{\pm}},\lambda_{H_{\alpha}^{\pm}})$ can be found in Appendix~\ref{app-form}. And for reasons similar to those of $h{\rightarrow} MZ$, we ignore the direct contributions.

But in the process of $h{\rightarrow} M\gamma$ $[ M=J/\psi, \Upsilon(nS) ]$, the direct contributions cannot simply be ignored, because its value is significantly larger than that from indirect contributions. The decay rate $\Gamma \left ( h{\rightarrow}M\gamma   \right ) $ is given by

\begin{eqnarray}
\Gamma \left ( h{\rightarrow}M\gamma   \right )=| A_\mathrm{dir}+A_\mathrm{ind}| ^2,
\end{eqnarray}
where
\begin{eqnarray}
&&A_\mathrm{dir} = \frac{1}{2} \sqrt{\Phi} e e_Q \kappa_{q} \sum_{n=0,2,4,\alpha_s} U_{f_V^\perp}(\mu, \mu_0) f_V^\perp(\mu_0) \int_0^1 dx \, T_H(x, \mu) \phi_V^{\perp(n)}(x, \mu),\label{eq5}\\
&&A_\mathrm{ind} = -\sqrt{\Phi} \frac{e_Q}{|e_Q|} \frac{\sqrt{24\pi m_h}}{m_M} \left[ \frac{\Gamma(M {\rightarrow} \ell^+ \ell^-) \Gamma(h {\rightarrow} \gamma\gamma)}{\alpha(0)} \right]^{\frac{1}{2}}.\label{eq4}
\end{eqnarray}
and
\begin{equation}
\int_0^1 dx\, T_H (x,\mu) \phi_V^\perp{}^{(n)} (x, \mu)
= 
\sum_{n_1=0}^\infty 
\sum_{n_2=0} ^\infty \hat T_H (n_1,\mu) 
U_{n_1 n_2} (\mu, \mu_0) \hat \phi_V^\perp{}^{(n)} (n_2, \mu_0). 
\end{equation}
\begin{eqnarray}
	\Phi = \frac{1}{2m_h} \frac{m_M (m_h^2 - m_M^2)}{2\pi m_h^2}.
\end{eqnarray}

The definitions of the parameters evolution matrix $U_{f_V^\perp}(\mu, \mu_0)$, the decay constant $f_V^\perp(\mu_0)$, the perturbative hard part $T_H(x, \mu)$, light-cone distribution amplitude $\phi_V^{\perp(n)}(x, \mu)$, and evolution matrix $U_{n_1 n_2} (\mu, \mu_0)$ can be found in Refs.~\cite{Brambilla:2019fmu,Batra:2022wsd,Bodwin:2016edd}.

\begin{table*}
	\begin{tabular*}{\textwidth}{@{\extracolsep{\fill}}lllll@{}}
		\hline
		$M$ & $m_M$/{\rm GeV} & $\Gamma(M {\rightarrow} \ell^+ \ell^-)$/{\rm keV} \\
		\hline
		$J/\psi$ & 3.0969 & 5.529$\pm$ 0.03 \\
		$\Upsilon \left(1S\right)$ & 9.4604 & 1.291$\pm$ 0.043 \\
		$\Upsilon \left(2S\right)$ & 10.0234 & 0.6108$\pm$ 0.05 \\
		$\Upsilon \left(3S\right)$ & 10.3551 & 0.443$\pm$ 0.4 \\
		\hline
	\end{tabular*}
	\caption{The meson masses $m_M$ and the leptonic decay rates $\Gamma(M {\rightarrow} \ell^+ \ell^-)$ are all from ~\cite{ParticleDataGroup:2024cfk}.}
	\label{tab-l}
\end{table*} 

We take the values of $m_M$ and $\Gamma(M {\rightarrow} \ell^+ \ell^-)$ listed in Table~\ref{tab-l} as inputs. The average experimental data on the branch ratios are~\cite{ParticleDataGroup:2024cfk}
\begin{eqnarray}
	&&{\rm Br}(h{\rightarrow} \rho\gamma)< (1.04)\times 10^{-3},\nonumber\\
	&&{\rm Br}(h{\rightarrow} \omega\gamma )< (5.5)\times 10^{-4},\nonumber\\
	&&{\rm Br}(h{\rightarrow} \phi\gamma )< (5)\times 10^{-4},\nonumber\\
	&&{\rm Br}(h{\rightarrow} J/\psi\gamma )< (2.0)\times 10^{-4},\nonumber\\
	&&{\rm Br}(h{\rightarrow} \Upsilon (1S)\gamma )< (2.5)\times 10^{-4},\nonumber\\
	&&{\rm Br}(h{\rightarrow} \Upsilon (2S)\gamma )< (4.2)\times 10^{-4},\nonumber\\
	&&{\rm Br}(h{\rightarrow} \Upsilon (3S)\gamma )< (3.4)\times 10^{-4}.
	\label{eqBD}
\end{eqnarray}

Similarly, we define $\mu_{hM\gamma}$ $[M=\rho, \omega, \phi, J/\psi, \Upsilon(nS)]$ as follows:
\begin{eqnarray}
	\mu_{hM\gamma} &\approx& \frac{\Gamma_{\rm{NP}}(h{\rightarrow} gg)}{\Gamma_{\rm{SM}}(h{\rightarrow} gg)} \cdot \frac{\Gamma_{\rm{NP}}(h{\rightarrow} M\gamma)/\Gamma_{\rm{NP}}^{h}}{\Gamma_{\rm{SM}}(h{\rightarrow} M\gamma)/\Gamma_{\rm{SM}}^{h}} \nonumber\\
	&=& \frac{\Gamma_{\rm{SM}}^{h}}{\Gamma_{\rm{NP}}^{h}} \cdot \frac{\Gamma_{\rm{NP}}(h{\rightarrow} gg)}{\Gamma_{\rm{SM}}(h{\rightarrow} gg)} \cdot \frac{\Gamma_{\rm{NP}}(h{\rightarrow} M\gamma)}{\Gamma_{\rm{SM}}(h{\rightarrow} M\gamma)}.
\end{eqnarray}

\section{Numerical analysis\label{sec4}}
There are many free parameters in the FDM. In order to obtain a transparent numerical results, we make some assumptions on parameter space without losing the generality. For the free parameters in the scalar sector of the FDM, we take $v_1=v_2=v_{1,2}$, $\lambda_4'=\lambda_4''=\lambda_4/2$, $\lambda_5'=\lambda_5''=\lambda_5/2$, $\lambda_6'=\lambda_6''=\lambda_6/2$, and all parameters to be real for simplicity. The Higgs properties are affected negligibly by the gauge coupling constants $g_F$ and $g_{YF}$ and also considering the constraints from collider direct searches for new gauge bosons or scalars~\cite{Cao:2025zwn}; we take $g_F=0.20$, $g_{YF}=0.10$ and $v_\chi\geq4\;{\rm TeV}$ in the following analysis. Considering the requirements of tree-level perturbative unitary bounds condition, we adopt the following parameter ranges~\cite{Yang:2024znv}:
\begin{eqnarray}
	&&|\lambda_1|,\;|\lambda_2|,\;|\lambda_3|,\;|\lambda_\chi|\leq\frac{4\pi}{3},\;|{\rm Re}(\lambda_{10})|\leq4\pi,\nonumber\\
	&&|\lambda_4'+\lambda_4''|,\;|\lambda_5'+\lambda_5''|,\;|\lambda_6'+\lambda_6''|,\;|\lambda_7|,\;|\lambda_8|,\;|\lambda_9|\leq 8\pi.\label{eqPU}
\end{eqnarray}

And the vacuum stability requirement the scalar potential should be bounded by the conditions~\cite{Cao:2025zwn}
\begin{eqnarray}
&&\lambda_1> 0,\quad \lambda_2> 0,\quad \lambda_3> 0,\quad \lambda_\chi > 0,\quad \left | \lambda_{10} \right |^2\le (L_{13} +2 \sqrt{\lambda_1 \lambda_3})(L_{23} +2 \sqrt{\lambda_2 \lambda_3}),\nonumber\\
&&L_{12} +2 \sqrt{\lambda_1 \lambda_2}\ge 0,\quad L_{13} +2 \sqrt{\lambda_1 \lambda_3}\ge 0,\quad L_{23} +2 \sqrt{\lambda_2 \lambda_3}\ge 0,\nonumber\\
&&\lambda_7 +2 \sqrt{\lambda_1 \lambda_\chi}\ge 0,\quad \lambda_8 +2 \sqrt{\lambda_2 \lambda_\chi}\ge 0,\quad \lambda_9 +2 \sqrt{\lambda_3 \lambda_\chi}\ge 0,\nonumber\\
&&\sqrt{\lambda_1 \lambda_2 \lambda_\chi} +\frac{L_{12} \sqrt{\lambda_\chi}+\lambda_7 \sqrt{\lambda_2}+\lambda_8 \sqrt{\lambda_1}}{2}+\frac{1}{2} \sqrt{(L_{12}+2\sqrt{\lambda_1 \lambda_2})(\lambda_7+2\sqrt{\lambda_1 \lambda_\chi})(\lambda_8+2\sqrt{\lambda_2 \lambda_\chi})}\ge 0,\nonumber\\
&&\sqrt{\lambda_1 \lambda_3 \lambda_\chi} +\frac{L_{13} \sqrt{\lambda_\chi}+\lambda_7 \sqrt{\lambda_3}+\lambda_9 \sqrt{\lambda_1}}{2}+\frac{1}{2} \sqrt{(L_{13}+2\sqrt{\lambda_1 \lambda_3})(\lambda_7+2\sqrt{\lambda_1 \lambda_\chi})(\lambda_9+2\sqrt{\lambda_3 \lambda_\chi})}\ge 0,\nonumber\\
&&\sqrt{\lambda_2 \lambda_3 \lambda_\chi} +\frac{L_{23} \sqrt{\lambda_\chi}+\lambda_8\sqrt{\lambda_3}+\lambda_9 \sqrt{\lambda_2}}{2}+\frac{1}{2} \sqrt{(L_{23}+2\sqrt{\lambda_2 \lambda_3})(\lambda_8+2\sqrt{\lambda_2 \lambda_\chi})(\lambda_9+2\sqrt{\lambda_3 \lambda_\chi})} \ge 0,\nonumber\\
&&\sqrt{\lambda_1 \lambda_2 \lambda_3} +\frac{L_{12} \sqrt{\lambda_3}+\tilde{L}_{13}\sqrt{\lambda_2}+\tilde{L}_{23}\sqrt{\lambda_1}}{2}+\frac{1}{2} \sqrt{(L_{12}+2\sqrt{\lambda_1 \lambda_2})(\tilde{L}_{13}+2\sqrt{\lambda_1 \lambda_3})(\tilde{L}_{23}+2\sqrt{\lambda_2 \lambda_3})} \ge 0.\nonumber\\
\label{eqVS}
\end{eqnarray}
where
\begin{eqnarray}
&&L_{12}=\lambda_4'+\text{min}(0,\lambda_4''),\quad L_{13}=\lambda_5'+\text{min}(0,\lambda_5''),\quad L_{23}=\lambda_6'+\text{min}(0,\lambda_6''),\nonumber\\
&&\tilde{L}_{13}=L_{13}-\left | \lambda_{10} \right | \sqrt{\frac{L_{13} +2 \sqrt{\lambda_1 \lambda_3}}{L_{23} +2 \sqrt{\lambda_2 \lambda_3}} },\nonumber\\
&&\tilde{L}_{23}=L_{23}-\left | \lambda_{10} \right | \sqrt{\frac{L_{23} +2 \sqrt{\lambda_2 \lambda_3}}{L_{13} +2 \sqrt{\lambda_1 \lambda_3}} }.
\label{eqLC}
\end{eqnarray}

For relevant parameters in the SM, we choose~\cite{ParticleDataGroup:2024cfk}
\begin{eqnarray}
&&\alpha_s(m_{Z})=0.118,\quad  m_t=172.57\;{\rm GeV},\quad  m_{c}=1.273\;{\rm GeV},\quad  m_{Z}=91.1880\;{\rm GeV},
\nonumber\\
&&\alpha(m_{Z})=1/127.93,\quad\,  m_b=4.183\;{\rm GeV},\quad\,  m_\tau=1.777\;{\rm GeV},  m_{W}=80.3692\;{\rm GeV}.
\end{eqnarray}

\begin{figure}
	\setlength{\unitlength}{1mm}
	\centering
	\includegraphics[width=2.3in]{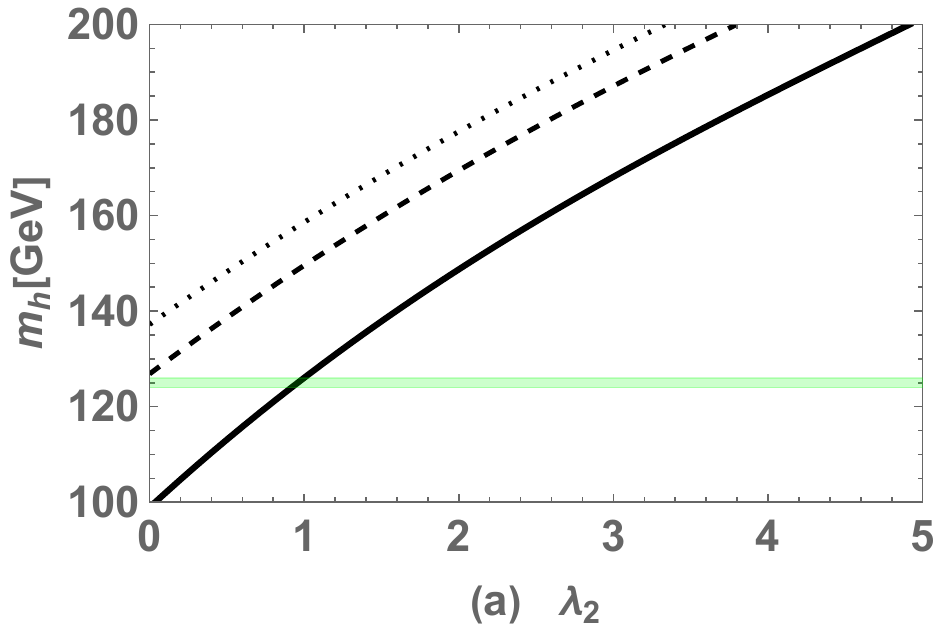}
	\vspace{0.2cm}
	\includegraphics[width=2.3in]{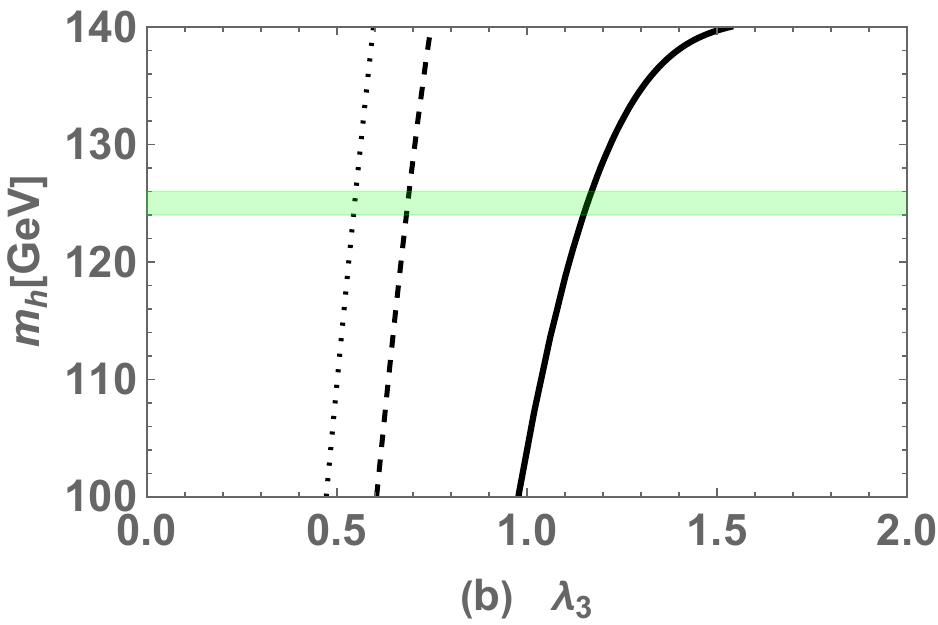}
	\vspace{0.2cm}
	\includegraphics[width=2.3in]{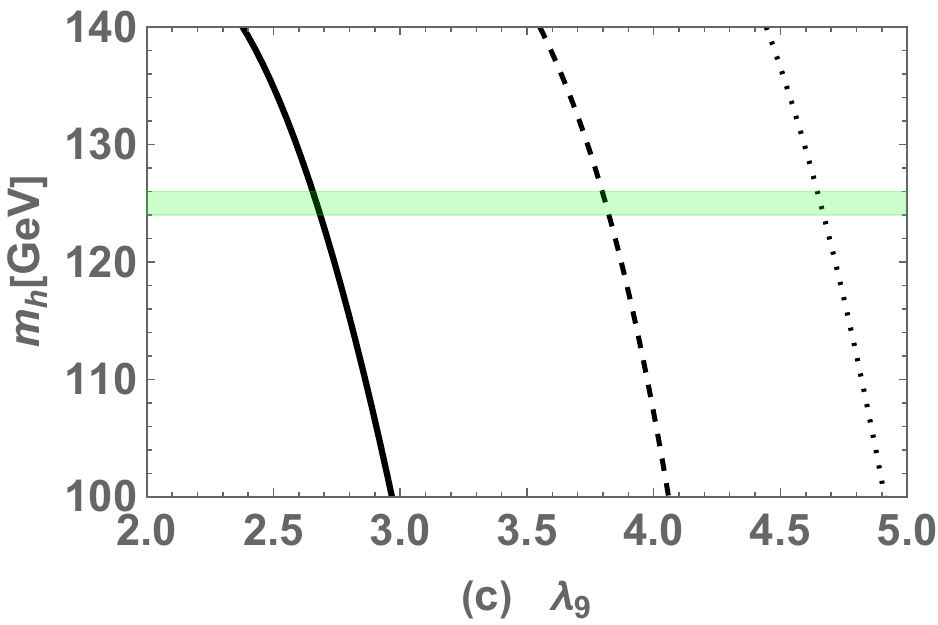}
	\vspace{0.2cm}
	\includegraphics[width=2.3in]{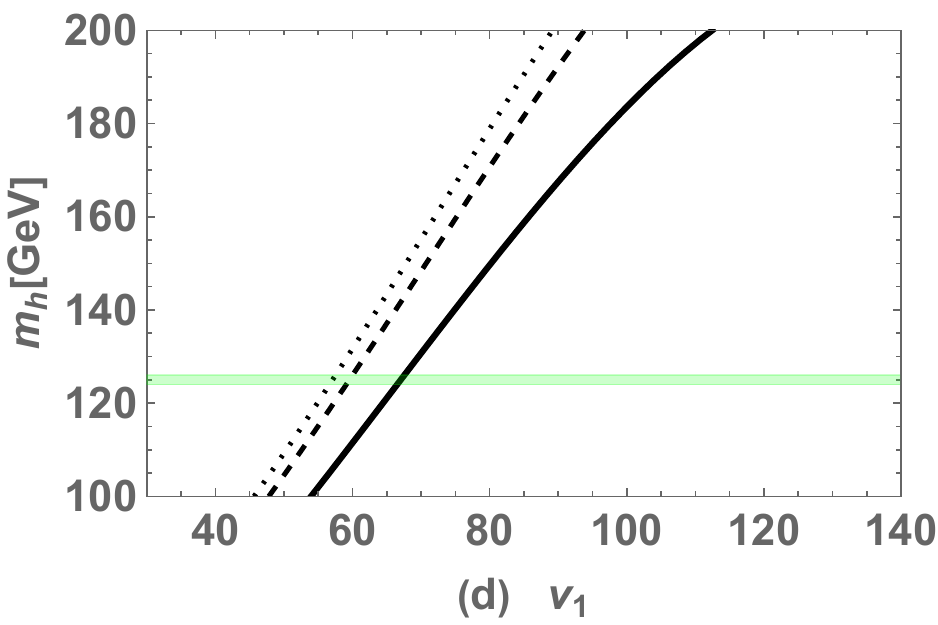}
	\vspace{0cm}
	\caption[]{The plots show $m_{h}$ versus $\lambda_2$ (a), $m_{h}$ versus $\lambda_9$ (b),  $m_{h}$ versus $\lambda_2$ (c), and $m_{h}$ versus $v_1$ (d), where the solid, dashed and dotted curves represent the results for $\lambda_\chi=2,\;3,\;4$ respectively. The light green areas indicate the range of $124\;{\rm GeV}<m_{h}<126\;{\rm GeV}$.}
	\label{fig-V7}
\end{figure}
In the FDM, the mass of the lightest Higgs boson can be obtained by diagonalizing the squared mass matrix in Eq.~(\ref{eq4h}). Among the three Higgs doublets, $\Phi _3$ is identified as the SM-like Higgs boson. As indicated by Eq.~(\ref{eqmh}), $\lambda_{3}$ exerts a significant influence on $m_{h}$. Furthermore, the mixing between $S_3$ and $S_\chi $ implies that both $\lambda _9 $ and $\lambda _\chi$
play a crucial role in determining $m_{h}$. In our chosen parameter space, $v_1$ not only affect the Yukawa couplings directly, but also affect the mixing effect of three Higgs doublets. It indicates that $v_1$ plays a crucial role in determining the numerical results. To investigate the effects of $\lambda _2 $, $\lambda _3 $, $\lambda _9 $, $S_\chi $, and $v_1$ on $m_{h}$, we take $\;\lambda_1=\lambda_2=\lambda_4=\lambda_5=\lambda_6=\lambda_7=\lambda_8=2,\;\lambda_{10}=-2,\; $and $\kappa=-1\;{\rm TeV}$, and plot $m_{h}$ versus $\lambda _2 $, $\lambda _3 $, $\lambda _9 $, and $S_\chi $ in Figs.\ref{fig-V7} (a)-(d) respectively, where the light green zone represents the range of $124\;{\rm GeV}< m_h< 126\;{\rm GeV}$ and the solid, dashed, and dotted curves denotes the results for $\lambda_\chi=2,\;3,\;4$ respectively. These plots show that $m_{h}$ increases with higher values of $\lambda_3$, $\lambda_\chi$, $\lambda_2$, and $v_1$ and decreases with higher values of $\lambda_9$. Furthermore, the parameters $\lambda_2$, $\lambda_3$, $\lambda_9$, $v_1$, and $\lambda_\chi$ have a significant impact on the theoretical predictions of $m_{h}$.
\begin{table*}
	\begin{tabular*}{\textwidth}{@{\extracolsep{\fill}}lllll@{}}
		\hline
		Parameters&Min&Max\\
		\hline
		$v_1/{\rm GeV}$&30&142&\\
		$\lambda_i$ $\;(i=2,3,9,\chi)$&0&5\\
		$\lambda_{10}$&-4&0\\
        $v_\chi/{\rm TeV}$&4& 10\\
		\hline
	\end{tabular*}
	\caption{Scanning parameters for scalar potential coefficient on the Higgs decays.}
	\label{tab-scan}
\end{table*}
\begin{table*}
	\begin{tabular*}{\textwidth}{@{\extracolsep{\fill}}lllll@{}}
		\hline
		Observables&$O_i^{\rm th}$&$O_i^{\rm exp}$&$\sigma_i^{\rm exp}$\\
		\hline
		$\mu_{hZZ^*}$&0.988&1.02&0.08\\
		$\mu_{hWW^*}$&0.988&1.00&0.08\\
		$\mu_{h\tau\bar{\tau}}$&0.984&0.91&0.09\\
		$\mu_{hb\bar{b}}$&0.998&0.99&0.12\\
		$\mu_{h\gamma\gamma}$&1.094&1.10&0.06\\
		$\mu_{hZ\gamma}$&1.0253&2.2&0.7\\
		$m_h$&125.20&125.20&0.11\\
		\hline
	\end{tabular*}
	\caption{The results obtained for the best fit point corresponding to $\chi ^2=3.94$.}
	\label{tab-chi}
\end{table*}

To explore the combined effects of $\lambda _2 $, $\lambda _3 $, $\lambda _9 $, $\lambda _{10} $, $\lambda_\chi $, and $v_1$ on the Higgs signal strengths, we fix $\;\lambda_1=\lambda_2=\lambda_4=\lambda_5=\lambda_6=\lambda_7=\lambda_8=2,\; $and $ \kappa=-1\;{\rm TeV}$, and scan the parameter space listed in Table \ref{tab-scan}. In addition, in this work, we perform a $\chi^2$ test to explore the best fit describing 125 GeV Higgs mass and the signal strength in the model. Generally, the $\chi^2$ function can be constructed as
\begin{eqnarray}
	&&\chi^2=\sum_1 \Big(\frac{O_i^{\rm th}-O_i^{\rm exp}}{\sigma_i^{\rm exp}}\Big)^2,
\end{eqnarray}
where $O_i^{\rm th}$ denotes the $i$th observable computed theoretically, $O_i^{\rm exp}$ is the corresponding experimental value and $\sigma_i^{\rm exp}$ is the uncertainty in $O_i^{\rm exp}$, $O_i^{\rm th}$, $\sigma_i^{\rm exp}$ and $O_i^{\rm exp}$ are listed in Table~\ref{tab-chi} and Eq.~(\ref{eqstu}). The parameters obtaining the best fit are
\begin{eqnarray}
	&&\lambda _2=0.907,\quad  \lambda _3=2.201,\quad  \lambda _9=3.885,\nonumber\\
	&&\lambda _{10}=-3.135,\quad  \lambda _\chi =4.807,\quad  v_1=128.267,\quad  v_\chi=7689.94.	\label{eq.best}
\end{eqnarray}

\begin{figure}
	\setlength{\unitlength}{1mm}
	\centering
	\includegraphics[width=2.1in]{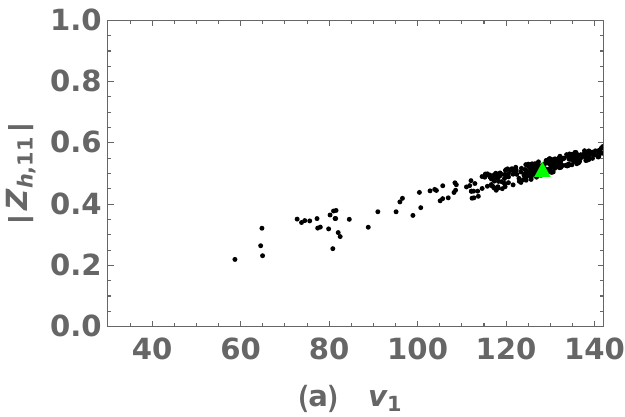}
	\vspace{0.2cm}
	\includegraphics[width=2.1in]{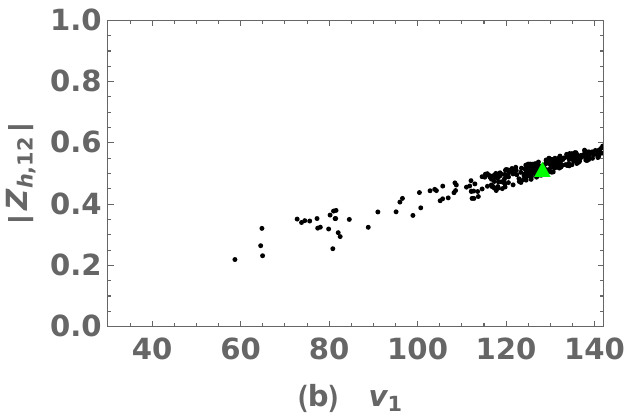}
	\vspace{0.2cm}
	\includegraphics[width=2.1in]{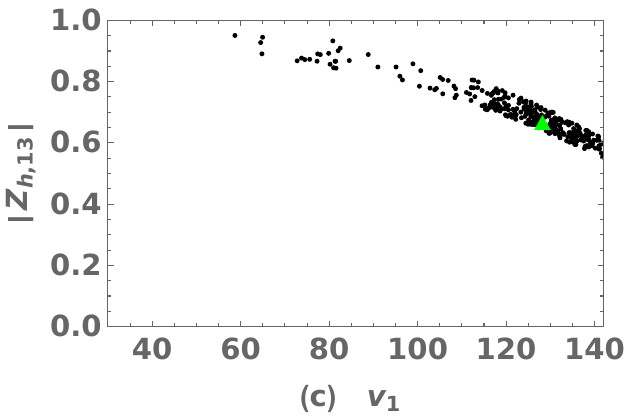}
	\vspace{0cm}
	\caption[]{Scanning the parameter space in Table~\ref{tab-scan} keeping $\;\lambda_1=\lambda_2=\lambda_4=\lambda_5=\lambda_6=\lambda_7=\lambda_8=2,\; $and $ \kappa=-1\;{\rm TeV}$, and the results of $v_1-\left |Z_{h,11}  \right | $ (a), $v_1-\left |Z_{h,12}  \right | $ (b), and $v_1-\left |Z_{h,13}  \right | $ (c) are plotted, where the green triangle denotes the best fit point $\chi ^2=3.94$.}
	\label{fig-V6}
\end{figure}

Additionally, to explore the mixing effects among the three Higgs doublets, we scan the parameter space in Table~\ref{tab-scan}, keeping $\;\lambda_1=\lambda_2=\lambda_4=\lambda_5=\lambda_6=\lambda_7=\lambda_8=2,\; $and $  \kappa=-1\;{\rm TeV}$ to examine the variation of $\left|Z_{h,11}\right|$, $\left|Z_{h,12}\right|$, $\left|Z_{h,13}\right|$ with respect to $v_1$ in Fig.~\ref{fig-V6} (a)-(c) respectively, where green triangle denotes the best fit point. We take $v_3 > v_{1,2}$ in the chosen parameter space, which indicates that $\Phi_3$ dominates the SM-like Higgs, i.e., $|Z_{h,13}|>|Z_{h,11}|\approx |Z_{h,12}|$. With the increasing of $ v_{1,2}$, the contributions from $\Phi_1$ and $\Phi_2$ to the lightest SM-like Higgs play more important roles, and the ones from $\Phi_1$, $\Phi_2$, and $\Phi_3$ are comparable when $v_3 \approx v_{1,2}\approx 142.2~\rm GeV$. All these facts can be seen obviously in Fig.~\ref{fig-V6}. 

 Then, based on the points obtained in Fig.~\ref{fig-V6}, we will explore the impact of $\lambda _2 $, $\lambda _3 $, $\lambda _9 $, $\lambda _{10} $, $\lambda_\chi $, $v_\chi$ and $v_1$ on the signal strength in the following subsections. 

\subsection{$h{\rightarrow}\gamma\gamma,gg$, $h{\rightarrow} VV^*$ ($V=Z,W$), $h{\rightarrow} f\bar{f}$ ($f=c,b,\tau$), and $h{\rightarrow} Z\gamma$\label{sec4.1}}

\begin{figure}
	\setlength{\unitlength}{1mm}
	\centering
	\includegraphics[width=2.1in]{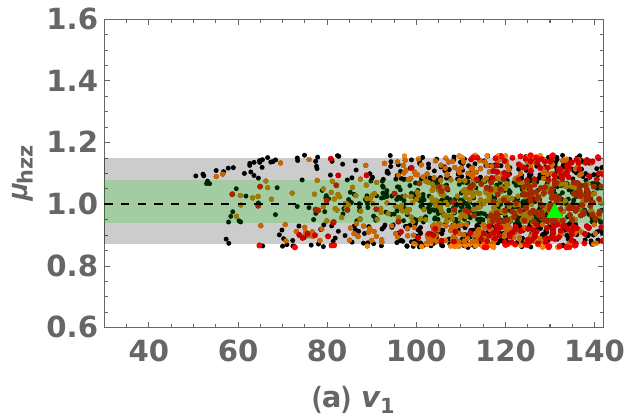}
	\vspace{0.2cm}
	\includegraphics[width=2.1in]{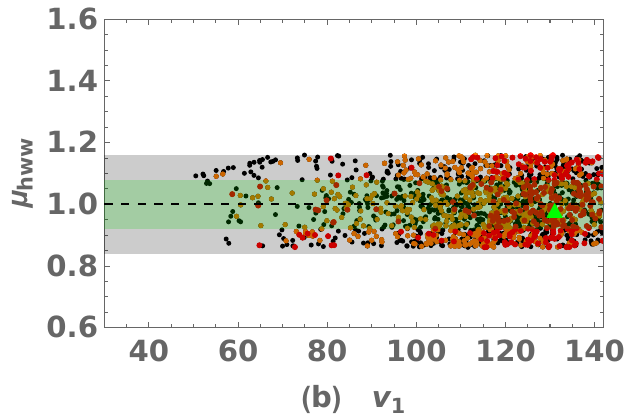}
	\vspace{0.2cm}
	\includegraphics[width=2.1in]{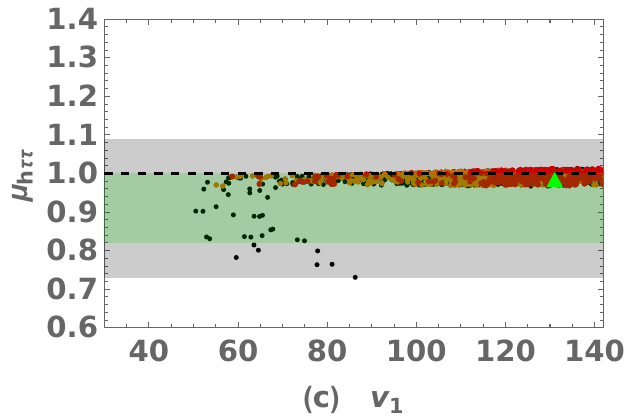}
	\vspace{0.2cm}
	\includegraphics[width=2.1in]{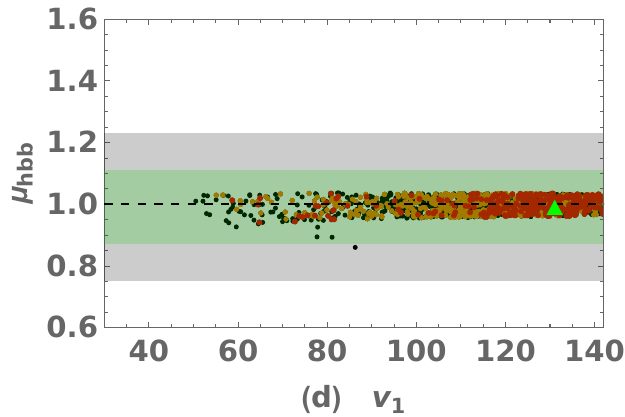}
	\vspace{0.2cm}
	\includegraphics[width=2.1in]{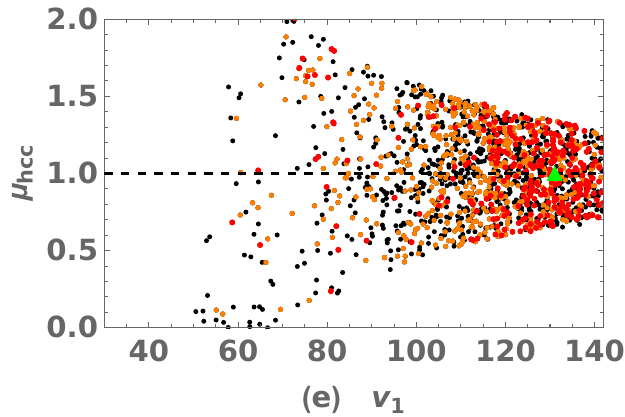}
	\vspace{0.2cm}
	\includegraphics[width=2.1in]{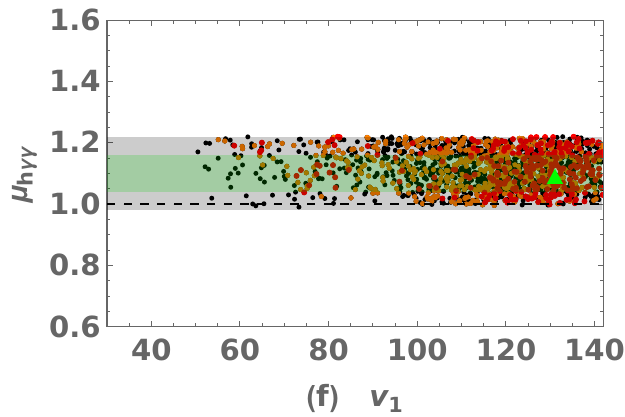}
	\vspace{0.2cm}
	\includegraphics[width=2.1in]{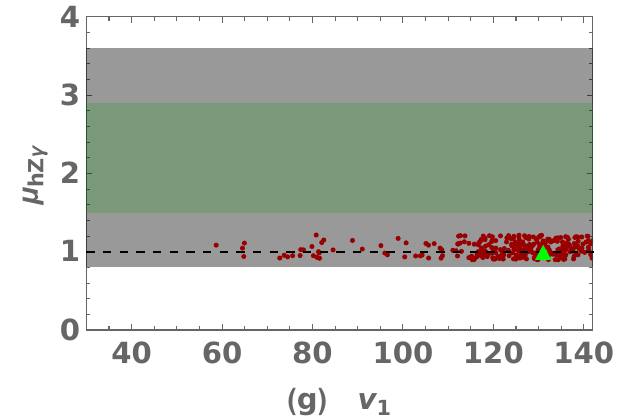}
	\vspace{0cm}
	\caption[]{Keeping $\;\lambda_1=\lambda_2=\lambda_4=\lambda_5=\lambda_6=\lambda_7=\lambda_8=2,\; $and $ \kappa=-1\;{\rm TeV}$ and scanning the parameter space in Table~\ref{tab-scan}. $\mu_{hZZ^*}$ (a),  $\mu_{hWW^*}$ (b), $\mu_{h\tau\bar{\tau}}$ (c), $\mu_{hb\bar{b}}$ (d), $\mu_{hc\bar{c}}$ (e), $\mu_{h\gamma\gamma}$ (f) and $\mu_{hZ\gamma}$ (g) versus $v_1$ are plotted, respectively. The gray areas represent the $2\sigma $ confidence interval, and the green areas represent the $1\sigma $ confidence interval, the horizontal dashed line represents a signal strength of 1, where the green triangle denotes the best fit point $\chi ^2=3.94$.}
	\label{fig-V1}
\end{figure}

We focus on the precisely measured signal strengths $\mu_{h\gamma\gamma}$, $\mu_{hVV^*}$, $\mu_{hf\bar{f}}$and $\mu_{hZ\gamma}$ first, and the results of $\mu_{hZZ^*}$ (a),  $\mu_{hWW^*}$ (b), $\mu_{h\tau\bar{\tau}}$ (c), $\mu_{hb\bar{b}}$ (d), $\mu_{hc\bar{c}}$ (e), $\mu_{h\gamma\gamma}$ (f) and $\mu_{hZ\gamma}$ (g) versus $v_1$ are plotted in Fig.~\ref{fig-V1}, where these plots are obtained by scanning the parameter space listed in Table~\ref{tab-scan} and keeping $\;\lambda_1=\lambda_2=\lambda_4=\lambda_5=\lambda_6=\lambda_7=\lambda_8=2,\;\kappa=-1\;{\rm TeV}$ and $124\,{\rm GeV} < m_h < 126\,{\rm GeV}$. In addition, we employ the conditions of several FCNC processes in Sec.~\ref{subFCNC} and electroweak precision observables in Sec.~\ref{subEWPM} described as additional constraints. All signal strengths lie within the 2$\sigma$ confidence intervals. Fig.~\ref{fig-V1} depicts the 1$\sigma$ (light green region) and 2$\sigma$ (gray region) confidence intervals. The dashed line indicates a signal strength of 1. The best fit point is denoted by the green triangle. The black dots denote points satisfying only the FCNC constraints, the orange dots correspond to points that fulfill both the FCNC and signal strength constraints and red dots correspond to the cases that are consistent with all three constraints. Notably, a significant region of parameter space with $v_1 > 60$ GeV accommodates signal strengths consistent with experimental constraints even as $v_3 \approx v_{1,2}\approx 142.2~\rm GeV$. This demonstrates the viability of the chosen parameter space in reproducing the observed data. In the process in Fig.~\ref{fig-V1}(e), due to the Yukawa coupling between Higgs and charm quark with respect to $v_1$ sensitively, the overall trend shows a noticeable change. 
   
In the best fit point in Fig.~\ref{fig-V1}(g), the branch ratio is $Br[h{\rightarrow} Z\gamma]=\ ( 1.56\pm0.06 \ ) \times 10^{-3}$, and the error in the branch ratio mainly stems from the uncertainty associated with the full width of the Higgs boson. The prediction of the FDM for the branch ratio is notably below the measured value that our results do not fall within the $1\sigma$ confidence interval range. While the FDM prediction differs significantly from experimental results, it is too early to dismiss the SM or FDM.  The discrepancy may stem from experimental uncertainties rather than a fundamental flaw in either model. Future, higher-precision measurements from the HL-LHC and potential Higgs factories (CEPC, FCC-ee) are crucial to resolve this issue~\cite{Sang:2024vqk}.
 
\subsection{$h{\rightarrow} {MZ}$}

To help to analyze the results of the processes $h{\rightarrow} {MZ}$, we define the ratio
\begin{table}
	\begin{tabular}{| c | c | c | c | c | c |}
		\hline
		\quad & \bm{${\rho}$} & \bm{$\omega$} & \bm{$\phi$} & \rm{J}/\bm{$\psi$} & \bm{$\Upsilon(1S)$} \\
		\hline
		\bm{$F_{\parallel ind,NP}^{MZ}$} & $\makecell[c]{0.042\\+4.05\times10^{-4}C_{\gamma Z}}$ & $\makecell[c]{-0.01\\+1.2\times10^{-4}C_{\gamma Z}}$ & $\makecell[c]{-0.039\\-2\times10^{-4}C_{\gamma Z}}$ & $\makecell[c]{0.040\\+7.3\times10^{-4}C_{\gamma Z}}$ & $\makecell[c]{-0.114\\-6\times10^{-4}C_{\gamma Z}}$ \\
		\hline
		\bm{$F_{\parallel ind,SM}^{MZ}$} & $\makecell[c]{0.041}$ & $\makecell[c]{-0.01}$ & $\makecell[c]{-0.038}$ & $\makecell[c]{0.039}$ & $\makecell[c]{-0.11}$ \\
		\hline
		\bm{$F_{\perp ind,NP}^{MZ}$} & $\makecell[c]{0.042\\+1.10C_{\gamma Z}}$ & $\makecell[c]{-0.01\\+0.32C_{\gamma Z}}$ & $\makecell[c]{-0.039\\-0.31C_{\gamma Z}}$ & $\makecell[c]{0.040\\+0.12C_{\gamma Z}}$ & $\makecell[c]{-0.114\\-0.011C_{\gamma Z}}$ \\
		\hline
		\bm{$F_{\perp ind,SM}^{MZ}$} & $-2.63$ & $-0.79$ & $0.71$ & $-0.26$ & $-0.088$ \\
		\hline
		\bm{$\frac{8r_{M}r_{Z}}{(1-r_{Z}-r_{M})^2}$} & $7.35\times10^{-4}$ & $7.59\times10^{-4}$ & $1.29\times10^{-3}$ & $1.19\times10^{-2}$ & 0.114 \\
		\hline
		\bm{$R_{M}-1$} & $\makecell[c]{-0.74 + 0.132\times\\\left |  -2.37 + C_{\gamma Z}^{NP} \right |^2}$ & $\makecell[c]{-0.013+ 0.135\times\\\left |  -2.473 + C_{\gamma Z}^{NP} \right |^2}$ & $\makecell[c]{-0.29+ 0.058\times\\\left |  -2.24 + C_{\gamma Z}^{NP} \right |^2}$ & $\makecell[c]{-0.295+ 0.079\times\\\left |  -1.93 + C_{\gamma Z}^{NP} \right |^2}$ & $\makecell[c]{-0.172+ 0.001\times\\\left |  13.32 + C_{\gamma Z}^{NP} \right |^2}$\\
		\hline
	\end{tabular}
	\caption{The indirect contributions for the decay width of $h{\rightarrow} MZ$, where $C_{\gamma Z}=C_{\gamma Z}^{SM}+C_{\gamma Z}^{NP}$, with $C_{\gamma Z}^{SM}\simeq-2.43$. }
	\label{tabmeson}
\end{table}
\begin{eqnarray}
	R_{M} = \frac{\Gamma_{U(1)_F}(h{\rightarrow} MZ)}{\Gamma_{\rm{SM}}(h{\rightarrow} MZ)},
\end{eqnarray}
\begin{figure}
	\setlength{\unitlength}{1mm}
	\centering
	\includegraphics[width=2.1in]{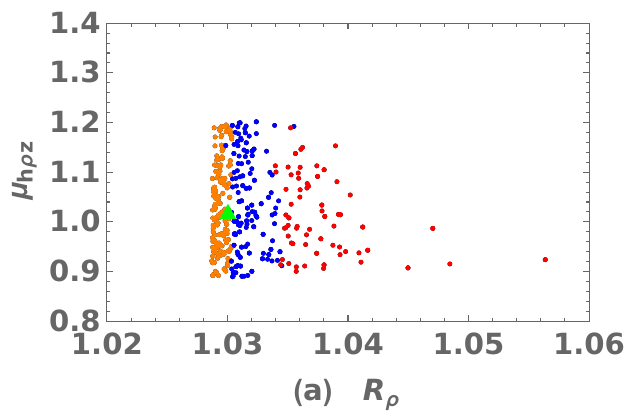}
	\vspace{0.2cm}
	\includegraphics[width=2.1in]{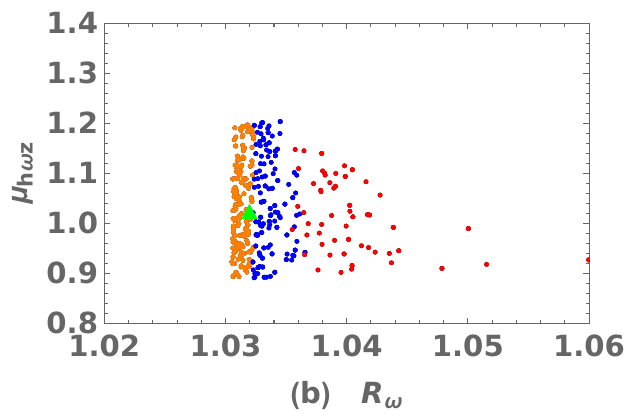}
	\vspace{0.2cm}
	\includegraphics[width=2.1in]{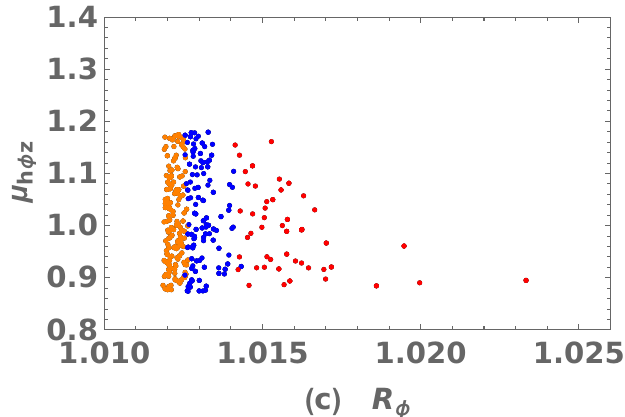}
	\vspace{0.2cm}
	\includegraphics[width=2.1in]{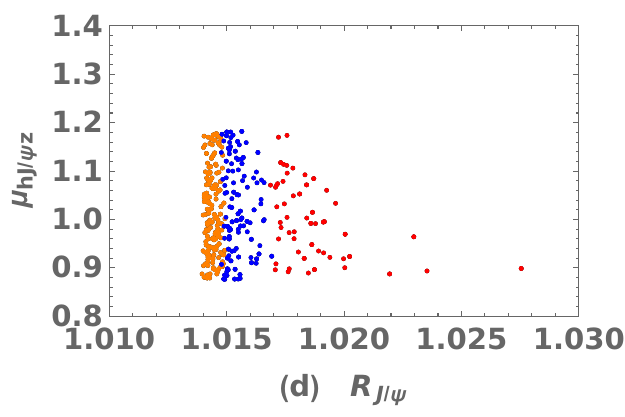}
	\vspace{0.2cm}
	\includegraphics[width=2.1in]{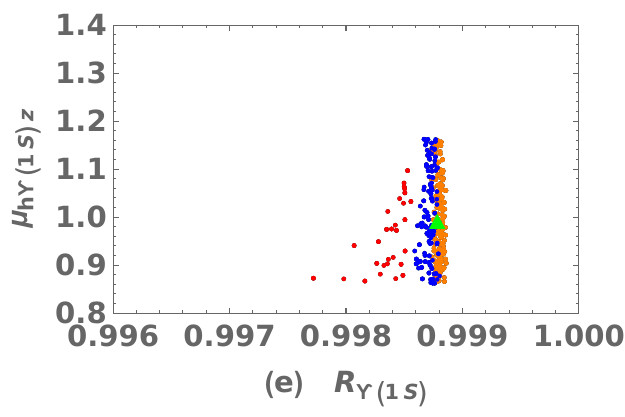}
	\vspace{0.2cm}
	\includegraphics[width=2.1in]{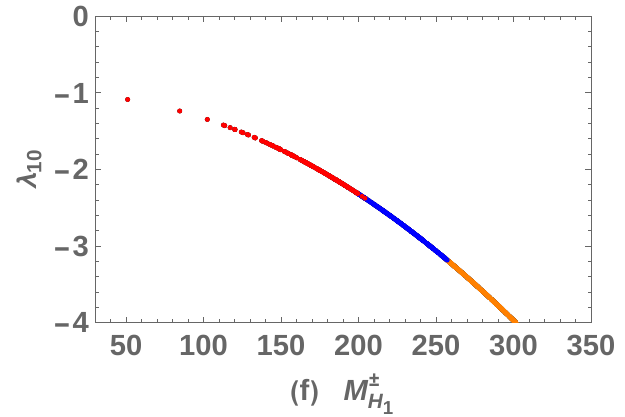}
	\vspace{0cm}
	\caption[]{Using the points obtained in Fig.~\ref{fig-V1},  $\mu_{hMZ}$ versus $R_{M}$ are plotted. And the relationship between $\lambda_{10}$ and $M_{H_1^{\pm}}$ is plotted (f),  where the orange dots in the figure represent points in the range of $\leq 1.03$ for $R_{\rho}$, the blue dots in the figure represent points in the range of $\left(1.03, 1.33 \right)$ for $R_{\rho}$, and the red dots in the figure represent points in the range of $\geq 1.033$ for $R_{\rho}$. The green triangles represent the best fit point $\chi ^2=3.94$.}
	\label{fig-V5}
\end{figure}
and the results of the corresponding calculations are listed in Table \ref{tabmeson}.

Taking the points obtained in Fig.~\ref{fig-V1}, we plot $\mu_{hMZ}$ versus $R_{M}$ in Fig.~\ref{fig-V5}, where $ M=\rho, \omega, \phi, J/\psi, \Upsilon(nS)$ for Figs.~\ref{fig-V5}(a)-(e) respectively. The orange dots in the figure represent points in the range of $\left(-4, -3\right)$ for $\lambda_{10}$, the blue dots represent points in the range of $\left(-3, -2\right)$ for $\lambda_{10}$, and the red dots represent points in the range of $\left(-2, 0\right)$ for $\lambda_{10}$. The green triangle represents the best fit point. In Fig.~\ref{fig-V5}(a) or \ref{fig-V5}(c), the graph indicates that as $\lambda_{10}$ increases, $R_{\rho}$ or $R_{\phi }$ increases, and the value of $\mu_{h\rho Z}$ or $\mu _{h\phi Z}$ gradually decreases. When $\lambda_{10}$ is within the range of $\left(-4, -3\right)$, $\mu_{h\rho Z}$ and $\mu _{h\phi Z}$ can deviate significantly from the SM, exceeding 20\%. In Fig.~\ref{fig-V5}(b), the distribution of $\mu_{h\omega Z}$ and $R_{\omega}$ is very similar to those in Fig.~\ref{fig-V5}(a). The calculation in Table~\ref{tabmeson} shows that the only variable in the change process of $R_M-1$ for these two processes is $C_{\gamma Z}^{NP}$, with coefficients of 0.132 and 0.135 which differ very slightly, indicating that the small difference between $R_\rho$ and $R_\omega$. For a similar reason, the results in Fig.~\ref{fig-V5}(d) are similar to those in Fig.~\ref{fig-V5}(c). Figure.~\ref{fig-V5}(e) shows $R_{\Upsilon(1S) }$ versus $\mu _{h\Upsilon(1S) Z}$. In Table \ref{tabmeson}, $\frac{F_{\parallel ind,NP}^{MZ}}{F_{\parallel ind,SM}^{MZ}} $ $\approx $1, so the ratio of $\frac{F_{\perp ind,NP}^{MZ}}{F_{\perp ind,SM}^{MZ}} $ is the main influencing factor of $R_{\Upsilon(1S) }$. The value of $R_{\Upsilon(1S) }$ approaching 1 indicates that the values of FDM and SM are almost the same. This is because the quality of $m_{\Upsilon (1S)}$ and the value of $f_\Upsilon $ are relatively high, leading to a significant contribution from SM. 

Figures.~\ref{fig-V5}(a)-\ref{fig-V5}(e) show obviously that $\lambda_{10}$ affects all results of $\mu_{hMZ}$ significantly, because the dominant contributions to these processes in the $R_{M}$ come from the charged Higgs, and $\lambda_{10}$ affects the lightest charged Higgs mass $M_{H_1^{\pm}}$ acutely as shown in Fig.~\ref{fig-V5}(f). 

\subsection{$h{\rightarrow} M\gamma$}

\begin{figure}
	\setlength{\unitlength}{1mm}
	\centering
	\includegraphics[width=2.1in]{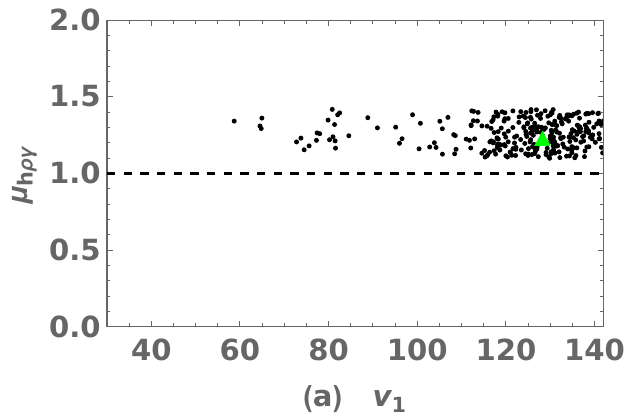}
	\vspace{0.2cm}
	\includegraphics[width=2.1in]{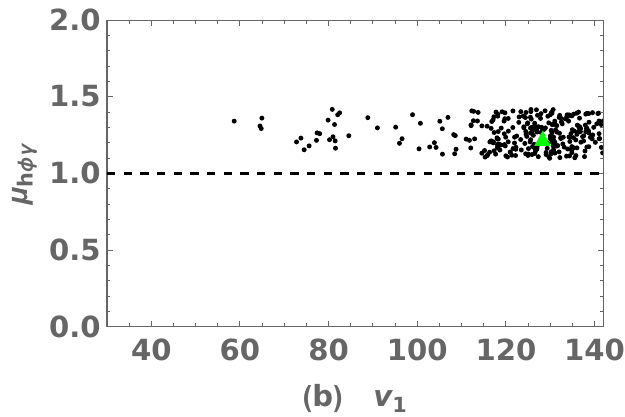}
	\vspace{0.2cm}
	\includegraphics[width=2.1in]{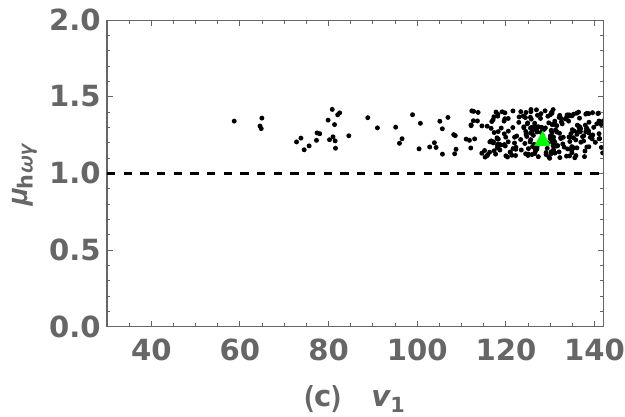}
	\vspace{0.2cm}
	\includegraphics[width=2.1in]{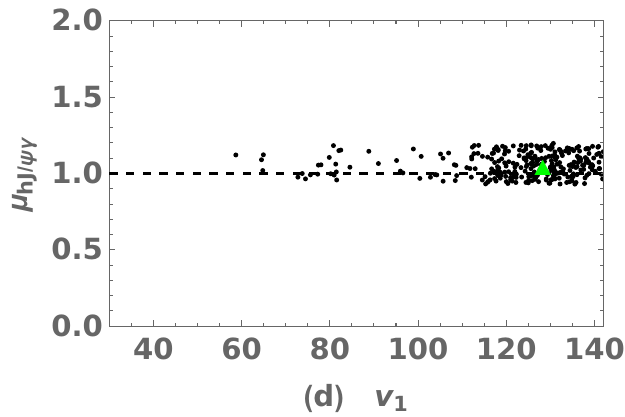}
	\vspace{0.2cm}
	\includegraphics[width=2.1in]{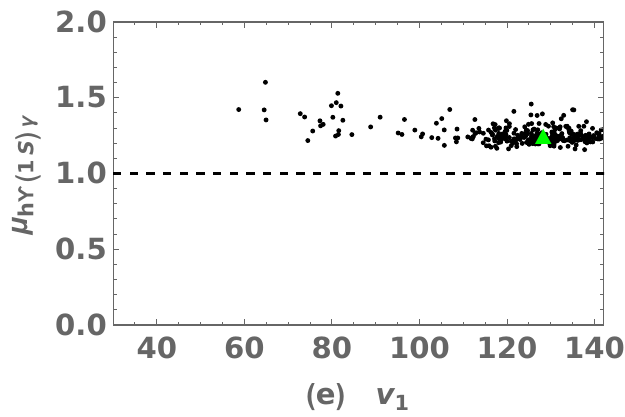}
	\vspace{0.2cm}
	\includegraphics[width=2.1in]{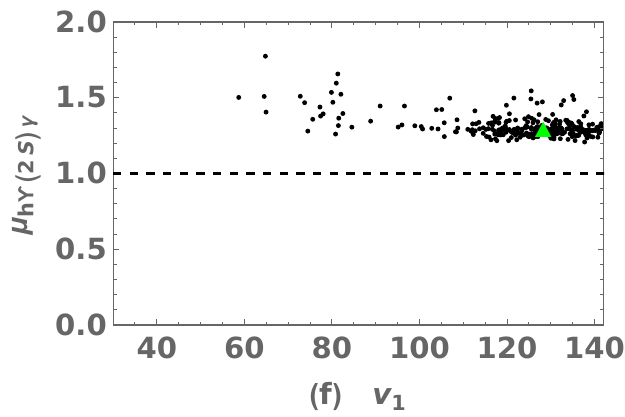}
	\vspace{0.2cm}
	\includegraphics[width=2.1in]{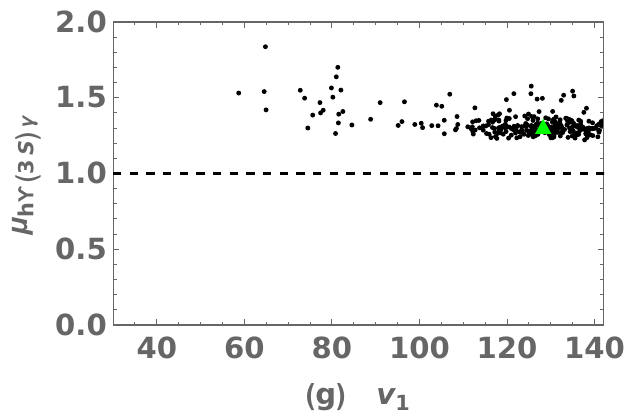}
	\vspace{0cm}
	\caption[]{Scanning the parameter space in Table \ref{tab-scan}, and the results of $\mu _{\rho  \gamma }-v_1$ (a), $\mu _{\phi  \gamma }-v_1$ (b), $\mu _{\omega   \gamma }-v_1$ (c), $\mu _{J/\psi  \gamma }-v_1$ (d), and $\mu _{\Upsilon(nS)  \gamma }-v_1$ (e)-(g) are plotted, and the green triangles represent the best fit point $\chi ^2=3.94$.}
	\label{fig-V4}
\end{figure}

The parameter space, scanned using data points from Fig.~\ref{fig-V1}, is marked by a green triangle indicating the best fit point from previous analysis and a dashed line representing the contour where the signal strength equals 1. Figures.~\ref{fig-V4}(a)-\ref{fig-V4}(c) exhibit signal strengths that deviate by 40\% from SM predictions. This deviation arises from the indirect contributions of the new scalar sector and strongly suggests the potential presence of NP. Figure.~\ref{fig-V4}(d) shows a relatively small deviation (up to 20\%) in $\mu _{J/\psi  \gamma }$ due to less significant cancellation between direct and indirect contributions. Conversely, Figs.~\ref{fig-V4}(e)-\ref{fig-V4}(g) show substantial enhancements (up to 50\%) in  $\mu _{h\Upsilon (nS)\gamma }$ signal strengths, despite the large masses and small branching ratios. This sensitivity of SM predictions to the interplay of direct and indirect contributions provides a valuable opportunity to probe for NP.

The numerical results for $m_M$ and $\Gamma(M {\rightarrow} \ell^+ \ell^-)$ are taken from Table\ref{tab-l}. The amplitudes of the direct and indirect contributions are presented in Table \ref{tab:dir-ind}, and the decay widths and branching ratios are further computed and summarized in Table\ref{tab:decayrates}. These results are consistent with the experimental values given in Eq. (\ref{eqBD}), with all parameters determined based on the best fit point obtained from prior analysis. For $h{\rightarrow} J/\psi+\gamma$, $A_\mathrm{ind}$ is larger by an order of magnitude compared to $A_\mathrm{dir}$. For $\Gamma _{\Upsilon (nS)+\gamma }$ ($n$=1,2,3), as $n$ increases, the absolute value of $A_\mathrm{dir}$ gets closer and closer to $A_\mathrm{ind}$; this leads to the corresponding decay width rapidly decreasing.

\begin{table*}
	\begin{tabular*}{\textwidth}{@{\extracolsep{\fill}}lllll@{}}
		\hline
		$M$& $A_\mathrm{ind}\times 10^5 \, (\text{GeV}^{1/2})$&$A_\mathrm{dir}\times 10^5 \, (\text{GeV}^{1/2})$\\
		\hline
		$J/\psi$&$-11.81$&$0.6295+0.06488 i$\\
		$\Upsilon(1S)$&$3.26$&$ -2.6688-0.2857 i$\\
		$\Upsilon(2S)$&$2.178$&$-1.8941-0.1968 i$\\
		$\Upsilon(3S)$&$1.825 $&$-1.614-0.165 i$\\
		\hline
	\end{tabular*}
	\caption{Numerical results for the amplitudes $A_\mathrm{ind}$ and $A_\mathrm{dir}$. }
	\label{tab:dir-ind}
\end{table*} 

\begin{figure}
	\setlength{\unitlength}{1mm}
	\centering
	\includegraphics[width=2.6in]{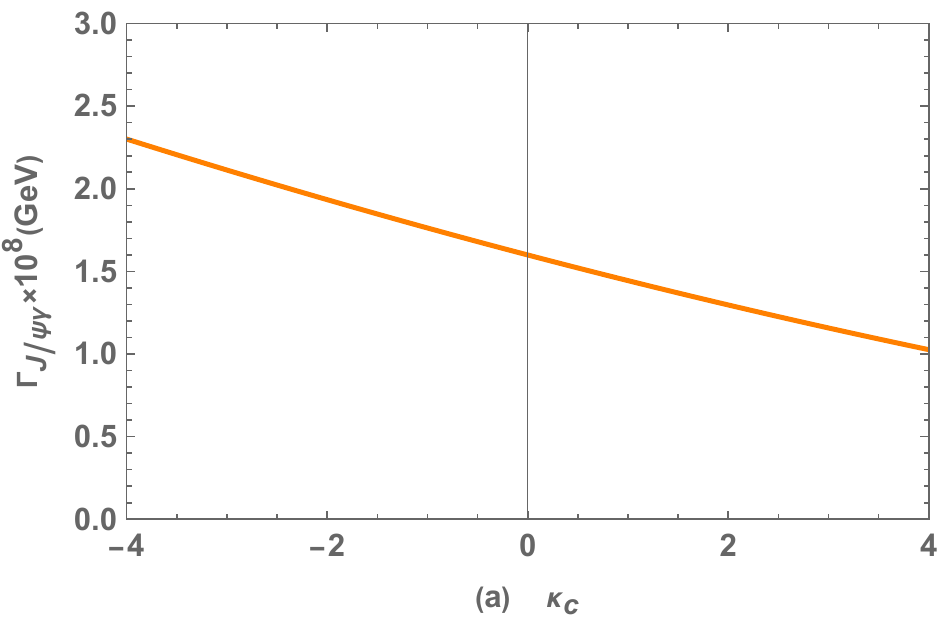}
	\vspace{0.2cm}
	\includegraphics[width=2.6in]{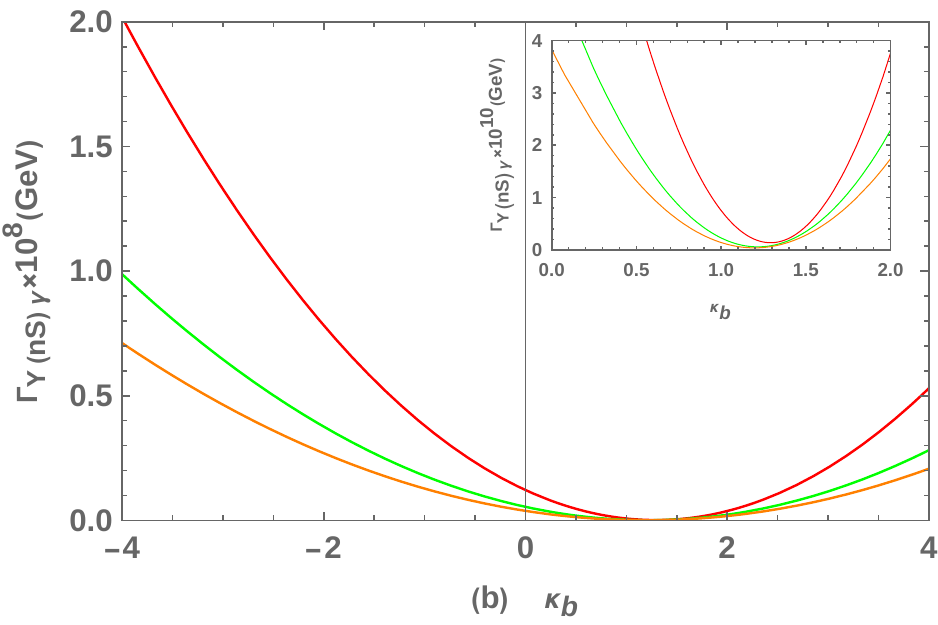}
	\vspace{0.2cm}
	\caption[]{The decay rate $\Gamma_{J/\Psi + \gamma} = \Gamma (h {\rightarrow} J/\Psi + \gamma)$ is determined by scaling the Higgs charm coupling with a factor $\kappa_c$, while the decay rates $\Gamma_{\Upsilon(nS) + \gamma} = \Gamma (h {\rightarrow} \Upsilon(nS) + \gamma)$ for $n = 1, 2, 3$ are computed by scaling the Higgs bottom coupling with a factor $\kappa_b$, represented by the red, green, and orange solid lines respectively. In the SM prediction, we have $\kappa_c =\kappa_b= 1$.}
	\label{Figv5}
\end{figure}

We calculate $\Gamma (h {\rightarrow} M + \gamma)$ through $|\kappa _qA_\mathrm{dir} +
A_\mathrm{ind}|^2$ to study the deviation of the Yukawa coupling $\kappa_{q}$ from the SM; the $c\bar{c} \gamma $ and $b\bar{b} \gamma $ vertexes are highly constrained by QED, so typically modifications to the direct contribution amplitude can be made only by modifying $hc\bar{c}$ and $hb\bar{b}$ vertexes. As shown in Fig.~\ref{Figv5}(a), as $\kappa _c$ increases from -4 to 4, the decay width of $\Gamma _{J/\Psi +\gamma } $ decreases gradually, which helps us narrow down the range of $\kappa _c$. The FDM decay width is larger ($\kappa _c<1$) or smaller ($\kappa _c>1$) than the SM's, primarily due to the dominance of $|A_\mathrm{ind}|$ over $|A_\mathrm{dir}|$, and their opposite signs. However, direct contributions are negligible (≈5\%) in the $h{\rightarrow} J/\Psi +\gamma$ process, resulting in a near-linear relationship. Figure.~\ref{Figv5}(b) shows the $h{\rightarrow} \Upsilon(nS) \gamma $ decay width as a concave function of $\kappa _b$ (for all $nS$), with a minimum near $\kappa _b=1$. This concavity facilitates experimental constraints on $\kappa _b$. Significant cancellation between direct and indirect contributions occurs in the $\Gamma _{\Upsilon (3S)+\gamma }$ process, particularly in the SM ($\kappa _b=1$). This cancellation reduces the decay width, increasing experimental challenges. However, the FDM predicts a larger decay width than the SM, improving its experimental prospects.

\begin{table*}
	\begin{tabular*}{\textwidth}{@{\extracolsep{\fill}}lllll@{}}
		\hline
		$M$& $\Gamma_{SM}(h {\rightarrow} M + \gamma) \, (\text{GeV})$&$\text{Br}_{SM}(h {\rightarrow} M + \gamma)$& $\Gamma_{NP}(h {\rightarrow} M + \gamma) \, (\text{GeV})$&$\text{Br}_{NP}(h {\rightarrow} M + \gamma)$\\
		\hline
		$J/\psi$&$1.253 \times 10^{-8}$&$3.047 \times 10^{-6}$&$1.445 \times 10^{-8}$&$3.45 \times 10^{-6}$\\
		$\Upsilon(1S)$&$4.316 \times 10^{-11}$&$1.049 \times 10^{-8}$&$7.54 \times 10^{-11}$&$1.80 \times 10^{-8}$\\
		$\Upsilon(2S)$&$1.193 \times 10^{-11}$&$2.90 \times 10^{-9}$&$ 2.292 \times 10^{-11}$&$5.47 \times 10^{-9}$\\
		$\Upsilon(3S)$&$ 7.123 \times 10^{-12}$&$1.73 \times 10^{-9}$&$1.413 \times 10^{-11}$&$3.37 \times 10^{-9}$\\
		\hline
	\end{tabular*}
	\caption{Decay rates and branching ratios for $h {\rightarrow} M + \gamma$.}
	\label{tab:decayrates}
\end{table*}

Table~\ref{tab:decayrates} (calculated at $\kappa_{q}=1$) shows that the FDM increases both decay widths and branching ratios for all four processes, with some branching ratios nearly doubled. The consistency of these branching ratios with experimental data strongly suggests the presence of NP.

\section{Conclusion\label{sec5}}
In the framework of the FDM, the redefined scalar sector can have a significant impact on the properties of the 125 GeV Higgs boson, and, in order to explain the fermion mass hierarchy problem, the FDM also redefines the Yukawa couplings, which will impact the couplings between the Higgs and fermions. Under stringent experimental constraints on the signal strengths $\mu_{\gamma\gamma,Z\gamma}$, $\mu_{VV^\ast}$  ($V=Z,W$), and $\mu_{f\bar{f}}$ $(f=b,c,\tau)$ in the scalar sector, $\bar B \to X_s\gamma$ and $B_s^0 \to \mu^+\mu^-$ in the B meson sector, $t \to c h$ and $t \to u h$ in the top quark sector, the charged lepton flavor violation decays $\tau \to 3e$, $\tau \to 3\mu$, and $\mu \to 3e$, as well as the S, T, U parameters, the properties of the Higgs boson are determined via fits to these observables. Our results confirm that the FDM can explain the experimentally observed signal strengths, FCNCs, and the electroweak precision observables and identify the appropriate parameter space.

We present an in-depth analysis of $h{\rightarrow} {MZ}$ and $h{\rightarrow} {M\gamma}$, where $M$ is vector mesons $[\rho, \omega, \phi, J/\psi, \Upsilon(nS)]$. Our findings indicate an inverse relationship between $\lambda _{10}$ and $R_M$, mediated by the lightest charged Higgs boson mass: Decreasing $\lambda _{10}$ and increasing the lightest charged Higgs boson mass result in a smaller $R_M$. At low values of $R_M$, $\mu_{hMZ}$ ($M=\rho, \omega, \phi, J/\psi$) exhibits a deviation of up to 20\% above the SM; however, $\mu_{h\Upsilon(nS)Z}$ and $R_{\Upsilon(nS)}$ have an inverse relationship. These provide reference value for probing at the possible future high-energy colliders. The FDM partially reduces the cancellation between direct and indirect contributions of $h{\rightarrow} {M\gamma}$ processes in the SM, the signal strength can exceed the SM by up to 50\%, and the strength of the Higgs effective couplings also show a significant impact on the related decay widths. In addition, the parameter $v_{1,2}$ significantly impacts theoretical predictions of signal strengths, primarily by altering the mixing strength of the three Higgs doublet fields and other Yukawa couplings.

\section*{ACKNOWLEDGMENTS}

The work has been supported by the National Natural Science Foundation of China (NNSFC) with Grants No. 12075074 and No. 12235008, Hebei Natural Science Foundation for Distinguished Young Scholars with Grant No. A2022201017 and No. A2023201041, Natural Science Foundation of Guangxi Autonomous Region with Grant No. 2022GXNSFDA035068, and the youth top-notch talent support program of the Hebei Province.

\appendix

\section{FORM FACTORS\label{app-form}}

The form factors are
\begin{eqnarray}
&&A_0(x)=-(x-g(x))/x^2, \nonumber\\
&&A_{1/2}(x)=2\Big[x+(x-1)g(x)\Big]/x^2, \nonumber\\
&&A_1(x)=-\Big[2x^2+3x+3(2x-1)g(x)\Big]/x^2.
\end{eqnarray}
\begin{eqnarray}
&& F_{sf}(\tau)=\tau^{-1}\left [ 1+(1-\tau^{-1})g(\tau) \right ] \nonumber\\
&& F_{pf}(\tau)=\tau^{-1}g(\tau).
\end{eqnarray}
with
\begin{equation}
	\begin{aligned}
		g(x) &= \left\{
		\begin{array}{ll}
			\arcsin^2\sqrt{x}, & \text{if } x \leq 1; \\
			-\frac{1}{4}\left[\ln\left(\frac{1+\sqrt{1-1/x}}{1-\sqrt{1-1/x}}\right) - i\pi\right]^2, & \text{if } x > 1.
		\end{array}
		\right. \\
		f(x) &= x(1-x)g'(x),
	\end{aligned}
\end{equation}
%\begin{eqnarray}
%&&g(x)=\left\{\begin{array}{l}\arcsin^2\sqrt{x}, \hspace{2.6cm} x\le1;   \\
%-{1\over4}\Big[\ln{1+\sqrt{1-1/x}\over1-\sqrt{1-1/x}}-i\pi\Big]^2, \quad x>1, \end{array}\right \nonumber\\
%&&f\left ( x \right ) =x \left ( 1-x \right ) g{}' \left ( x \right ) .
%\end{eqnarray}
and
\begin{eqnarray}
	F(x) &=& -(1-x^2) \left( \frac{47}{2} x^2 - \frac{13}{2} + \frac{1}{x^2} \right) - 3 (1 - 6x^2 + 4x^4) \ln x \nonumber\\
	&& \hspace{1.45cm} + \frac{3 (1 - 8x^2 + 20x^4)}{\sqrt{4x^2 - 1}} \cos^{-1} \left( \frac{3x^2 - 1}{2x^3} \right),
\end{eqnarray}
\begin{eqnarray}
m_{Q}\left(\mu_{Q}^{2}\right){\rightarrow} 0:\quad\left\{\begin{array}{l}
	\operatorname{Re} C_{sf}{\rightarrow}-\frac{1}{18}\left[\log^{2}(4x)-\pi^{2}\right]-\frac{2}{3}\log(4x)+2\log\frac{\mu_{Q}^{2}}{m_{Q}^{2}} ,\mu_{Q}=M_h/2\\
	\operatorname{Im} C_{sf}{\rightarrow}\frac{\pi}{3}\left[\frac{1}{3}\log(4x)+2\right],
\end{array}\right.
\end{eqnarray}
\begin{eqnarray}
	 &&A_{f}(\tau, x)= \frac{3 \tau}{2(1-x)}\left\{1-\frac{2 x}{1-x}[f(\tau)-f(\tau / x)]+\left(1-\frac{\tau}{1-x}\right)[g(\tau)-g(\tau / x)]\right\}, \nonumber\\
	&&B_{f}(\tau, x)= \frac{\tau}{1-x}[g(\tau)-g(\tau / x)],\nonumber \\
	&&A_{W}^{\gamma \gamma}(\tau, x)= \frac{2+3 \tau}{1-x}\left\{1-\frac{2 x}{1-x}[f(\tau)-f(\tau / x)]\right\}+\frac{3 \tau}{(1-x)^{2}}\left(2-\tau-\frac{8 x}{3}\right)[g(\tau)-g(\tau / x)],\nonumber \\
	&&A_{W}^{\gamma Z}(\tau, x)= \frac{1}{1-x}\left[1-2 s_{W}^{2}+\left(\frac{5}{2}-3 s_{W}^{2}\right) \tau\right]\left\{1-\frac{2 x}{1-x}[f(\tau)-f(\tau / x)]\right\},\nonumber \\
	&& +\frac{\tau}{(1-x)^{2}}\left[\left(\frac{5}{2}-3 s_{W}^{2}\right)(2-\tau)-2 x\left(3-4 s_{W}^{2}\right)\right][g(\tau)-g(\tau / x)],
\end{eqnarray}
\begin{eqnarray}
	&&A_{1 / 2}(x, \lambda)=I_{1}(x, \lambda)-I_{2}(x, \lambda), \nonumber \\
	&&A_{1}(x, \lambda)=c_{W}\left\{4\left(3-\frac{s_{W}^{2}}{c_{W}^{2}}\right) I_{2}(x, \lambda)+\left[\left(1+\frac{2}{x}\right) \frac{s_{W}^{2}}{c_{W}^{2}}-\left(5+\frac{2}{x}\right)\right] I_{1}(x, \lambda)\right\},  \nonumber\\
	&&A_{0}(x, \lambda)=I_{1}(x, \lambda),  \nonumber\\
	&&I_{1}(x, \lambda)=\frac{x \lambda}{2(x-\lambda)}+\frac{x^{2} \lambda^{2}}{2(x-\lambda)^{2}}\left[f\left(x^{-1}\right)-f\left(\lambda^{-1}\right)\right]+\frac{x^{2} \lambda}{(x-\lambda)^{2}}\left[g\left(x^{-1}\right)-g\left(\lambda^{-1}\right)\right], \nonumber\\
	&&I_{2}(x, \lambda)=\frac{x \lambda}{2(x-\lambda)}\left[f\left(x^{-1}\right)-f\left(\lambda^{-1}\right)\right].
\end{eqnarray}

\section{THE RELATED COUPLING VERTEXES\label{app-coupling}}

The related coupling vertexes are
\begin{eqnarray}
	C_{hW_\sigma ^+W_\mu ^-}=\frac{i}{2} g_{2}^{2} \left( v_{\phi_{1}} Z_{i 1}^{H} + v_{\phi_{2}} Z_{i 2}^{H} + v_{\phi_{3}} Z_{i 3}^{H} \right) \left( g_{\sigma\mu} \right),\left( i=1 \right)
\end{eqnarray}

\begin{eqnarray}
	C_{hZ_\sigma Z_\mu}=&& \frac{i}{2} \left( v_{\phi_1} \left( - \left( 2g_F + g_{YF} \right) \sin \theta'_W + \left( 2g_{FY} + g_1 \right) \cos \theta'_W \sin \theta_W + g_2 \cos \theta_W \cos \theta'_W \right)^2 Z_{i1}^H \right. \nonumber\\
	&& + v_{\phi_2} \left( - \left( -2g_F + g_{YF} \right) \sin \theta'_W + \left( -2g_{FY} + g_1 \right) \cos \theta'_W \sin \theta_W + g_2 \cos \theta_W \cos \theta'_W \right)^2 Z_{i2}^H \nonumber\\
	&& + g_2^2 v_{\phi_3} \cos^2 \theta_W \cos^2 \theta'_W Z_{i3}^H + g_1^2 v_{\phi_3} \cos^2 \theta_W \sin^2 \theta_W Z_{i3}^H \nonumber\\
	&& + g_1 g_2 v_{\phi_3} \cos^2 \theta'_W \sin 2\theta_W Z_{i3}^H - 2g_{YF} g_2 v_{\phi_3} \cos \theta_W \cos \theta'_W \sin \theta'_W Z_{i3}^H \nonumber\\
	&& - 2g_1 g_{YF} v_{\phi_3} \cos \theta'_W \sin \theta_W \sin \theta'_W Z_{i3}^H + g_{YF}^2 v_{\phi_3} \sin^2 \theta'_W Z_{i3}^H \nonumber\\
	&& + 16g_F^2 v_{\chi} v_{\chi} \cos^2 \theta'_W \sin^2 \theta_W Z_{i4}^H - 32g_{FY} v_{\chi} \cos \theta'_W \sin \theta_W \sin \theta'_W Z_{i4}^H \nonumber\\
	&& \left. + 16g_F^2 v_{\chi} \sin^2 \theta'_W Z_{i4}^H \right) \left( g_{\sigma \mu} \right), \quad (i=1,\sin \theta'_W=0,\cos \theta'_W=1)
\end{eqnarray}

\begin{eqnarray}
	C_{hd_j\bar d_k}=&&-i\frac{1}{\sqrt{2}}\delta_{\alpha\beta}\left(U_{R, i 2}^{d,*}\left(Y_{d 12} U_{L, j 1}^{d,*} Z_{k 3}^H+Y_{d32} U_{L, j 3}^{d,*} Z_{k2}^H\right)+U_{R, i 1}^{d,*}\left(Y_{d21} U_{L, j 2}^{d,*} Z_{k 3}^H+Y_{d31} U_{L, j 3}^{d,*} Z_{k1}^H\right)\right.\nonumber\\
	&&\left.+U_{R, i 3}^{d,*}\left(Y_{d 13} U_{L, j 1}^{d,*} Z_{k 1}^H+Y_{d23} U_{L, j 2}^{d,*} Z_{k2}^H+Y_{d33} U_{L, j 3}^{d,*} Z_{k3}^H\right)\right)\left(\frac{1-\gamma_5}{2}\right)\nonumber\\
	&&+i\frac{1}{\sqrt{2}}\delta_{\alpha\beta}\left(Y_{d 13}^* U_{R, j 3}^d U_{L, i 1}^d Z_{k1}^H+Y_{d31}^* U_{R, j 1}^d U_{L, i 3}^d Z_{k1}^H+Y_{d23}^* U_{R, j 3}^d U_{L, i 2}^d Z_{k2}^H+Y_{d32}^* U_{R, j 2}^d U_{L, i 3}^d Z_{k2}^H\right.\nonumber\\
	&&\left.+Y_{d12}^* U_{R, j 2}^d U_{L, i 1}^d Z_{k3}^H+Y_{d21}^* U_{R, j 1}^d U_{L, i 2}^d Z_{k3}^H+Y_{d33}^* U_{R, j 3}^d U_{L, i 3}^d Z_{k3}^H\right)\left(\frac{1+\gamma_5}{2}\right),\left( i=1 \right)
\end{eqnarray}

\begin{eqnarray}
	C_{hu_j\bar u_k}=&&-i\frac{1}{\sqrt{2}}\delta_{\alpha\beta}\left(U^{u,*}_{R,i2}\left(Y_{u12}U^{u,*}_{L,j1}Z^H_{k3}+Y_{u32}U^{u,*}_{L,j3}Z^H_{k1}\right)+U^{u,*}_{R,i1}\left(Y_{u21}U^{u,*}_{L,j2}Z^H_{k3}+Y_{u31}U^{u,*}_{L,j3}Z^H_{k2}\right)\right.\nonumber\\
	&&\left.+U^{u,*}_{R,i3}\left(Y_{u13}U^{u,*}_{L,j1}Z^H_{k2}+Y_{u23}U^{u,*}_{L,j2}Z^H_{k1}+Y_{u33}U^{u,*}_{L,j3}Z^H_{k3}\right)\right)\left(\frac{1-\gamma_5}{2}\right)\nonumber\\
	&&+i\frac{1}{\sqrt{2}}\delta_{\alpha\beta}\left(Y^*_{u23}U^{u}_{R,j3}U^{u}_{L,i2}Z^H_{k1}+Y^*_{u32}U^{u}_{R,j2}U^{u}_{L,i3}Z^H_{k1}+Y^*_{u13}U^{u}_{R,j3}U^{u}_{L,i1}Z^H_{k2}+Y^*_{u31}U^{u}_{R,j1}U^{u}_{L,i3}Z^H_{k2}\right.\nonumber\\
	&&\left.+Y^*_{u12}U^{u}_{R,j2}U^{u}_{L,i1}Z^H_{k3}+Y^*_{u21}U^{u}_{R,j1}U^{u}_{L,i2}Z^H_{k3}+Y^*_{u33}U^{u}_{R,j3}U^{u}_{L,i3}Z^H_{k3}\right)\left(\frac{1+\gamma_5}{2}\right),\left( i=1 \right)
\end{eqnarray}

\begin{eqnarray}
	C_{e_i\bar{e}_jh}=&&-i\frac{1}{\sqrt{2}}\left(U_{R, i 2}^{e,*}\left(Y_{e 12} U_{L, j 1}^{e,*} Z_{k 3}^H+Y_{e 32} U_{L, j 3}^{e,*} Z_{k 2}^H\right)+U_{R, i 1}^{e,*}\left(Y_{e 21} U_{L, j 2}^{e,*} Z_{k 3}^H+Y_{e 31} U_{L, j 3}^{e,*} Z_{k 1}^H\right)\right.\nonumber\\
	&&\left.+U_{R, i 3}^{e,*}\left(Y_{e 13} U_{L, j 1}^{e,*} Z_{k 1}^H+Y_{e 23} U_{L, j 2}^{e,*} Z_{k 2}^H+Y_{e 33} U_{L, j 3}^{e,*} Z_{k 3}^H\right)\right)\left(\frac{1-\gamma_5}{2}\right)\nonumber\\
	&&+i\frac{1}{\sqrt{2}}\left(Y_{e 13}^{*} U_{R, j 3}^{e} U_{L, i 1}^{e} Z_{k 1}^H+Y_{e 31}^{*} U_{R, j 1}^{e} U_{L, i 3}^{e} Z_{k 1}^H+Y_{e 23}^{*} U_{R, j 3}^{e} U_{L, i 2}^{e} Z_{k 2}^H+Y_{e 32}^{*} U_{R, j 2}^{e} U_{L, i 3}^{e} Z_{k 2}^H\right.\nonumber\\
	&&\left.+Y_{e 12}^{*} U_{R, j 2}^{e} U_{L, i 1}^{e} Z_{k 3}^H+Y_{e 21}^{*} U_{R, j 1}^{e} U_{L, i 2}^{e} Z_{k 3}^H+Y_{e 33}^{*} U_{R, j 3}^{e} U_{L, i 3}^{e} Z_{k 3}^H\right)\left(\frac{1+\gamma_5}{2}\right),\left( i=1 \right)
\end{eqnarray}

\begin{eqnarray}
	C_{hH_j^+H_k^-}=&&\frac{i}{2}\left(-2\lambda_5^{\prime} v_{\phi_3} Z_{i 3}^H Z_{j 1}^+ Z_{k 1}^+-2\lambda_7 v_\chi Z_{i 4}^H Z_{j 1}^+ Z_{k 1}^+-\sqrt{2}\kappa Z_{i 4}^H Z_{j 2}^+ Z_{k 1}^+\right.\nonumber\\
	&&\left.-\lambda_5^{\prime\prime} v_{\phi_1} Z_{i 3}^H Z_{j 3}^+ Z_{k 1}^+-v_{\phi_2}\lambda_{10}^* Z_{i 3}^H Z_{j 3}^+ Z_{k 1}^+-\sqrt{2}\kappa^* Z_{i 4}^H Z_{j 1}^+ Z_{k 2}^+\right.\nonumber\\
	&&\left.-2\lambda_6^{\prime} v_{\phi_3} Z_{i 3}^H Z_{j 2}^+ Z_{k 2}^+-2\lambda_8 v_\chi Z_{i 4}^H Z_{j 2}^+ Z_{k 2}^+-\lambda_6^{\prime\prime} v_{\phi_2} Z_{i 3}^H Z_{j 3}^+ Z_{k 2}^+\right.\nonumber\\
	&&\left.-v_{\phi_1}\lambda_{10}^* Z_{i 3}^H Z_{j 3}^+ Z_{k 2}^+-\lambda_5^{\prime\prime} v_{\phi_1} Z_{i 3}^H Z_{j 1}^+ Z_{k 3}^+-\lambda_{10} v_{\phi_2} Z_{i 3}^H Z_{j 1}^+ Z_{k 3}^+\right.\nonumber\\
	&&\left.-\lambda_{10} v_{\phi_1} Z_{i 3}^H Z_{j 2}^+ Z_{k 3}^+-\lambda_6^{\prime\prime} v_{\phi_2} Z_{i 3}^H Z_{j 2}^+ Z_{k 3}^+-4\lambda_3 v_{\phi_3} Z_{i 3}^H Z_{j 3}^+ Z_{k 3}^+ -2\lambda_9 v_\chi Z_{i 4}^H Z_{j 3}^+ Z_{k 3}^+\right.\nonumber\\
	&&\left.-Z_{i 1}^H\left(Z_{j 3}^+\left(2\lambda_5^{\prime} v_{\phi_1} Z_{k 3}^++\lambda_5^{\prime\prime} v_{\phi_3} Z_{k 1}^++v_{\phi_3}\lambda_{10}^* Z_{k 2}^+\right)\right.\right.\nonumber\\
	&&\left.+Z_{j 2}^+\left(2\lambda_4^{\prime} v_{\phi_1} Z_{k 2}^++\lambda_{10} v_{\phi_3} Z_{k 3}^++\lambda_4^{\prime\prime} v_{\phi_2} Z_{k 1}^+\right)\right.\nonumber\\
	&&\left.\left.+Z_{j 1}^{+}\left(4\lambda_1 v_{\phi_1} Z_{k 1}^++\lambda_4^{\prime\prime} v_{\phi_2} Z_{k 2}^++\lambda_5^{\prime\prime} v_{\phi_3} Z_{k 3}^+\right)\right)\right.\nonumber\\
	&&\left.-Z_{i 2}^H\left(Z_{j 3}^+\left(2\lambda_6^{\prime} v_{\phi_2} Z_{k 3}^++\lambda_6^{\prime\prime} v_{\phi_3} Z_{k 2}^++v_{\phi_3}\lambda_{10}^* Z_{k 1}^+\right)\right.\right].\left( i=1 \right)
\end{eqnarray}


\begin{thebibliography}{99}
%\cite{ParticleDataGroup:2024cfk}
\bibitem{ParticleDataGroup:2024cfk}
S.~Navas \textit{et al.} [Particle Data Group],
%``Review of particle physics,''
Phys. Rev. D \textbf{110} (2024) no.3, 030001
%doi:10.1103/PhysRevD.110.030001
%219 citations counted in INSPIRE as of 15 Oct 2024

%\cite{ATLAS:2023oaq}
\bibitem{ATLAS:2023oaq}
G.~Aad \textit{et al.} [ATLAS],
%``Combined Measurement of the Higgs Boson Mass from the H\textrightarrow{}\ensuremath{\gamma}\ensuremath{\gamma} and H\textrightarrow{}ZZ*\textrightarrow{}4\ensuremath{\ell} Decay Channels with the ATLAS Detector Using s=7, 8, and 13~TeV pp Collision Data,''
Phys. Rev. Lett. \textbf{131} (2023) no.25, 251802
%doi:10.1103/PhysRevLett.131.251802
[arXiv:2308.04775 [hep-ex]].
%35 citations counted in INSPIRE as of 14 Oct 2024

\bibitem{ATLAS1}
ATLAS Collaboration, ATLAS-CONF-2012-135;
ATLAS Collaboration, \emph{Phys. Lett.} {\bf B 718} (2012) 369 [arXiv:1207.0210];
ATLAS Collaboration, \emph{ JHEP} {\bf 09} (2012) 070 [arXiv:1206.5971];
ATLAS Collaboration,  ATLAS-CONF-2013-014;
ATLAS Collaboration, ATLAS-CONF-2013-034;
ATLAS Collaboration,   ATLAS-CONF-2013-040.

\bibitem{CMS1}
CMS Collaboration,  CMS-PAS-HIG-12-025;
CMS Collaboration,  CMS-PAS-HIG-13-004;
CMS Collaboration,  CMS-PAS-HIG-13-005.

%\cite{Zhang:2013hga}
\bibitem{Zhang:2013hga}
H.~B.~Zhang, T.~F.~Feng, F.~Sun, K.~S.~Sun, J.~B.~Chen and S.~M.~Zhao,
%``125 GeV Higgs boson decays in the \ensuremath{\mu} from \ensuremath{\nu} supersymmetric standard model,''
Phys. Rev. D \textbf{89} (2014) no.11, 115007
%doi:10.1103/PhysRevD.89.115007
[arXiv:1307.3607 [hep-ph]].
%16 citations counted in INSPIRE as of 14 Oct 2024

%\cite{deBlas:2016ojx}
\bibitem{deBlas:2016ojx}
J.~de Blas, M.~Ciuchini, E.~Franco, S.~Mishima, M.~Pierini, L.~Reina and L.~Silvestrini,
%``Electroweak precision observables and Higgs-boson signal strengths in the Standard Model and beyond: present and future,''
JHEP \textbf{12} (2016), 135
%doi:10.1007/JHEP12(2016)135
[arXiv:1608.01509 [hep-ph]].
%184 citations counted in INSPIRE as of 12 Jan 2025

%\cite{Kim:2024yrk}
\bibitem{Kim:2024yrk}
V.~T.~Kim,
%``Higgs Boson and Naturalness Problem in the Standard Model,''
Phys. Part. Nucl. \textbf{55} (2024) no.1, 156-159
%doi:10.1134/S1063779624010088
%0 citations counted in INSPIRE as of 13 Jan 2025

%\cite{Batra:2022wsd}
\bibitem{Batra:2022wsd}
A.~Batra, S.~Mandal and R.~Srivastava,
%``$h \to \Upsilon \gamma$ Decay: Smoking Gun Signature of Wrong-Sign $hb\bar{b}$ Coupling,''
[arXiv:2209.01200 [hep-ph]].
%4 citations counted in INSPIRE as of 10 Nov 2024

%\cite{Steingasser:2024hqi}
\bibitem{Steingasser:2024hqi}
T.~Steingasser,
%``Higgs criticality in and beyond the SM,''
PoS \textbf{CORFU2023} (2024), 150
%doi:10.22323/1.463.0150
[arXiv:2405.02415 [hep-ph]].
%5 citations counted in INSPIRE as of 13 Jan 2025

%\cite{Yang:2024kfs}
\bibitem{Yang:2024kfs}
J.~L.~Yang, H.~B.~Zhang and T.~F.~Feng,
%``A mechanism relating the fermionic mass hierarchy to the flavor mixing,''
Phys. Lett. B \textbf{853} (2024), 138677
%doi:10.1016/j.physletb.2024.138677
[arXiv:2404.15990 [hep-ph]].
%2 citations counted in INSPIRE as of 14 Oct 2024

%\cite{Yang:2024duo}
\bibitem{Yang:2024duo}
J.~L.~Yang and J.~Li,
%``Neutrinos in the flavor-dependent $U(1)_F$ model,''
[arXiv:2411.01744 [hep-ph]].
%0 citations counted in INSPIRE as of 05 Dec 2024

%\cite{Yang:2024znv}
\bibitem{Yang:2024znv}
J.~L.~Yang, H.~B.~Zhang and T.~F.~Feng,
%``The flavor-dependent $U(1)_F$ model,''
Eur. Phys. J. C \textbf{84} (2024) no.6, 616
%doi:10.1140/epjc/s10052-024-12958-5
[arXiv:2405.17807 [hep-ph]].
%2 citations counted in INSPIRE as of 14 Oct 2024

%\cite{Canetti:2012kh}
\bibitem{Canetti:2012kh}
L.~Canetti, M.~Drewes, T.~Frossard and M.~Shaposhnikov,
%``Dark Matter, Baryogenesis and Neutrino Oscillations from Right Handed Neutrinos,''
Phys. Rev. D \textbf{87} (2013), 093006
%doi:10.1103/PhysRevD.87.093006
[arXiv:1208.4607 [hep-ph]].
%395 citations counted in INSPIRE as of 12 Jan 2025
%\cite{Abada:2007ux}
\bibitem{Abada:2007ux}
A.~Abada, C.~Biggio, F.~Bonnet, M.~B.~Gavela and T.~Hambye,
%``Low energy effects of neutrino masses,''
JHEP \textbf{12} (2007), 061
%doi:10.1088/1126-6708/2007/12/061
[arXiv:0707.4058 [hep-ph]].
%419 citations counted in INSPIRE as of 12 Jan 2025

%\cite{Belanger:2013xza}
\bibitem{Belanger:2013xza}
G.~Belanger, B.~Dumont, U.~Ellwanger, J.~F.~Gunion and S.~Kraml,
%``Global fit to Higgs signal strengths and couplings and implications for extended Higgs sectors,''
Phys. Rev. D \textbf{88} (2013), 075008
%doi:10.1103/PhysRevD.88.075008
[arXiv:1306.2941 [hep-ph]].
%313 citations counted in INSPIRE as of 12 Jan 2025
%\cite{Aiko:2022gmz}
\bibitem{Aiko:2022gmz}
M.~Aiko, S.~Kanemura and K.~Sakurai,
%``Radiative corrections to decay branching ratios of the CP-odd Higgs boson in two Higgs doublet models,''
Nucl. Phys. B \textbf{986} (2023), 116047
%doi:10.1016/j.nuclphysb.2022.116047
[arXiv:2207.01032 [hep-ph]].
%8 citations counted in INSPIRE as of 12 Jan 2025

%\cite{Banerjee:2020tqc}
\bibitem{Banerjee:2020tqc}
A.~Banerjee and G.~Bhattacharyya,
%``Probing the Higgs boson through Yukawa force,''
Nucl. Phys. B \textbf{961} (2020), 115261
%doi:10.1016/j.nuclphysb.2020.115261
[arXiv:2006.01164 [hep-ph]].
%10 citations counted in INSPIRE as of 12 Jan 2025

%\cite{ATLAS:2015xst}
\bibitem{ATLAS:2015xst}
G.~Aad \textit{et al.} [ATLAS],
%``Evidence for the Higgs-boson Yukawa coupling to tau leptons with the ATLAS detector,''
JHEP \textbf{04} (2015), 117
%doi:10.1007/JHEP04(2015)117
[arXiv:1501.04943 [hep-ex]].
%442 citations counted in INSPIRE as of 12 Jan 2025
%\cite{ATLAS:2014aga}
\bibitem{ATLAS:2014aga}
G.~Aad \textit{et al.} [ATLAS],
%``Observation and measurement of Higgs boson decays to WW$^*$ with the ATLAS detector,''
Phys. Rev. D \textbf{92} (2015) no.1, 012006
%doi:10.1103/PhysRevD.92.012006
[arXiv:1412.2641 [hep-ex]].
%401 citations counted in INSPIRE as of 12 Jan 2025
%\cite{BhupalDev:2014bir}
\bibitem{BhupalDev:2014bir}
P.~S.~Bhupal Dev and A.~Pilaftsis,
%``Maximally Symmetric Two Higgs Doublet Model with Natural Standard Model Alignment,''
JHEP \textbf{12} (2014), 024
[erratum: JHEP \textbf{11} (2015), 147]
%doi:10.1007/JHEP12(2014)024
[arXiv:1408.3405 [hep-ph]].
%275 citations counted in INSPIRE as of 12 Jan 2025

%\cite{Park:2024apx}
\bibitem{Park:2024apx}
J.~Park [CMS],
%``Searches for flavor-changing neutral currents and charged lepton flavor violation in association with top quarks at ATLAS and CMS,''
PoS \textbf{EPS-HEP2023} (2024), 294
%doi:10.22323/1.449.0294
%0 citations counted in INSPIRE as of 12 Oct 2025

%\cite{Allanach:2023uxz}
\bibitem{Allanach:2023uxz}
B.~Allanach and A.~Mullin,
%``Plan B: new Z' models for b {\textrightarrow} s{\ensuremath{\ell}}$^{+}${\ensuremath{\ell}}$^{−}$ anomalies,''
JHEP \textbf{09} (2023), 173
%doi:10.1007/JHEP09(2023)173
[arXiv:2306.08669 [hep-ph]].
%29 citations counted in INSPIRE as of 12 Oct 2025

%\cite{Thielmann:2023hor}
\bibitem{Thielmann:2023hor}
O.~Thielmann,
%``Search for flavour-changing neutral current interactions in the top-quark Higgs boson sector in multi-lepton final states with the ATLAS detector at the LHC at s = 13 TeV,''
doi:10.25926/BUW/0-167.
%0 citations counted in INSPIRE as of 12 Oct 2025

%\cite{Duy:2024txy}
\bibitem{Duy:2024txy}
N.~T.~Duy, D.~T.~Huong, D.~Van Loi and P.~Van Dong,
%``Flavor-changing phenomenology in a U(1) model,''
Eur. Phys. J. C \textbf{85} (2025) no.9, 1053
%doi:10.1140/epjc/s10052-025-14803-9
[arXiv:2410.15635 [hep-ph]].
%1 citations counted in INSPIRE as of 12 Oct 2025

%\cite{Oliveira:2022vjo}
\bibitem{Oliveira:2022vjo}
V.~Oliveira and C.~A.~d.~S.~Pires,
%``Flavor changing neutral current processes and family discrimination in 3-3-1 models,''
J. Phys. G \textbf{50} (2023) no.11, 115002
%doi:10.1088/1361-6471/acf1b7
[arXiv:2208.00420 [hep-ph]].
%19 citations counted in INSPIRE as of 12 Oct 2025


\bibitem{HVZ-sm}
S.~Alte, M.~K\"onig and M.~Neubert,
%``Exclusive Weak Radiative Higgs Decays in the Standard Model and Beyond,''
JHEP \textbf{12} (2016), 037
%doi:10.1007/JHEP12(2016)037
[arXiv:1609.06310 [hep-ph]].
%36 citations counted in INSPIRE as of 03 Dec 2024

\bibitem{HVZ-zhao}S.-M. Zhao, T.-F. Feng, J.-B. Chen, J.-J. Feng, G.-Z. Ning, H.-B. Zhang, \emph{Phys. Rev.} {\bf D 97} (2018) 095043.

\bibitem{HVZ-sm1}A. L. Kagan, G. Perez, F. Petriello, Y. Soreq, S. Stoynev and J. Zupan, \emph{Phys. Rev. Lett.} {\bf 114} (2015) 101802, arXiv:1406.1722.
\bibitem{HVZ-sm2}G. T. Bodwin, H. S. Chung, J. H. Ee, J. Lee and F. Petriello, \emph{Phys. Rev. D} {\bf 90} (2014) 113010, arXiv:1407.6695.

%\cite{Konig:2015qat}
\bibitem{Konig:2015qat}
M.~K\"onig and M.~Neubert,
%``Exclusive Radiative Higgs Decays as Probes of Light-Quark Yukawa Couplings,''
JHEP \textbf{08} (2015), 012
%doi:10.1007/JHEP08(2015)012
[arXiv:1505.03870 [hep-ph]].
%123 citations counted in INSPIRE as of 14 Nov 2024

%\cite{Konig:2016xhe}
\bibitem{Konig:2016xhe}
M.~K\"onig,
%``Very rare, exclusive, hadronic decays in QCD factorization,''
EPJ Web Conf. \textbf{129} (2016), 00014
%doi:10.1051/epjconf/201612900014
%1 citations counted in INSPIRE as of 05 Dec 2024

%\cite{dEnterria:2023wjq}
\bibitem{dEnterria:2023wjq}
D.~d'Enterria and V.~D.~Le,
%``Rare and exclusive few-body decays of the Higgs, Z, W bosons, and the top quark,''
%doi:10.1088/1361-6471/ad3c59
[arXiv:2312.11211 [hep-ph]].
%11 citations counted in INSPIRE as of 05 Dec 2024

%\cite{CMS:2020ggo}
\bibitem{CMS:2020ggo}
A.~M.~Sirunyan \textit{et al.} [CMS],
%``Search for decays of the 125 GeV Higgs boson into a Z boson and a $\rho$ or $\phi$ meson,''
JHEP \textbf{11} (2020), 039
%doi:10.1007/JHEP11(2020)039
[arXiv:2007.05122 [hep-ex]].
%24 citations counted in INSPIRE as of 05 Dec 2024
%\cite{CMS:2022fsq}

%\cite{Isidori:2013cla}
\bibitem{Isidori:2013cla}
G.~Isidori, A.~V.~Manohar and M.~Trott,
%``Probing the nature of the Higgs-like Boson via $h \to V \mathcal{F}$ decays,''
Phys. Lett. B \textbf{728} (2014), 131-135
%doi:10.1016/j.physletb.2013.11.054
[arXiv:1305.0663 [hep-ph]].
%114 citations counted in INSPIRE as of 15 Mar 2025

\bibitem{QCD1}G. P. Lepage and S. J. Brodsky, \emph{Phys. Lett.} {\bf B 87}, 359 (1979).
\bibitem{QCD2}A. V. Efremov and A. V. Radyushkin, \emph{Phys. Lett.} {\bf B 94}, 245 (1980).
\bibitem{QCD3}V. L. Chernyak and A. R. Zhitnitsky, \emph{Phys. Rep.} {\bf 112}, 173 (1984).

%\cite{CMS:2023gjz}
\bibitem{CMS:2023gjz}
A.~Hayrapetyan \textit{et al.} [CMS],
%``Measurements of inclusive and differential cross sections for the Higgs boson production and decay to four-leptons in proton-proton collisions at $ \sqrt{s} $ = 13 TeV,''
JHEP \textbf{08} (2023), 040
%doi:10.1007/JHEP08(2023)040
[arXiv:2305.07532 [hep-ex]].
%26 citations counted in INSPIRE as of 05 Dec 2024

%\cite{ATLAS:2022qef}
\bibitem{ATLAS:2022qef}
G.~Aad \textit{et al.} [ATLAS],
%``Measurement of the total and differential Higgs boson production cross-sections at $ \sqrt{s} $ = 13 TeV with the ATLAS detector by combining the H \textrightarrow{} ZZ$^{*}$\textrightarrow{} 4\ensuremath{\ell} and H \textrightarrow{} \ensuremath{\gamma}\ensuremath{\gamma} decay channels,''
JHEP \textbf{05} (2023), 028
%doi:10.1007/JHEP05(2023)028
[arXiv:2207.08615 [hep-ex]].
%41 citations counted in INSPIRE as of 05 Dec 2024

%\cite{Modak:2016cdm}
\bibitem{Modak:2016cdm}
T.~Modak, J.~C.~Rom\~ao, S.~Sadhukhan, J.~P.~Silva and R.~Srivastava,
%``Constraining wrong-sign $hbb$ couplings with $h {\rightarrow} \Upsilon \gamma$,''
Phys. Rev. D \textbf{94} (2016) no.7, 075017
%doi:10.1103/PhysRevD.94.075017
[arXiv:1607.07876 [hep-ph]].
%29 citations counted in INSPIRE as of 02 Dec 2024

%\cite{Carlson:2021tes}
\bibitem{Carlson:2021tes}
B.~Carlson, T.~Han and S.~C.~I.~Leung,
%``Higgs boson to charm quark decay in vector boson fusion plus a photon,''
Phys. Rev. D \textbf{104} (2021) no.7, 073006
%doi:10.1103/PhysRevD.104.073006
[arXiv:2105.08738 [hep-ph]].
%10 citations counted in INSPIRE as of 10 Jan 2025

\bibitem{Jia:2024ini}
Y.~Jia, Z.~Mo and J.~Y.~Zhang,
%``Two-loop QCD corrections to Higgs radiative decay to vector quarkonium,''
[arXiv:2408.17448 [hep-ph]].
%0 citations counted in INSPIRE as of 02 Dec 2024

%\cite{Bodwin:2013gca}
\bibitem{Bodwin:2013gca}
G.~T.~Bodwin, F.~Petriello, S.~Stoynev and M.~Velasco,
%``Higgs boson decays to quarkonia and the $H\bar{c}c$  coupling,''
Phys. Rev. D \textbf{88} (2013) no.5, 053003
%doi:10.1103/PhysRevD.88.053003
[arXiv:1306.5770 [hep-ph]].
%201 citations counted in INSPIRE as of 24 Feb 2025

%\cite{CMS:2024hhg}
\bibitem{CMS:2024hhg}
A.~Hayrapetyan \textit{et al.} [CMS],
%``Search for rare decays of the Z and Higgs bosons to a J/$\psi$ or $\psi$(2S) meson and a photon in proton-proton collisions at $\sqrt{s}$ = 13 TeV,''
[arXiv:2411.15000 [hep-ex]].
%1 citations counted in INSPIRE as of 24 Feb 2025

%\cite{Coyle:2019hvs}
\bibitem{Coyle:2019hvs}
N.~M.~Coyle, C.~E.~M.~Wagner and V.~Wei,
%``Bounding the charm Yukawa coupling,''
Phys. Rev. D \textbf{100} (2019) no.7, 073013
%doi:10.1103/PhysRevD.100.073013
[arXiv:1905.09360 [hep-ph]].
%21 citations counted in INSPIRE as of 03 Dec 2024

%\cite{ATLAS:2022rej}
\bibitem{ATLAS:2022rej}
G.~Aad \textit{et al.} [ATLAS],
%``Searches for exclusive Higgs and Z boson decays into a vector quarkonium state and a photon using 139~fb$^{-1}$ of ATLAS $\sqrt{s}=13$~TeV proton\textendash{}proton collision data,''
Eur. Phys. J. C \textbf{83} (2023) no.9, 781
%doi:10.1140/epjc/s10052-023-11869-1
[arXiv:2208.03122 [hep-ex]].
%24 citations counted in INSPIRE as of 02 Dec 2024

%\cite{ATLAS:2023von}
\bibitem{ATLAS:2023von}
[ATLAS],
%``Summary of ATLAS searches for Higgs and vector boson decays to a meson and a photon,''
ATL-PHYS-PUB-2023-004.
%0 citations counted in INSPIRE as of 24 Feb 2025

%\cite{Choi:2021nql}
\bibitem{Choi:2021nql}
S.~Y.~Choi, J.~S.~Lee and J.~Park,
%``Decays of Higgs bosons in the Standard Model and beyond,''
Prog. Part. Nucl. Phys. \textbf{120} (2021), 103880
%doi:10.1016/j.ppnp.2021.103880
[arXiv:2101.12435 [hep-ph]].
%10 citations counted in INSPIRE as of 14 Oct 2024

%\cite{Spira:1995rr}
\bibitem{Spira:1995rr}
M.~Spira, A.~Djouadi, D.~Graudenz and P.~M.~Zerwas,
``Higgs boson production at the LHC,''
Nucl.\ Phys.\ B {\bf 453} (1995) 17
%doi:10.1016/0550-3213(95)00379-7
[hep-ph/9504378].
%%CITATION = doi:10.1016/0550-3213(95)00379-7;%%
%1386 citations counted in INSPIRE as of 23 Mar 2020

%\cite{Djouadi:1991tka}
\bibitem{Djouadi:1991tka}
A.~Djouadi, M.~Spira and P.~M.~Zerwas,
``Production of Higgs bosons in proton colliders: QCD corrections,''
Phys.\ Lett.\ B {\bf 264} (1991) 440.
%doi:10.1016/0370-2693(91)90375-Z.
%%CITATION = doi:10.1016/0370-2693(91)90375-Z;%%
%958 citations counted in INSPIRE as of 23 Mar 2020

%\cite{Baikov:2006ch}
\bibitem{Baikov:2006ch}
P.~A.~Baikov and K.~G.~Chetyrkin,
``Top Quark Mediated Higgs Boson Decay into Hadrons to Order $\alpha_s^5$,''
Phys.\ Rev.\ Lett.\  {\bf 97} (2006) 061803
%doi:10.1103/PhysRevLett.97.061803
[hep-ph/0604194].
%%CITATION = doi:10.1103/PhysRevLett.97.061803;%%
%92 citations counted in INSPIRE as of 23 Mar 2020

%\cite{Chetyrkin:1998mw}
\bibitem{Chetyrkin:1998mw}
K.~G.~Chetyrkin, B.~A.~Kniehl, M.~Steinhauser and W.~A.~Bardeen,
``Effective QCD interactions of CP odd Higgs bosons at three loops,''
Nucl.\ Phys.\ B {\bf 535} (1998) 3
%doi:10.1016/S0550-3213(98)00594-X
[hep-ph/9807241].
%%CITATION = doi:10.1016/S0550-3213(98)00594-X;%%
%65 citations counted in INSPIRE as of 23 Mar 2020

%\cite{Chetyrkin:1997iv}
\bibitem{Chetyrkin:1997iv}
K.~G.~Chetyrkin, B.~A.~Kniehl and M.~Steinhauser,
``Hadronic Higgs decay to order alpha-s**4,''
Phys.\ Rev.\ Lett.\  {\bf 79} (1997) 353
%doi:10.1103/PhysRevLett.79.353
[hep-ph/9705240].
%%CITATION = doi:10.1103/PhysRevLett.79.353;%%
%241 citations counted in INSPIRE as of 23 Mar 2020

%\cite{Muhlleitner:2006wx}
\bibitem{Muhlleitner:2006wx}
M.~Muhlleitner and M.~Spira,
``Higgs Boson Production via Gluon Fusion: Squark Loops at NLO QCD,''
Nucl.\ Phys.\ B {\bf 790} (2008) 1
%doi:10.1016/j.nuclphysb.2007.08.011
[hep-ph/0612254].
%%CITATION = doi:10.1016/j.nuclphysb.2007.08.011;%%
%100 citations counted in INSPIRE as of 24 Mar 2020

%\cite{Bonciani:2007ex}
\bibitem{Bonciani:2007ex}
R.~Bonciani, G.~Degrassi and A.~Vicini,
``Scalar particle contribution to Higgs production via gluon fusion at NLO,''
JHEP {\bf 0711} (2007) 095
%doi:10.1088/1126-6708/2007/11/095
[arXiv:0709.4227 [hep-ph]].
%%CITATION = doi:10.1088/1126-6708/2007/11/095;%%
%124 citations counted in INSPIRE as of 24 Mar 2020

%\cite{Djouadi:1990aj}
\bibitem{Djouadi:1990aj}
A.~Djouadi, M.~Spira, J.~J.~van der Bij and P.~M.~Zerwas,
``QCD corrections to gamma gamma decays of Higgs particles in the intermediate mass range,''
Phys.\ Lett.\ B {\bf 257} (1991) 187.
%doi:10.1016/0370-2693(91)90879-U.
%%CITATION = doi:10.1016/0370-2693(91)90879-U;%%
%138 citations counted in INSPIRE as of 24 Mar 2020

%\cite{Djouadi:1996pb}
\bibitem{Djouadi:1996pb}
A.~Djouadi, V.~Driesen, W.~Hollik and J.~I.~Illana,
``The Coupling of the lightest SUSY Higgs boson to two photons in the decoupling regime,''
Eur.\ Phys.\ J.\ C {\bf 1} (1998) 149
%doi:10.1007/BF01245805
[hep-ph/9612362].
%%CITATION = doi:10.1007/BF01245805;%%
%123 citations counted in INSPIRE as of 24 Mar 2020

%\cite{Spira:2016ztx}
\bibitem{Spira:2016ztx}
M.~Spira,
``Higgs Boson Production and Decay at Hadron Colliders,''
Prog. Part. Nucl. Phys. \textbf{95} (2017), 98-159
%doi:10.1016/j.ppnp.2017.04.001
[arXiv:1612.07651 [hep-ph]].
%54 citations counted in INSPIRE as of 20 Oct 2020

%\cite{Liu:2020nsm}
\bibitem{Liu:2020nsm}
C.~X.~Liu, H.~B.~Zhang, J.~L.~Yang, S.~M.~Zhao, Y.~B.~Liu and T.~F.~Feng,
%``Higgs boson decay $h{\rightarrow} Z\gamma$ and muon magnetic dipole moment in the $\mu\nu$SSM,''
JHEP \textbf{04} (2020), 002
%doi:10.1007/JHEP04(2020)002
[arXiv:2002.04370 [hep-ph]].
%20 citations counted in INSPIRE as of 14 Oct 2024

%\cite{Braaten:1980yq}
\bibitem{Braaten:1980yq}
E.~Braaten and J.~P.~Leveille,
``Higgs Boson Decay and the Running Mass,''
Phys.\ Rev.\ D {\bf 22} (1980) 715.
%doi:10.1103/PhysRevD.22.715.
%%CITATION = doi:10.1103/PhysRevD.22.715;%%
%352 citations counted in INSPIRE as of 20 Mar 2020
%\cite{Kataev:1993be}
\bibitem{Kataev:1993be}
A.~L.~Kataev and V.~T.~Kim,
``The Effects of the QCD corrections to Gamma (H0 $\to$ b anti-b),''
Mod.\ Phys.\ Lett.\ A {\bf 9} (1994) 1309.
%doi:10.1142/S0217732394001131.
%%CITATION = doi:10.1142/S0217732394001131;%%
%89 citations counted in INSPIRE as of 20 Mar 2020

%\cite{Melnikov:1995yp}
\bibitem{Melnikov:1995yp}
K.~Melnikov,
``Two loop O(N(f) alpha-s**2) correction to the decay width of the
Higgs boson to two massive fermions,''
Phys.\ Rev.\ D {\bf 53} (1996) 5020
%doi:10.1103/PhysRevD.53.5020
[hep-ph/9511310].
%%CITATION = doi:10.1103/PhysRevD.53.5020;%%
%22 citations counted in INSPIRE as of 20 Mar 2020

%\cite{Larin:1995sq}
\bibitem{Larin:1995sq}
S.~A.~Larin, T.~van Ritbergen and J.~A.~M.~Vermaseren,
``The Large top quark mass expansion for Higgs boson decays into bottom quarks and into gluons,''
Phys.\ Lett.\ B {\bf 362} (1995) 134
%doi:10.1016/0370-2693(95)01192-S
[hep-ph/9506465].
%%CITATION = doi:10.1016/0370-2693(95)01192-S;%%
%75 citations counted in INSPIRE as of 20 Mar 2020

%\cite{Herren:2017osy}
\bibitem{Herren:2017osy}
F.~Herren and M.~Steinhauser,
``Version 3 of RunDec and CRunDec,''
Comput. Phys. Commun. \textbf{224} (2018), 333-345
%doi:10.1016/j.cpc.2017.11.014
[arXiv:1703.03751 [hep-ph]].
%241 citations counted in INSPIRE as of 14 Oct 2024

%\cite{Chetyrkin:2015mxa}
\bibitem{Chetyrkin:2015mxa}
K.~G.~Chetyrkin, J.~H.~K\"uhn, M.~Steinhauser and C.~Sturm,
%``Massive Tadpoles: Techniques and Applications,''
Nucl. Part. Phys. Proc. \textbf{261-262} (2015), 19-30
%doi:10.1016/j.nuclphysbps.2015.03.003
[arXiv:1502.00509 [hep-ph]].
%7 citations counted in INSPIRE as of 14 Oct 2024

%\cite{Fleischer:1980ub}
\bibitem{Fleischer:1980ub}
J.~Fleischer and F.~Jegerlehner,
``Radiative Corrections to Higgs Decays in the Extended Weinberg-Salam Model,''
Phys.\ Rev.\ D {\bf 23} (1981) 2001.
%doi:10.1103/PhysRevD.23.2001.
%%CITATION = doi:10.1103/PhysRevD.23.2001;%%
%293 citations counted in INSPIRE as of 20 Mar 2020

%\cite{Bardin:1990zj}
\bibitem{Bardin:1990zj}
D.~Y.~Bardin, B.~M.~Vilensky and P.~K.~Khristova,
%``Calculation of the Higgs boson decay width into fermion pairs,''
Sov. J. Nucl. Phys. \textbf{53} (1991), 152-158
JINR-P2-90-287.
%118 citations counted in INSPIRE as of 13 Dec 2025

%\cite{Kniehl:1991ze}
\bibitem{Kniehl:1991ze}
B.~A.~Kniehl,
``Radiative corrections for $H \to$ f anti-f ($\gamma$) in the standard model,''
Nucl.\ Phys.\ B {\bf 376} (1992) 3.
%doi:10.1016/0550-3213(92)90065-J.
%%CITATION = doi:10.1016/0550-3213(92)90065-J;%%
%135 citations counted in INSPIRE as of 20 Mar 2020

%\cite{Djouadi:1991uf}
\bibitem{Djouadi:1991uf}
A.~Djouadi, D.~Haidt, B.~A.~Kniehl, P.~M.~Zerwas and B.~Mele,
``Higgs in the standard model,''
In *Munich/Annecy/Hamburg 1991, Proceedings, e+ e- collisions at 500-GeV, pt. A* 11-30.
%2 citations counted in INSPIRE as of 20 Mar 2020

%\cite{Kwiatkowski:1994cu}
\bibitem{Kwiatkowski:1994cu}
A.~Kwiatkowski and M.~Steinhauser,
``Corrections of order ${\cal O}(G_F \alpha_s m_t^2)$ to the
Higgs decay rate $\Gamma (H \to b \bar b)$,''
Phys.\ Lett.\ B {\bf 338} (1994) 66
Erratum: [Phys.\ Lett.\ B {\bf 342} (1995) 455]
%doi:10.1016/0370-2693(94)01527-J, 10.1016/0370-2693(94)91345-5
[hep-ph/9405308].
%%CITATION = doi:10.1016/0370-2693(94)01527-J, 10.1016/0370-2693(94)91345-5;%%
%41 citations counted in INSPIRE as of 20 Mar 2020

%\cite{Chetyrkin:1996wr}
\bibitem{Chetyrkin:1996wr}
K.~G.~Chetyrkin, B.~A.~Kniehl and M.~Steinhauser,
``Virtual top quark effects on the H $\to$ b anti-b decay
at next-to-leading order in QCD,''
Phys.\ Rev.\ Lett.\  {\bf 78} (1997) 594
%doi:10.1103/PhysRevLett.78.594
[hep-ph/9610456].
%%CITATION = doi:10.1103/PhysRevLett.78.594;%%
%52 citations counted in INSPIRE as of 20 Mar 2020

%\cite{Ge:2024rdr}
\bibitem{Ge:2024rdr}
Z.~f.~Ge, F.~Y.~Niu and J.~L.~Yang,
%``The origin of the 95~GeV excess in the flavor-dependent $U(1)_X$ model,''
Eur. Phys. J. C \textbf{84} (2024) no.5, 548
%doi:10.1140/epjc/s10052-024-12872-w
[arXiv:2405.07243 [hep-ph]].
%2 citations counted in INSPIRE as of 14 Oct 2024

%\cite{Yang:2018fvw}
\bibitem{Yang:2018fvw}
J.~L.~Yang, T.~F.~Feng, S.~M.~Zhao, R.~F.~Zhu, X.~Y.~Yang and H.~B.~Zhang,
%``Two loop electroweak corrections to $\bar B\rightarrow X_s\gamma$ and $B_s^0\rightarrow \mu^+\mu^-$ in the B-LSSM,''
Eur. Phys. J. C \textbf{78} (2018) no.9, 714
%doi:10.1140/epjc/s10052-018-6174-5
[arXiv:1803.09904 [hep-ph]].
%40 citations counted in INSPIRE as of 15 Oct 2025

%\cite{Yang:2018utw}
\bibitem{Yang:2018utw}
J.~L.~Yang, T.~F.~Feng, H.~B.~Zhang, G.~Z.~Ning and X.~Y.~Yang,
%``Top quark decays with flavor violation in the B-LSSM,''
Eur. Phys. J. C \textbf{78} (2018) no.6, 438
%doi:10.1140/epjc/s10052-018-5919-5
[arXiv:1806.01476 [hep-ph]].
%21 citations counted in INSPIRE as of 15 Oct 2025

%\cite{Hisano:1995cp}
\bibitem{Hisano:1995cp}
J.~Hisano, T.~Moroi, K.~Tobe and M.~Yamaguchi,
%``Lepton flavor violation via right-handed neutrino Yukawa couplings in supersymmetric standard model,''
Phys. Rev. D \textbf{53} (1996), 2442-2459
%doi:10.1103/PhysRevD.53.2442
[arXiv:hep-ph/9510309 [hep-ph]].
%809 citations counted in INSPIRE as of 15 Oct 2025

%\cite{Porod:2003um}
\bibitem{Porod:2003um}
W.~Porod,
%``SPheno, a program for calculating supersymmetric spectra, SUSY particle decays and SUSY particle production at e+ e- colliders,''
Comput. Phys. Commun. \textbf{153} (2003), 275-315
%doi:10.1016/S0010-4655(03)00222-4
[arXiv:hep-ph/0301101 [hep-ph]].
%1327 citations counted in INSPIRE as of 23 Oct 2025

%\cite{CMS:2022fsq}
\bibitem{CMS:2022fsq}
A.~Tumasyan \textit{et al.} \mbox{[CMS]},
%``Search for Higgs boson decays into Z and J/\ensuremath{\psi} and for Higgs and Z boson decays into J/\ensuremath{\psi} or Y pairs in pp collisions at s=13~TeV,''
Phys. Lett. B \textbf{842} (2023), 137534
%doi:10.1016/j.physletb.2022.137534
[arXiv:2206.03525 [hep-ex]].
%20 citations counted in INSPIRE as of 02 Dec 2024

\bibitem{experiment2}A.M.Sirunyan et al.(CMS Collaboration), \emph{JHEP} {\bf 11} (2020) 039, arXiv:2007.05122.

%\bibitem{100TeV}M. L. Mangano, Physics at the FCC-hh: a 100 TeV pp collider, CERN-2017-003-M (CERN, Geneva, 2017).

%\bibitem{HL-LHC}G. Apollinari, I. B\'{e}jar Alonso, O. Br\"{u}ning, M. Lamont, L. Rossi, High-Luminosity Large Hadron Collider (HL-LHC): Preliminary Design Report, CERN-2015-005 (CERN, Geneva, 2015).

%\cite{Liu:2020mev}
\bibitem{Liu:2020mev}
C.~X.~Liu, H.~B.~Zhang, J.~L.~Yang, S.~M.~Zhao and T.~F.~Feng,
%``Lightest Higgs boson decays $h\rightarrow MZ$ in the $\mu$ from $\nu$ supersymmetric standard model,''
JHEP \textbf{05} (2023), 134
%doi:10.1007/JHEP05(2023)134
[arXiv:2012.14786 [hep-ph]].
%2 citations counted in INSPIRE as of 21 Oct 2024

%\cite{Jegerlehner:2011mw}
\bibitem{Jegerlehner:2011mw}
F.~Jegerlehner,
%``Electroweak effective couplings for future precision experiments,''
Nuovo Cim. C \textbf{034S1} (2011), 31-40
%doi:10.1393/ncc/i2011-11011-0
[arXiv:1107.4683 [hep-ph]].
%103 citations counted in INSPIRE as of 21 Oct 2024

%\bibitem{QCD4}M. Spira, A. Djouadi and P. M Zerwas, \emph{Phys. Lett. } {\bf B 276}, 350 (1992).
\bibitem{QCD5}Yuval Grossman, Matthias K$\ddot{o}$nig and Matthias Neubert, \emph{JHEP} {\bf 1504} (2015) 101.

\bibitem{phi-function3}M. Beneke, G. Buchalla, M. Neubert, and C. T. Sachrajda, \emph{Nucl. Phys.} {\bf B 591}, 313 (2000).

\bibitem{signal}A. Arbey, A. Deandrea, F. Mahmoudi, and A. Tarhini, \emph{Phys. Rev. D} {\bf 87}, 115020 (2013).

%\cite{Dong:2022bkd}
\bibitem{Dong:2022bkd}
H.~Dong, P.~Sun and B.~Yan,
%``Probing the H\ensuremath{\gamma}\ensuremath{\gamma} coupling via Higgs boson exclusive decay into quarkonia plus a photon at the HL-LHC,''
Phys. Rev. D \textbf{106} (2022) no.9, 095013
%doi:10.1103/PhysRevD.106.095013
[arXiv:2208.05153 [hep-ph]].
%4 citations counted in INSPIRE as of 05 Dec 2024

%\cite{Bodwin:2016edd}
\bibitem{Bodwin:2016edd}
G.~T.~Bodwin, H.~S.~Chung, J.~H.~Ee and J.~Lee,
%``New approach to the resummation of logarithms in Higgs-boson decays to a vector quarkonium plus a photon,''
Phys. Rev. D \textbf{95} (2017) no.5, 054018
%doi:10.1103/PhysRevD.95.054018
[arXiv:1603.06793 [hep-ph]].
%33 citations counted in INSPIRE as of 02 Dec 2024

%\cite{Brambilla:2019fmu}
\bibitem{Brambilla:2019fmu}
N.~Brambilla, H.~S.~Chung, W.~K.~Lai, V.~Shtabovenko and A.~Vairo,
%``Order $v^4$ corrections to Higgs boson decay into $J/\psi + \gamma$,''
Phys. Rev. D \textbf{100} (2019) no.5, 054038
%doi:10.1103/PhysRevD.100.054038
[arXiv:1907.06473 [hep-ph]].
%20 citations counted in INSPIRE as of 15 Nov 2024

%\cite{Cao:2025zwn}
\bibitem{Cao:2025zwn}
Z.~Cao, Z.~J.~Yang, J.~L.~Yang and T.~F.~Feng,
%``Probing the light charged Higgs boson, pseudoscalar Higgs boson, and $Z^\prime$ boson in the $U(1)_F$ model at the LHC,''
[arXiv:2510.22537 [hep-ph]].
%0 citations counted in INSPIRE as of 28 Oct 2025

%\cite{Sang:2024vqk}
\bibitem{Sang:2024vqk}
W.~L.~Sang, F.~Feng and Y.~Jia,
%``Next-to-leading-order electroweak correction to H\textrightarrow{}Z0\ensuremath{\gamma},''
Phys. Rev. D \textbf{110} (2024) no.5, L051302
%doi:10.1103/PhysRevD.110.L051302
[arXiv:2405.03464 [hep-ph]].
%1 citations counted in INSPIRE as of 03 Dec 2024


\end{thebibliography}
\end{document}